\documentclass[11pt,a4paper]{article}
\usepackage[textwidth=16cm,textheight=24cm]{geometry}

\usepackage{graphicx}
\usepackage{rotating}
\usepackage{tikz}
\usetikzlibrary{shapes,arrows}
\usepackage{array}
\usepackage{tabularx}
\usepackage{graphics,amsfonts,amssymb,amsmath,latexsym}
\usepackage{amssymb}
\usepackage{epstopdf}
\usepackage{subfigure}
 \PassOptionsToPackage{round, sort, numbers, compress}{natbib}

 \usepackage{multirow}
\usepackage{relsize}
\usepackage{upgreek}

\newcommand{\rmax} { r_\mathrm{max} }
\newcommand{\rmin} { r_\mathrm{min} }

\newcommand{\sigmad} {  \sigma_\mathrm{dev}  }
\newcommand{\sigmadkstar} { \sigma_\mathrm{dev}^* }

\newcommand{\pkstar} { P^* }

\newcommand{\epsd} { \varepsilon_\mathrm{dev} }
\newcommand{\Qd} { Q_\mathrm{dev} }

\newcommand{\sd} { s_\mathrm{dev} }
\newcommand{\Fdev} { F_\mathrm{dev} }

\newcommand{\Fv} { F_\mathrm{v} }

\newcommand{\sdmax} { s_\mathrm{dev}^{\mathrm{max}} }

\newcommand{\Fdmax} { F_\mathrm{dev}^{\mathrm{max}} }

\newcommand{\betaF} { {\beta_F}}

\newcommand{\coord} { {\mathrm\alpha\alpha}}

\newcommand{\nuzero} {\nu_{\mathrm{0}}}

\newcommand{\kstar} {k^*}

\newcommand{\epsiso} { \varepsilon_\mathrm{v} }

\newcommand{\epsalpha} { \varepsilon_\mathrm{\alpha\alpha} }
\newcommand{\epsrate} { \dot\varepsilon_\mathrm{\alpha\alpha} }

\newcommand{\epsisodot} { \dot\varepsilon_\mathrm{v} }
\newcommand{\epsddot} { \dot\varepsilon_\mathrm{dev} }

\newcommand{\strnrate} {{\mathrm{\dot{\boldsymbol{\mathrm E}}}}}
\newcommand{\strain} {{{{\mathbf{ E}}}}}

\newcommand{\Goct} { G^\mathrm{oct} }

\newcommand{\CAo} {a_\mathrm{I}}
\newcommand{\CAt} {a_\mathrm{II}}
\newcommand{\CGo} {g_\mathrm{I}}

\usepackage{bm}
\usepackage{color}
\usepackage{tikz}

\begin{document}

\begin{center}
\vspace*{2cm}

{\LARGE 
Macroscopic model with anisotropy based on micro-macro informations} \\

\vspace{.5cm}
{\large 

 N.\ Kumar, S.\ Luding, and  V.\ Magnanimo  

\vspace{.5cm}
       Multi Scale Mechanics (MSM), Faculty of Engineering Technology, MESA+,\\
       University of Twente, P.O. Box 217, 7500 AE Enschede, The Netherlands\\
\today
}
\end{center}

 \begin{abstract}
Physical experiments can characterize the elastic response of granular materials in terms of macroscopic state-variables, namely volume (packing) fraction and stress, while the microstructure is not accessible and thus neglected. Here, by means of numerical simulations, we analyze dense, frictionless, granular assemblies with the final goal to relate the elastic moduli to the fabric state, i.e., to micro-structural averaged contact network features as contact number density and anisotropy.

The particle samples are first isotropically compressed and later quasi-statically sheared under constant volume (undrained conditions). 
From various static, relaxed configurations at different shear strains, now infinitesimal strain steps are applied to ``measure'' the effective elastic response; we quantify the strain needed so that plasticity in the sample develops as soon as contact and structure rearrangements happen. Because of the anisotropy induced by shear, volumetric and deviatoric stresses and strains are cross-coupled via a single anisotropy modulus, which is proportional to the product of deviatoric fabric and bulk modulus (i.e. the isotropic fabric). Interestingly, the shear modulus of the material depends also on the actual stress state, along with the contact configuration anisotropy.

Finally, a constitutive model based on incremental evolution equations for stress and fabric is introduced. By using the previously measured dependence of the stiffness tensor (elastic moduli) on the microstructure, the theory is able to predict with good agreement the evolution of pressure, shear stress and deviatoric fabric (anisotropy) for an independent undrained cyclic shear test, including the response to reversal of strain.
 \end{abstract}


\section{Introduction}

Granular materials behave differently from usual solids or fluids and show 
peculiar mechanical properties 
like dilatancy, history dependence, ratcheting and anisotropy 
\cite{imole2013hydrostatic, tejchman2007effect, goncu2010constitutive, 
goncu2013effect, kumar2013effect, kumar2013evolution, sun2011constitutive, nicot2007micromechanical, shaebani2012unilateral, guo2013signature, zhao2013unique}. 
The behavior of these materials is highly non-linear and involves plasticity also for very small 
strain due to rearrangements of the elementary particles \cite{goddard1990nonlinear, darve1995yield, belyaeva1993nonlinear}. The concept of an initial purely elastic regime (small strain)
for granular assemblies is an issue still under debate in the mechanical and geotechnical communities. 
On the other hand, approaches that neglect the effect of elastic stored energy, 
i.e.,\ all the work done by the internal forces is dissipated, are also questionable.  
Features visible in experiments, like wave propagation, can hardly be described without considering an elastic regime.  
In a general picture, both the deformations at contact
and the irrecoverable rearrangements of the grains sum up to the total strain.  
The former represents the elastic, reversible contribution to the behavior of the material.
That is, for very small strain the response of a finite granular 
system in static equilibrium can be assumed to be linearly elastic 
\cite{nicot2007basic, shaebani2012unilateral, farago2000fluctuation, kuwano2002applicability}, as long as no irreversible rearrangements take place. 

Despite these arguments and the long-standing debate, 
basic features of the physics of granular elasticity are currently unresolved, 
like the determination of a proper set of state variables to describe the effective moduli.
Physical experiments carried out on sand and glass beads show that wave propagation 
in the aggregate depends upon the 
stress state and the volume fraction 
\cite{ezaoui2009experimental, khidas2012probing, wang2008mechanisms, jiang1997inherent, yimsiri2010analysis, kuwano2002applicability}. 
Recent works \cite{khidas2012probing, Ragione2012contact, guo2013signature, zhao2013unique, agnolin2007internal_1} 
show that along 
with the macroscopic properties (stress and volume fraction)
\cite{khidas2012probing, ezaoui2009experimental, zhang2010statistical}, 
also the fabric \cite{oda1972initial, luding2005anisotropy, shaebani2012unilateral, zhao2013unique,chung2013how} 
plays a crucial role, as it characterizes, on average, the geometric arrangement of contacts. 
Due to preparation and loading path, the microstructure of granular aggregates is often far from isotropic and this 
is at the origin of interesting features in those materials.
The mechanical behavior of anisotropic soils is a topic of current interest for both experimental and theoretical investigations. 
As one example, extensive experimental work of anisotropy has been carried out on laboratory-prepared (by careful `raining' or bedding) sand specimens \cite{wu1998rational, yoshhimine1998effects}. 
These and other studies show that the sample deformation characteristics 
depends highly on the orientation of the bedding plane with respect to the principal stress and strain axes
\cite{ezaoui2009experimental, oda1972initial, yoshhimine1998effects, yimsiri2010analysis, yang2008quantifying, Ragione2012contact}
On the other hand, 
when the material is sheared, anisotropy in the 
contact network develops, as related to the opening and closing of contacts, 
restructuring, and the creation and destruction of force-chains, affecting the material response.
\cite{belloti1996anisotropy, thornton2010quasistatic, zhao2013unique, kuwano2002applicability, wang2008mechanisms}.

Most standard constitutive models, involving elasticity and/or plasticity
have been applied to describe the incremental behavior of (an)isotropic granular solids - sometimes with success, 
but typically only in a limited range of parameters. 
In the majority of the models, the stress increment is related to the actual stress state of the granular system 
and its density. This is the case for hypoplasticity \cite{goddard2010parametric, khidas2012probing}, 
where a single non-linear tensorial equation relates the Jaumann stress-rate with strain-rate and stress tensors. 
Only few theories after the pioneer work by Cowin \cite{cowin1985relationship}, consider explicitly the influence of the micro-mechanic structure 
on the elastic stiffness, plastic flow-rule or noncoaxiality of stress and strain, see \cite{tejchman2007effect, sun2011constitutive, nassar2000micromechanically, dafalias2004simple, muhlhaus2010influence, chang1993micromechanical, chang1995estimates}
and references therein. 
The evolution of microstructure due to deformations is an essential part of a constitutive model for granular matter because it stores
the information how different paths
have affected the mechanical state of the system. In this
sense, fabric is a tensorial history variable. 
When included in the formulation, the effect of structure is often described by a fixed fabric tensor
normal to the bedding plane of deposited sands 
\cite{tejchman2007effect, dafalias2004simple, wu1998rational, li2002constitutive}. 
Recently Li \& Dafalias \cite{li2012anisotropic} have proposed a new framework 
(rather than a specific constitutive model)
by reconsidering the classical steady state theory by Roscoe \textit{et al.\ }\cite{roscoe1958yielding}, 
with a fabric tensor evolving towards a properly defined steady state value. 
This is supported by experimental \cite{yoshhimine1998effects} and extensive numerical works \cite{guo2013signature, imole2013hydrostatic, luding2005anisotropy, zhao2013unique, thornton2010quasistatic}.
In a similar fashion, the anisotropy model proposed in \cite{luding2011local, magnanimo2011local} postulates the split of isotropic and deviatoric stress, strain and fabric and 
includes the microstructure as a variable,
whose behavior is described by an evolution equation independent of stress. 
Refs.\ \cite{imole2013hydrostatic, kumar2013effect} predicts uniaxial simulation results under this assumption 
(independent evolution of stress and structure), where the simplified model well captures the qualitative behavior.

In this work we use the Discrete Element Method (DEM)
to study granular assemblies made of polydisperse frictionless particles and focus on their elastic behavior. 
By isolating elasticity we aim to distinguish the kinematics at the microscale that lead to either
macroscopic elasticity or plasticity. 
We analyze the role of microstructure, stress state
and volume fraction on the evolution of the elastic moduli, with the goal to
characterize them in terms of a unique, limited set of variables. 
In order to calculate the stiffness tensor, we apply small-strain probes to various equilibrium states 
along a volume conserving (undrained) shear deformation path.
In the case of a finite assembly of particles, in simulations,
an elastic regime can always be detected and the elastic stiffnesses can be measured 
by means of an actual, very small, strain perturbation \cite{magnanimo2008characterizing}. 
The purpose is to improve the understanding of elasticity in particle systems 
and to guide further developments for new constitutive
models.
As an example, the relation between moduli and fabric here is used in the anisotropic 
constitutive model, as proposed in \cite{luding2011local, magnanimo2011local}, 
to predict the macroscopic behavior during a more general deformation path, involving also strain reversal. 

This paper is organized as follows:
The simulation method and parameters used 
and the averaging definitions for scalar and tensorial quantities
are given in section\ \ref{sec:simmeth}. 
The preparation test procedures, and the results from the deviatoric simulation are explained in section\ \ref{sec:main}. 
Section\ \ref{sec:perturb} is devoted to the measurement of elastic moduli by means of 
small isotropic and deviatoric perturbations. 
There we present the evolution of the moduli with strain and link them to fabric and stress. 
Finally, section\ \ref{sec:predict} is devoted to theory, where we relate the evolution
of the microstructural anisotropy to that of stress and strain, as proposed in Refs.\ \cite {luding2011local,magnanimo2011local}. This displays the predictive
quality of the model, calibrated only for isochoric, uni-directional shear, when applied to an independent, cyclic shear test. 

\section{Numerical simulation}
\label{sec:simmeth}
The Discrete Element Method (DEM) \cite {alonsomarroquin2005role,luding2005anisotropy, imole2013hydrostatic} helps to study the 
deformation behavior of particle systems. At the basis of DEM are laws that relate the interaction force to the overlap (relative deformation) 
of two particles. Neglecting tangential forces, if all normal forces ${\mathbf{f}}_i$ 
acting on particle $i$, from all sources, are known, the problem is reduced to the integration of Newton's equations of motion for the translational degrees of freedom:

\begin{equation}
        \label{eq:force}
        \frac{\mathrm d{}}{\mathrm{d}t}\left(m_i {\bf v}_i \right) = \mathbf{f}_i +m_i\mathbf{g},
\end{equation}
with the mass $m_i$ of particle $i$, its position $\mathbf{r}_i$, velocity ${\bf v}_i$ ($ = \dot{\mathbf{r}}_{i}$) and the resultant force $\mathbf{f}_i = \sum_{c}{\mathbf{f}_i}^c$ acting 
on it due to contacts with other particles or with the walls, and the acceleration due to gravity, $\mathbf{g}$ (which is neglected in this study).
The force on particle $i$, from particle $j$, at contact $c$, has normal and tangential components, but the latter are disregarded in this study to focus on frictionless packings.

For the sake of simplicity, the linear visco-elastic contact model for the normal component of force
is used,
\begin{equation}
        \label{eq:contactmodel}
        f^{n} = k \delta + \gamma \dot\delta,
\end{equation}
where $k$ is the spring stiffness, $\gamma$ is the contact viscosity parameter, $\delta = \left(d_i + d_j \right)/2 - \left( \mathbf{r}_i - \mathbf{r}_j \right) \cdot \hat{\mathbf{n}}$ is the overlap
between two interacting species $i$ and $j$ with diameters $d_i$ and $d_j$,
$\hat{\mathbf{n}} = \left( \mathbf{r}_i - \mathbf{r}_j \right)/\left|\left( \mathbf{r}_i - \mathbf{r}_j \right)\right|$
and $\dot\delta$ is the relative velocity in the normal direction. 
In order to reduce dynamical effects and shorten relaxation times, an artificial viscous 
background dissipation force ${\bf f}_b = -\gamma_b {\bf v}_{i}$ proportional to the moving  velocity ${\bf v}_i$ of particle $i$ is added, resembling the damping due to a background medium, as e.g. a fluid.

The standard simulation parameters are, $N = 9261 (=21^3)$ particles with average radius $\langle r \rangle= 1$ [mm], density $\rho= 2000$ [kg/m\textsuperscript{3}], elastic stiffness 
$k = 10^8$ [kg/s\textsuperscript{2}], particle damping coefficient $\gamma= 1$ [kg/s], background dissipation $\gamma_b= 0.1$ [kg/s]. 
Note that the polydispersity of the 
system is quantified by the width $\left(w=\rmax/\rmin=3 \right)$ of a uniform size distribution \cite{goncu2010constitutive}, 
where  $\rmax$ and $\rmin$ are the radii  of the biggest and smallest particles respectively.  
 
The average time time scale is determined when two averaged size particle (with $r_\mathrm{avg} = \left\langle r \right \rangle = 1$) 
with mass $m_\mathrm{avg} = \rho\left( 4\pi r_\mathrm{avg}^3/3\right) = 8.377$ [$\mu$g] interact, 
and is given as 
$t_{c,\mathrm{avg}}= \pi / \sqrt {k/m'_\mathrm{avg} - (\gamma/\left(2m'_\mathrm{avg}\right))^2}$ =$0.6431$ [$\mu$s], where $m'_\mathrm{avg}=m_\mathrm{avg}/2$ is the reduced mass,
with restitution coefficient \\
$e_\mathrm{avg} =  \exp \left(-\gamma t_{c,\mathrm{avg}}/\left(2m'_\mathrm{avg}\right) \right)$ = $0.926$.
The fastest response time scale in the system is determined when two smallest particle with mass $m_\mathrm{small} = \rho\left( 4\pi \rmin^3/3\right) = 1.047$ [$\mu$g] interact, 
and is given as $t_{c,\mathrm{small}}= \pi / \sqrt {k/m_\mathrm{small}' - (\gamma/\left(2m_\mathrm{small}'\right))^2} =0.2279$ [$\mu$s], where $m_\mathrm{small}'=m_\mathrm{small}/2$ is the reduced mass,
with restitution coefficient $e_\mathrm{small} =$ $ \exp \left(-\gamma t_{c,\mathrm{small}}/\left(2m'_\mathrm{small}\right) \right) = 0.804$. 

\subsection{Coordination number and fraction of rattlers} 
\label{subsec:coord}
In order to link the macroscopic load carried by the sample with the active 
microscopic contact network, all particles that do not contribute 
to the force network are excluded. Frictionless particles with less than 4 contacts are thus `rattlers', 
since they cannot be mechanically stable and hence do not contribute to the 
contact or force network \cite{imole2013hydrostatic, goncu2010constitutive, kumar2013effect}. 
The classical definition of coordination number is $C = M/N$, 
where $M$ is the total number of contacts and $N = 9261$ is the total number of particles. 
The corrected coordination number is $C^* = {M_4}/{N_4},$ where, $M_4$ is the total number of contacts of 
the $N_4$ particles with at least 4 contacts. 
Moreover, we introduce here the reduced number of contacts $M_4^p$, where contacts related to rattlers are excluded twice, as they do not contribute to the stability 
of both the rattler and the particle in contact with it. 
Hence, $M_4^p = M_4 - M_1-M_2 - M_3 = M - 2\left( M_1 + M_2 + M_3 \right)$, where $M_1$, $M_2$ and $M_3$ 
are total number of contacts of particles with only 1, 2 and 3 contacts respectively. This leads to a 
modification in the definition of the corrected coordination number is $C^*_p = {M_4^p}/{N_4}$.
The fraction of rattlers  
is $\phi_r = \left(N-N_4\right)/N$, hence, $C = C^* \left( 1-\phi_r\right)$. 
The total volume of particles is $\sum_{\mathcal{P}=1}^N V_{\mathcal{P}} = 4 \pi N \langle r^3 \rangle$, 
where $\langle r^3 \rangle /3$ is the third moment of the size distribution \cite{goncu2010constitutive, kumar2013effect}  and  
volume fraction is defined as $\nu = \left(1/V\right)\sum_{\mathcal{P}=1}^N V_{\mathcal{P}}$, where $V$ is the volume of the periodic system. 

\subsection{Macroscopic (tensorial) quantities} 
\label{subsec:tensorial}
 Here, we focus on defining averaged tensorial macroscopic quantities 
 -- including strain-, stress- and fabric (structure) tensors -- that provide information about the state of the packing
and reveal interesting bulk features.

By speaking about the strain-rate tensor $\strnrate$, we refer to the external strain that we apply to the sample. 
The isotropic part of the infinitesimal strain tensor $\epsiso$ \cite{imole2013hydrostatic, goncu2010constitutive, kumar2013effect} is defined as:
\begin{equation}
\label{eq:straineq}
\delta\epsiso = -{\epsisodot}\mathrm {dt}= -\frac{\delta\varepsilon_{xx} + \delta\varepsilon_{yy} + \delta\varepsilon_{zz}}{3} = -\frac{1}{3}\mathrm{tr}(\delta\strain) =-\frac{1}{3}\mathrm{tr}(\strnrate)\mathrm {dt},
\end{equation}
where $\epsalpha$= $\epsrate \mathrm {dt}$ with $\coord$ = $xx$, $yy$ and $zz$ are the diagonal components 
of the tensor in the Cartesian $x-y-z$ reference system. The trace integral of $3\epsiso$ is
denoted as the volumetric strain $\varepsilon_{v}$, the true or logarithmic strain, i.e.,\ the volume change of the system, relative to the initial reference volume, $V_0$. 

On the other hand, from DEM simulations, one can measure the `static' stress in the system \cite{christoffersen1981micromechanical} as 
\begin{equation}
\pmb{\sigma}=\left({1}/{V}\right)\sum_{c\in V}\mathbf{l}^{c}\otimes\mathbf{f}^{c}, 
\end{equation}
average over all the contacts in the volume $V$ of the dyadic products between the contact force $\mathbf{f}^{c}$
and the branch vector $\mathbf{l}^{c}$, where the contribution of the kinetic fluctuation energy has been neglected 
\cite{luding2005anisotropy, imole2013hydrostatic}. 
The isotropic component of the stress is the pressure $P = \mathrm{tr}(\pmb{\sigma})/3$. 

In order to characterize the geometry/structure of the static aggregate at microscopic level, 
we will measure the fabric tensor, defined as 
\begin{equation}
\mathbf{F}=\frac{1}{V}\sum_{{\mathcal{P}}\in V}V^{{\mathcal{P}}}\sum_{c\in {\mathcal{P}}}\mathbf{n}^{c}\otimes\mathbf{n}^{c},
\label{eq:fabriceq}
\end{equation}
where $V^{\mathcal{P}}$ is the volume relative to particle ${\mathcal{P}}$, which lies inside the
averaging volume $V$, and $\mathbf{n}^{c}$ is the normal unit branch-vector
pointing from center of particle ${\mathcal{P}}$ to contact $c$ \cite{luding2005anisotropy, kumar2013evolution, zhang2010statistical}. 
We want to highlight that a different, convention for the fabric tensor involves only the orientation of contacts as follows \cite{satake1982fabric, oda1972initial, zhao2013unique}:
 \begin{equation}
\label{eq:scaledFabricdefn}
\mathbf{F}^o =\frac{1}{N_c} \sum_{c \in N_c} \mathbf{n}^{c}\otimes\mathbf{n}^{c} \,
\end{equation}
where $N_c$ is the total number of contacts in the system.
An approximated relationship between Eqs.\ (\ref{eq:fabriceq}) and (\ref{eq:scaledFabricdefn}) can be derived as:
\begin{equation}
\label{eq:fabricrelation}
\mathbf{F}^o \approx  \frac{3\mathbf{F}}{\mathrm{tr}({\mathbf F})}, 
\end{equation}
with $\mathrm{tr}({\mathbf F}^o)=1$. This relation is exactly equal for monodisperse assemblies but largely deviates for assemblies with high polydispersity (see further discussion in section \ref{sec:main}). 
The difference also becomes more significant when the jamming volume fraction \cite{makse2000packing, wang2013regime} is approached.
%
In the following, when not explicitly stated, we will refer to Eq.\ (\ref{eq:fabriceq}), since we
combine the effects of volume fraction and number/orientation of contacts, both relevant quantities when the 
elastic moduli are considered \cite{goncu2010constitutive}.

In a large volume with a given distribution of particle radii, the relation between the 
isotropic fabric, i.e.,\ the trace of $\mathbf{F}$, is proportional to the volume fraction $\nu$ and the coordination number $C$ Refs.\ \cite{imole2013hydrostatic, goncu2010constitutive, kumar2013effect} as 
\begin{equation}
\Fv = \mathrm{tr}({\mathbf F})=g_3 \nu C = g_3 \nu C^* \left( 1-\phi_r\right), 
\label{eq:Fveqn}
\end{equation}
where $C$, $C^*$ and $\phi_r$ have been introduced in previous section \ref{subsec:coord}
and $g_3\approx 1.22$ for polydispersity $w=3$, being only a weighted, non-dimensional moments of the size distribution \cite{goncu2010constitutive, shaebani2012influence, kumar2013effect}.


\subsection{Isotropic and Deviatoric parts}
We choose here to describe each symmetric second order tensor $\mathbf{Q}$, in terms of its isotropic part (first invariant)
and the second
$$J_2=\frac{1}{2}\left[(Q^{D}_{1})^{2}+(Q^{D}_{2})^{2}+(Q^{D}_{3})^{2}\right]$$
and third
$$J_3=\text{det}(\mathbf{Q}^{D})=Q^{D}_{1}Q^{D}_{2}Q^{D}_{3},$$
invariants of the deviator,
with $Q^{D}_{1}, Q^{D}_{2}$ and $Q^{D}_{3}$ eigenvalues of the deviatoric tensor $\mathbf{Q}^{D}=\mathbf{Q}-(\text{tr}(\mathbf{Q})/3)\mathbf{I}$.
We use the following definition (of the Euclidean or Frobenious norm) to
quantify with a single scalar the magnitude of the deviatoric part \cite{kumar2013evolution, kumar2013effect}
of $\mathbf{Q}$:
\begin{equation}
 \Qd = \mathrm{Fsgn}\left(\mathbf{Q} \right)\sqrt{2J_{2}}=\mathrm{Fsgn}\left(\mathbf{Q} \right)\sqrt{\frac{{\left(Q_{xx} -Q_{yy}\right)^2 + \left(Q_{yy} -Q_{zz}\right)^2 + \left(Q_{zz} -Q_{zz}\right)^2 +6\left(Q_{xy}^2 + Q_{yz}^2  + Q_{zx}^2 \right) }}{3}},
\label{eq:devQ}
\end{equation}
where $Q_{xx}$, $Q_{yy}$ and $Q_{zz}$ are its diagonal, and $Q_{xy}$, $Q_{yz}$ and $Q_{zx}$ its off-diagonal components
and the deviators $\epsd$, $\sigmad$ and $\Fdev$ refer to strain $\mathbf{E}$, stress 
$\pmb{\sigma}$ and fabric $\mathbf{F}$, respectively.
$\mathrm{Fsgn}\left(\mathbf{Q} \right)$ is the sign function that relates the tensorial quantity to be measured, $\mathbf{Q}$, with  
the reference-tensor that describes the (strain- or stress-controlled) path applied to the sample, $\mathbf{H}_{0}$:
\begin{equation*}
\mathrm{Fsgn}\left(\mathbf{Q} \right)=\mathrm{sgn}\left(\mathbf{H}_{0}:\mathbf{Q}\right).
\end{equation*}
For a given, complex deformation path, the reference tensor $\mathbf{H}_{0}$ must be chosen in a convenient way, in order 
to take into account both the actual loading path and/or the previous deformation history of the sample.
In the special case of undrained shear test, as introduced later in section \ref{sec:main}, we use as reference $\mathbf{H}_{0}=-\mathbf{\dot E}=(-1,1,0)$, where only the diagonal values are given, so that
$\mathrm{Fsgn}$ simplifies to
$$\mathrm{Fsgn}\left(\mathbf{Q} \right) = \mathrm{sgn}\left(Q_{yy} -Q_{xx} \right),$$
with $x-$wall expanding, $y-$wall compressing and $z-$wall non-mobile \cite{kumar2013effect}. 
We want to point out here that, during a deformation, the response of stress $\pmb{\sigma}$ and fabric $\mathbf{F}$ is opposite in sign to applied strain rate $\mathbf{\dot E}$. 
Unless mentioned explicitly, we will be using a sign convention for strain (isotropic $\delta\epsiso=-(1/3)\mathrm{tr}(\delta\strain)$ and deviatoric $\delta\epsd=-\delta E_\mathrm{dev}$), 
such that consistently a positive strain increment leads to a positive stress and fabric response.

Finally we note that in this work, we will use $\kstar=k/\left(2\langle r \rangle\right)$ to non-dimensionalize pressure $P$ and deviatoric stress $\sigmad$ to give $\pkstar$ and $\sigmadkstar$, respectively, and will be referring to deviatoric stress as shear stress.
\footnote[1]{It is important to point out that the rattlers are excluded in defining the (corrected) coordination number $C^*$. 
However dynamic rattler particles with $1\le M_p \le3$ contacts are included in the definitions of fabric and stress. 
We verified that during shear deformation, the maximum contribution in deviatoric stress due to rattlers is $0.03\%$, while in the case of 
deviatoric fabric the contribution rises to $0.5\%$. This is not surprising since only contacting particles contribute to the definitions 
of both stress and fabric and dynamic rattlers have a smaller weight for stress than for fabric, see Eq.\ (\ref{eq:fabriceq}). 
Note also that the number of rattlers decreases with increasing size of the particles \cite{kumar2013effect}.}

\section{Volume conserving (undrained) biaxial shear test }
\label{sec:main}

In this section, we first describe the sample preparation procedure 
and then the details of the numerical shear test. 

The initial configuration is such that spherical particles are randomly generated, 
with low volume fraction and rather large random velocities
in a periodic 3D box, such that they have sufficient space and time to exchange places and to randomize themselves.
This granular gas is then compressed isotropically, to a target volume 
fraction $\nuzero = 0.640$, sightly below the isotropic jamming volume fraction \cite{makse2000packing, imole2013hydrostatic, kumar2013effect}
$\nu_c\approx0.658$ and then relaxed to allow 
the particles to fully dissipate their potential energy \cite{imole2013hydrostatic, kumar2013effect}.
\footnote[2]{Note that the jamming volume fraction is given for a uniform radius distribution for polydispersity $w=3$. 
The results will be different if the distribution is different, e.g., when uniform surface or volume distributions are used. 
See Ref. \cite{kumar2013effect} for a detailed discussion on the evolution of jamming volume fractions with polydispersity for a uniform radius distribution.}
The relaxed state is then compressed (loading) isotropically from $\nuzero$ to a higher volume fraction of $\nu = 0.82$, 
and de-compressed back (unloading) to $\nuzero$ \cite{imole2013hydrostatic, kumar2013effect}. 

The preparation procedure, as described above  
provides many different initial configurations with volume fractions $\nu_i$, each one in mechanical equilibrium. 
Starting from various $\nu_i$ chosen from the unloading branch \cite{imole2013hydrostatic, kumar2013effect}, 
the samples are then sheared keeping the total volume constant, that is with a strain-rate tensor
\begin{eqnarray} 
\strnrate=\epsddot \left[ \begin{array}{ccc}
-1 & 0 & 0\\
0 &+1 & 0\\
0 & 0 & 0\\
\end{array} \right],
\end{eqnarray}
where $\epsddot = 28.39$ [$\mathrm{\mu s^{-1}}$] is the strain-rate (compression $>0$) amplitude applied to the moving $x-$ and $y-$walls, 
while the third $z-$wall is stationary. Our shear test, where the total volume is conserved during deformation, 
resembles the undrained test typical in geotechnical practice \cite{zhao2013unique}. 
The chosen deviatoric path is on the one hand similar to the pure shear situation, and on the other hand allows for simulation of the biaxial element test \cite{morgeneyer2003investigation,rock2008steady}
(with two walls static, while four walls are moving, in contrast to the more difficult isotropic compression, where all the six walls are moving). 
Pure shear is here used to identify constant volume deviatoric loading with principal strain axis keeping the same orientation as the geometry 
(cuboidal) of the system for the whole experiment. In this case, there is no rotation (vorticity) of the principal strain (rate) axis and no distortion/rotation of the sample due to shear deformation. 
Different types of volume conserving deviatoric deformations can be applied to shear the system, but very similar behavior has been observed \cite{imole2013hydrostatic}, in terms of shear stress.

\subsection{Evolution of stress}
\label {sec:stressresults}
\begin{figure}[!ht]
\centering
\subfigure[]{\includegraphics[scale=0.50,angle=270]{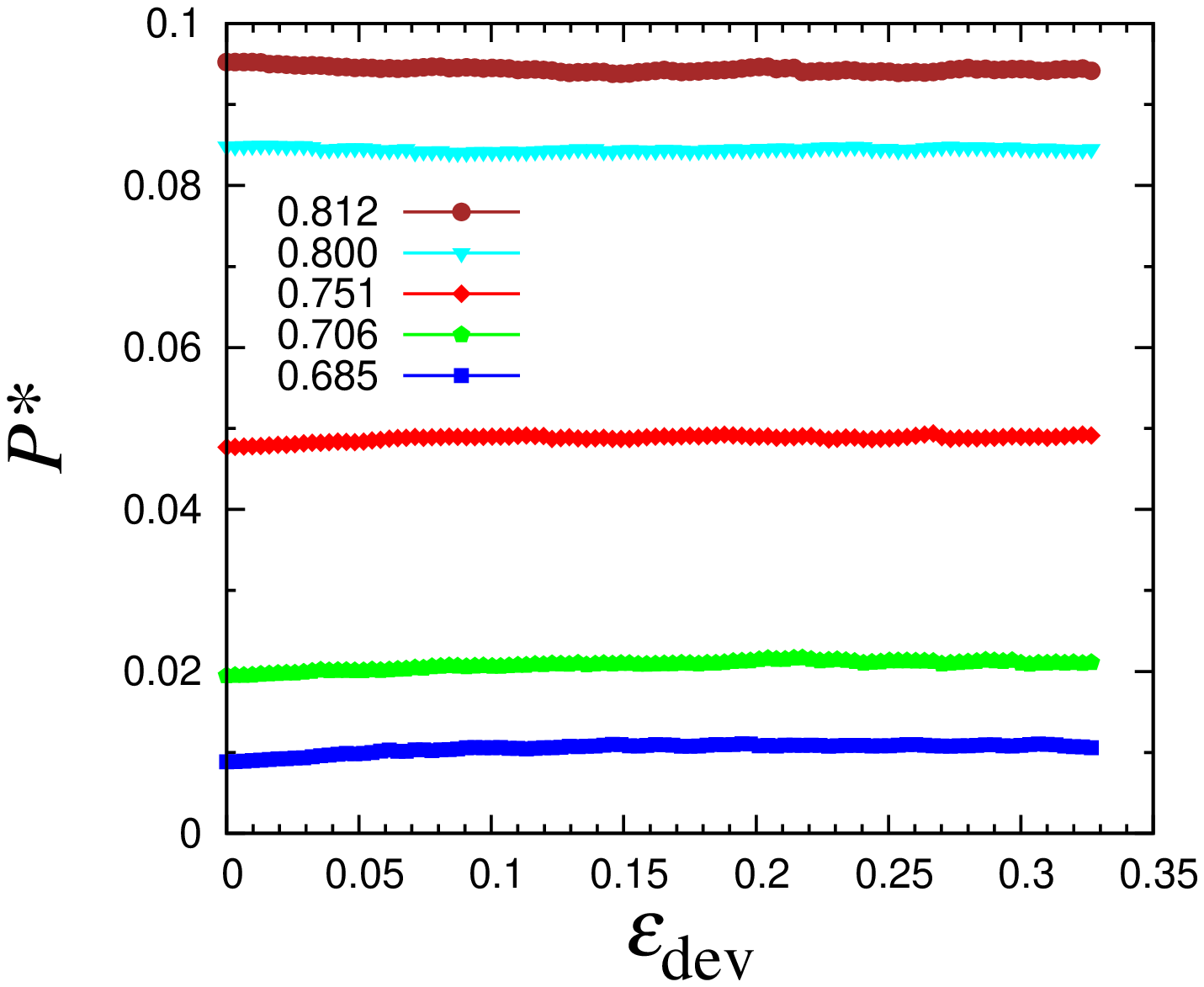}\label{pres}}
\subfigure[]{\includegraphics[scale=0.50,angle=270]{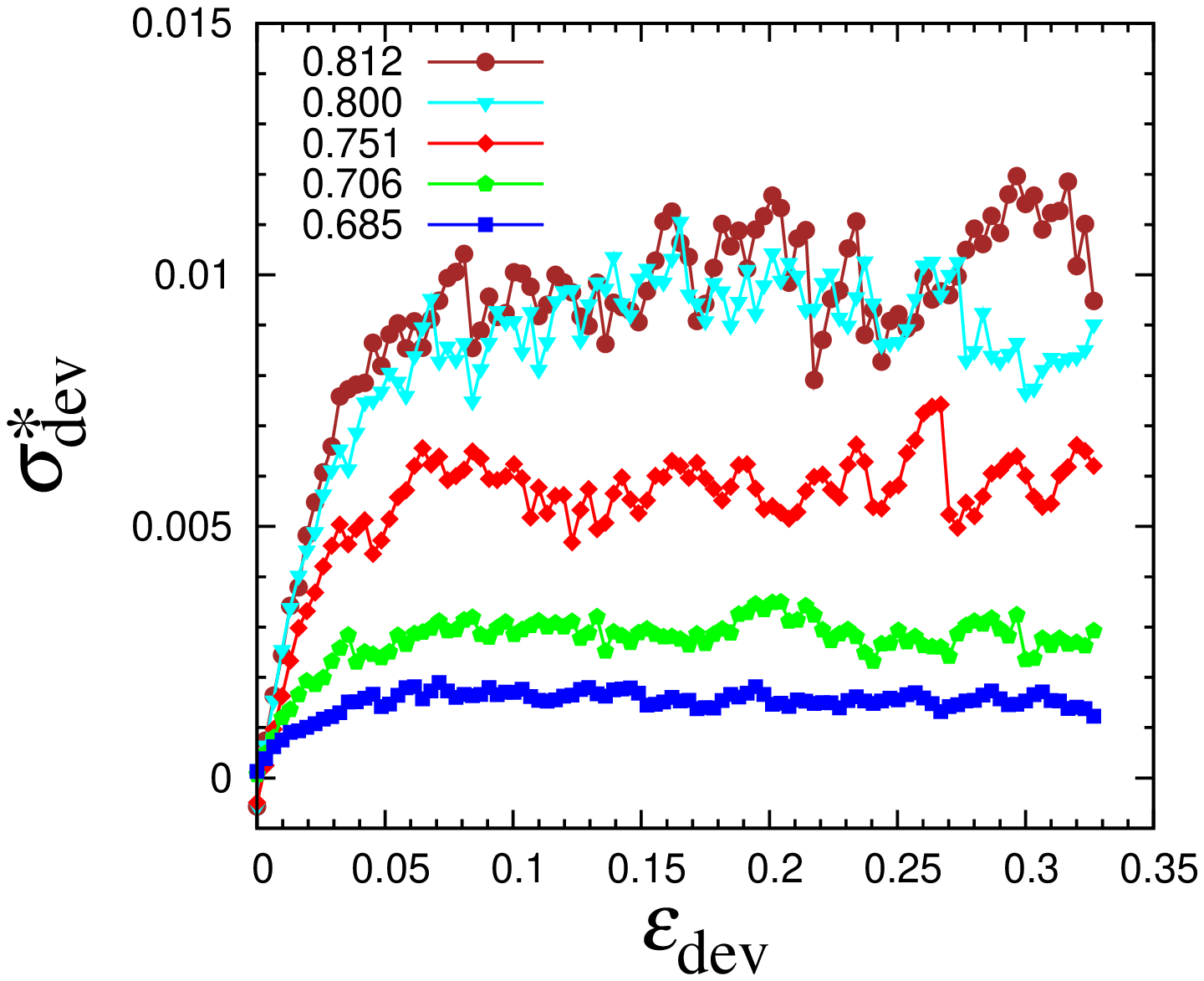}\label{tau}}
\caption{Evolution of non-dimensional (a) pressure $\pkstar$ and (b) shear stress $\sigmadkstar$ along the main strain path for the pure shear deformation mode for five different volume 
fractions, as given in the inset.}
\label{stress}
\end{figure}

\begin{figure}[!ht]
\centering
\subfigure[]{\includegraphics[scale=0.44,angle=270]{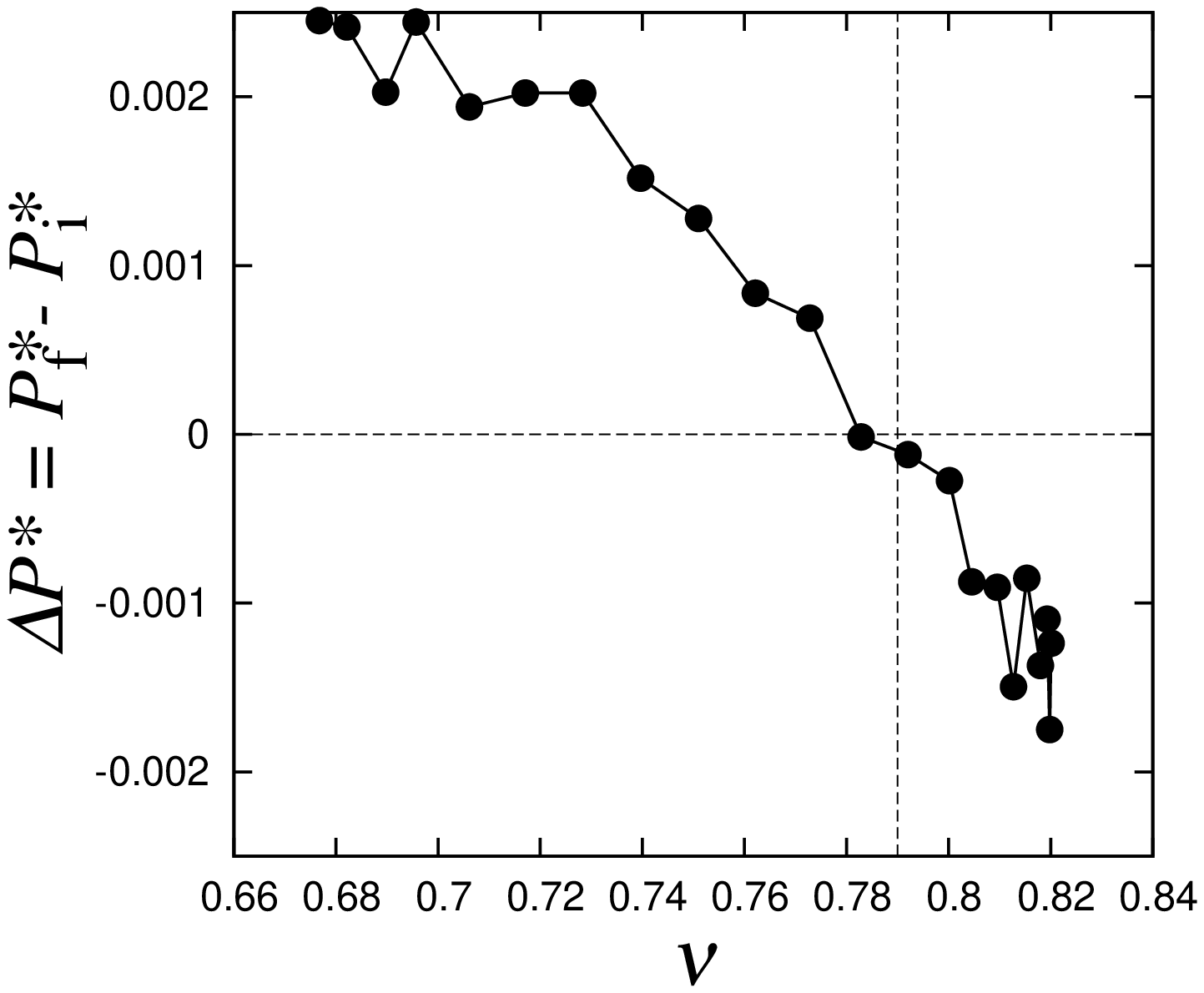}\label{dpwithnu}}
\subfigure[]{\includegraphics[scale=0.44,angle=270]{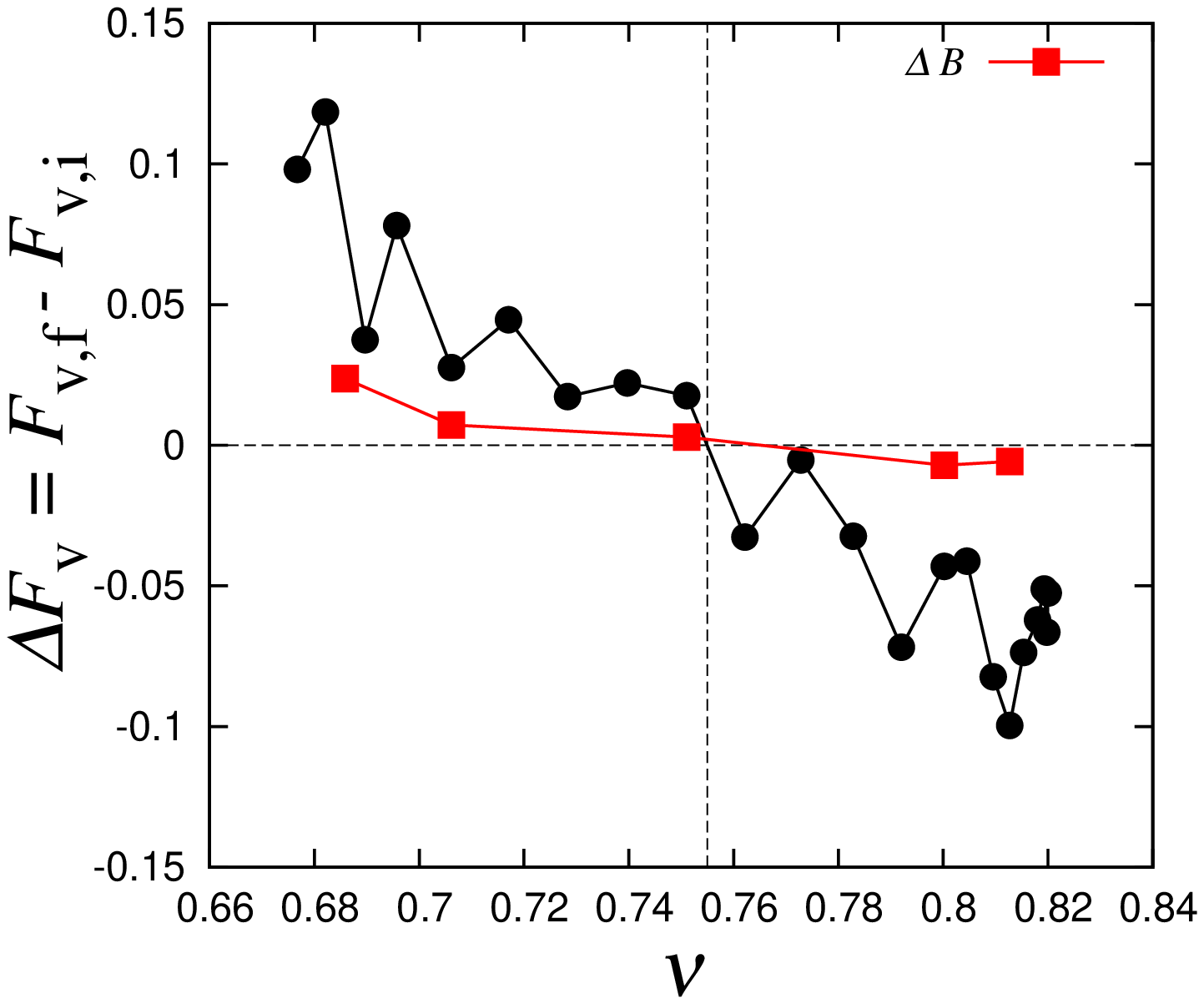}\label{dfvwithnu}}
\caption{Difference between the final and initial values in (a) non-dimensional pressure $\pkstar$ and 
(b) isotropic fabric $\Fv$ for the pure shear deformation mode for different volume fractions. Red `$\blacksquare$' represents the change in bulk modulus, as derived in section \ref{sec:perturbresults}.
Dashed lines in the plots represent the crossover when these quantities change sign.}
\label{changewithnu}
\end{figure}

The evolution of non-dimensional pressure $\pkstar$ with deviatoric strain $\epsd$ 
is presented in Fig.\ \ref{pres} during undrained shear tests for some exemplary volume fraction.
For frictionless systems analyzed here, only a slight variation of the pressure is observed
at the beginning of the test, due to the development of anisotropy in the sample, after which $\pkstar$ 
remains constant.\footnote[3]{We observe a much more pronounced change in pressure when friction is included in the calculation, 
in agreement with other studies, see e.g. \cite{guo2013signature}. These data are not shown here and are subject of ongoing research.}
Both the (small) initial pressure change and the final saturation value vary with
the vicinity of $\nu$ to the jamming volume fraction $\nu_{c}$. 
Interestingly, depending on the volume fraction, some of the samples show increase of the pressure (dilatancy) with respect to 
the initial value and some other decrease (compactancy), as shown in Fig.\ \ref{stress}. This 
supports the idea of a certain threshold value $\nu_d^p = 0.79$, as shown in Fig.\ \ref{dpwithnu}, 
where the pressure of the system changes behavior,
similarly to the switch between volumetric dilation and contraction visible in triaxial tests.

The evolution of the (non-dimensional) shear stress $\sigmadkstar$ during 
shear, as function of the deviatoric strain $\epsd$, is shown in 
Fig.\ \ref{tau}, for the same simulations as in Fig.\ \ref{pres}. 
The stress grows with applied strain until an asymptote (of maximum stress anisotropy) 
is reached where it remains fairly constant -- with slight fluctuations around the maximum $\sigmadkstar$ \cite{chung2013how}. 
The growth rate and the asymptote of $\sigmadkstar$, both increase with $\nu$. 

\subsection{Evolution of fabric}
\label {sec:fabricresults}

\begin{figure}[!ht]
\centering
\subfigure[]{\includegraphics[scale=0.50,angle=270]{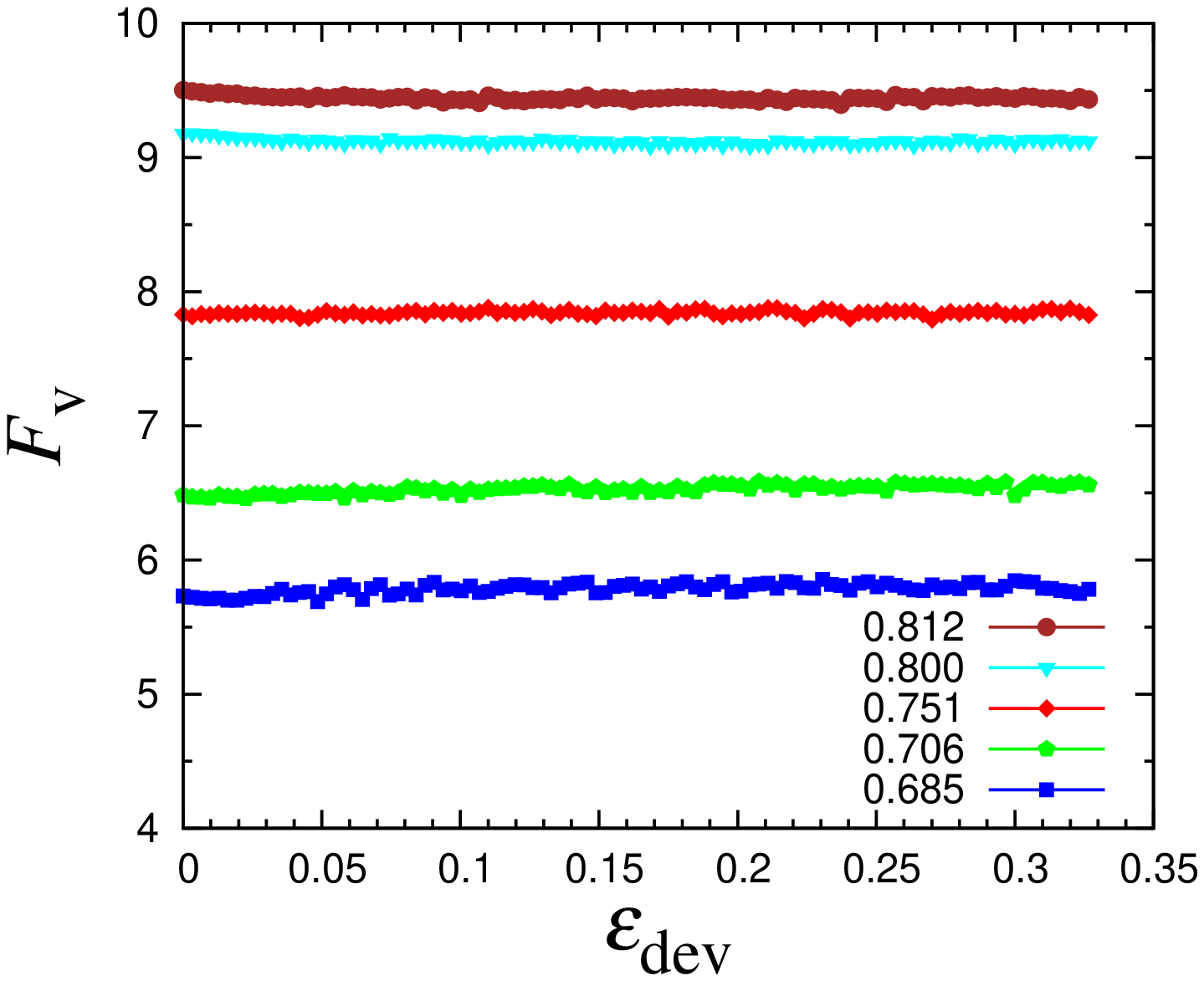}\label{fv}}
\subfigure[]{\includegraphics[scale=0.50,angle=270]{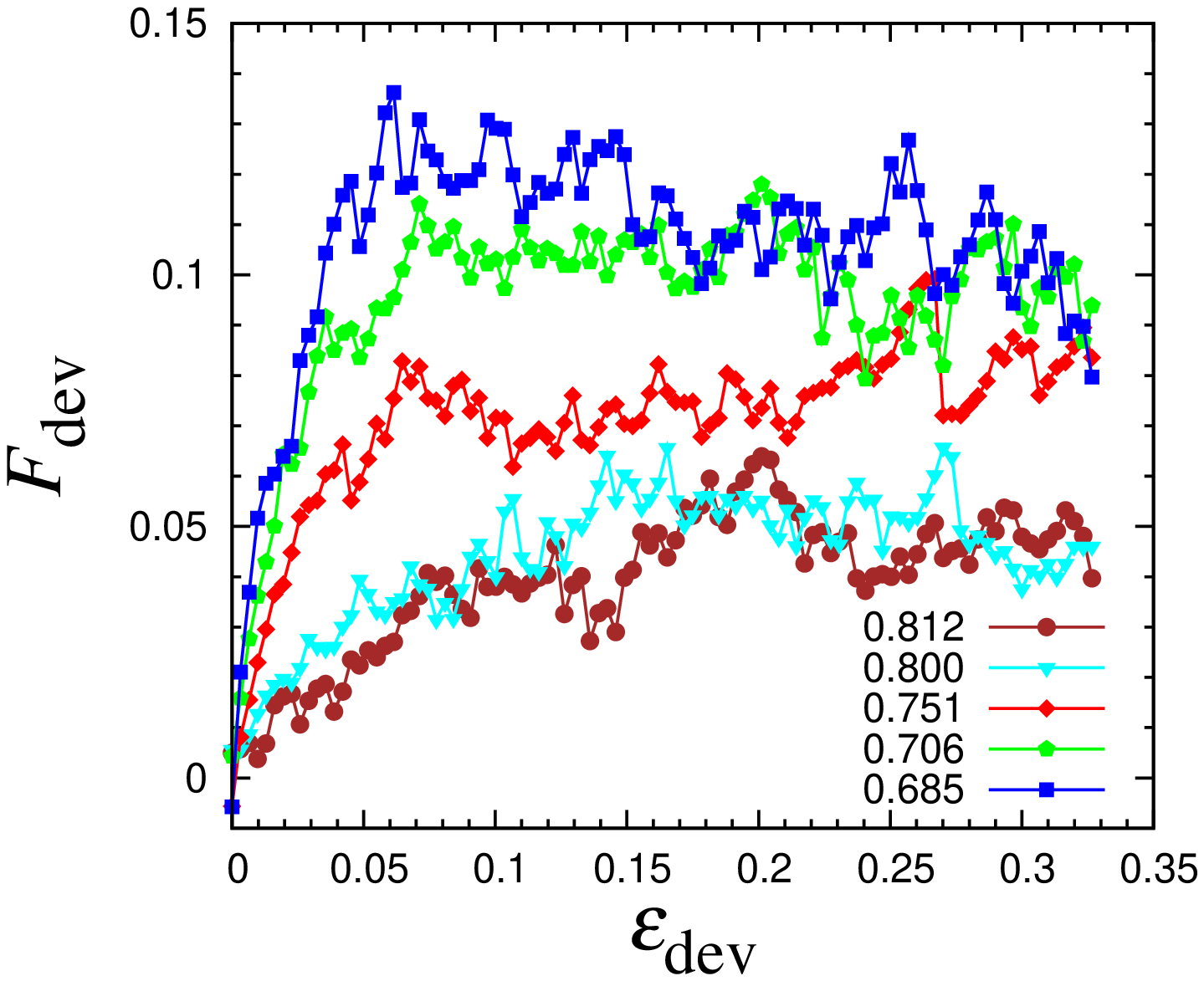}\label{fdev}}
\caption{Evolution of (a) isotropic fabric $\Fv$ and (b) deviatoric fabric $\Fdev$ along the main strain path for the pure shear deformation mode for five different volume 
fractions, as given in the inset.}
\label{fabric}
\end{figure}

Complementary to stress, in this subsection we study the evolution of the microstructure in the sample during the 
volume conserving shear test. Fig.\ \ref{fv} shows that the isotropic fabric $\Fv$
behaves in a very similar fashion as $\pkstar$, with a slight increase/decrease at the beginning, 
followed by saturation stage, whose value increases continuously with $\nu$.
Fig.\ \ref{dfvwithnu} shows that the difference between the initial value of $\Fv$ and its saturation value, 
changes sign when a certain volume fraction, $\nu_d^F = 0.755$, is reached. 
Note that $\nu_d^F \neq \nu_d^p$, that further confirms the independent evolution of $\pmb{F}$ and $\pmb{\sigma}$. 

From Eq.\ (\ref{eq:Fveqn}) $\Fv$ is proportional to the product of volume fraction $\nu$, that remains unchanged 
during deviatoric deformations, and coordination number $C$, that varies only slightly 
for sheared frictionless systems \cite{imole2013hydrostatic}. 
Note that as $C=C^*\left(1-\phi_r\right)$, knowing the (empirical) relations of $C^*$ and $\phi_r$ with volume fraction, 
as presented in Refs.\ \cite{imole2013hydrostatic, kumar2013effect}, 
we can fully describe the isotropic fabric state. 
In this study, we assume $\Fv$ to stay constant during the shear test. 
This assumption will be used later in section \ref{sec:predict} for the prediction of a cyclic shear test. 
However, the small changes in $\Fv$ or $\pkstar$ can be associated to a (small) change in the jamming volume fraction \cite{kumar2014memory}.

\begin{figure}[!ht]
\centering
\subfigure[]{\includegraphics[scale=0.34,angle=270]{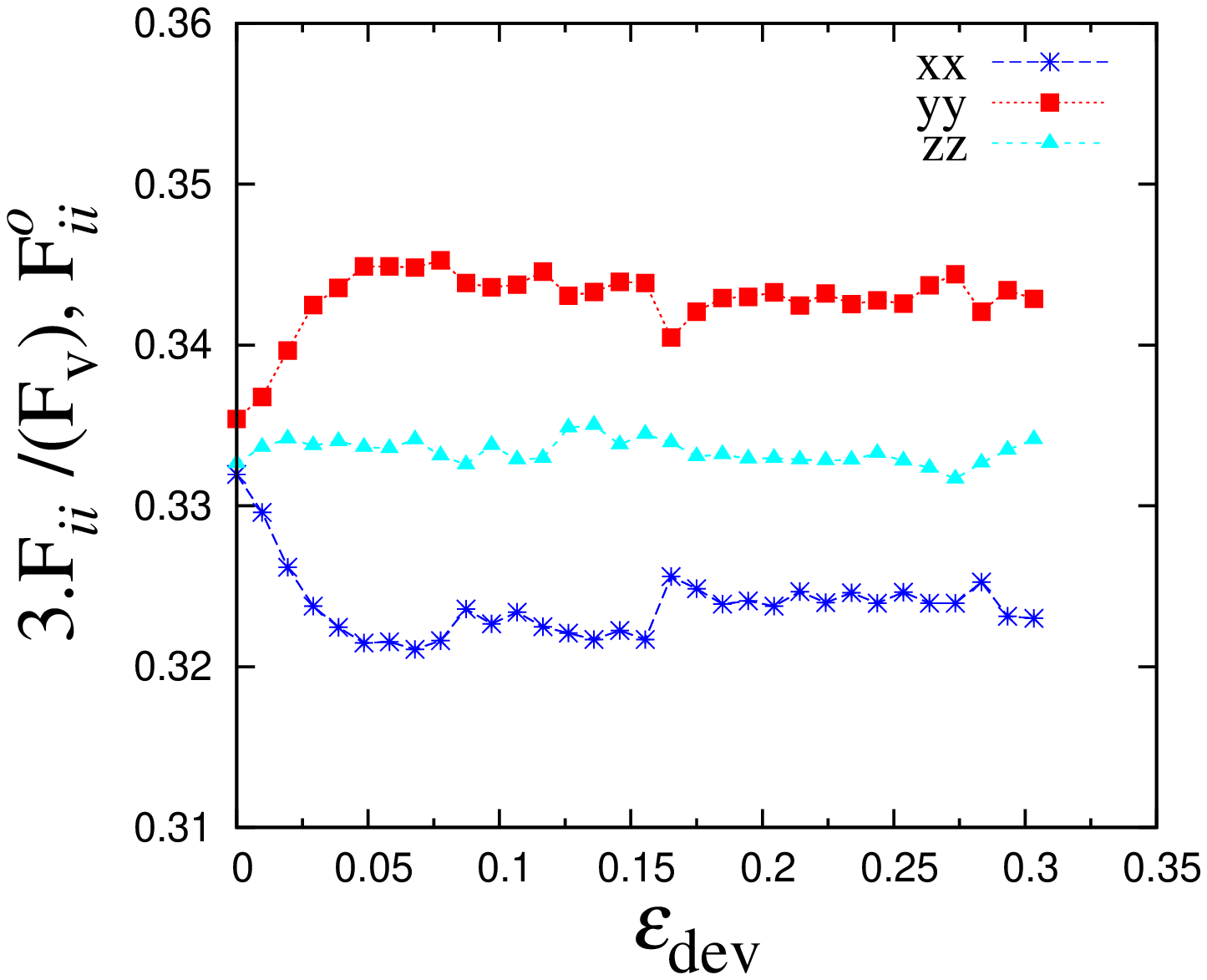}\label{FAB_collapse_w1}}
\subfigure[]{\includegraphics[scale=0.34,angle=270]{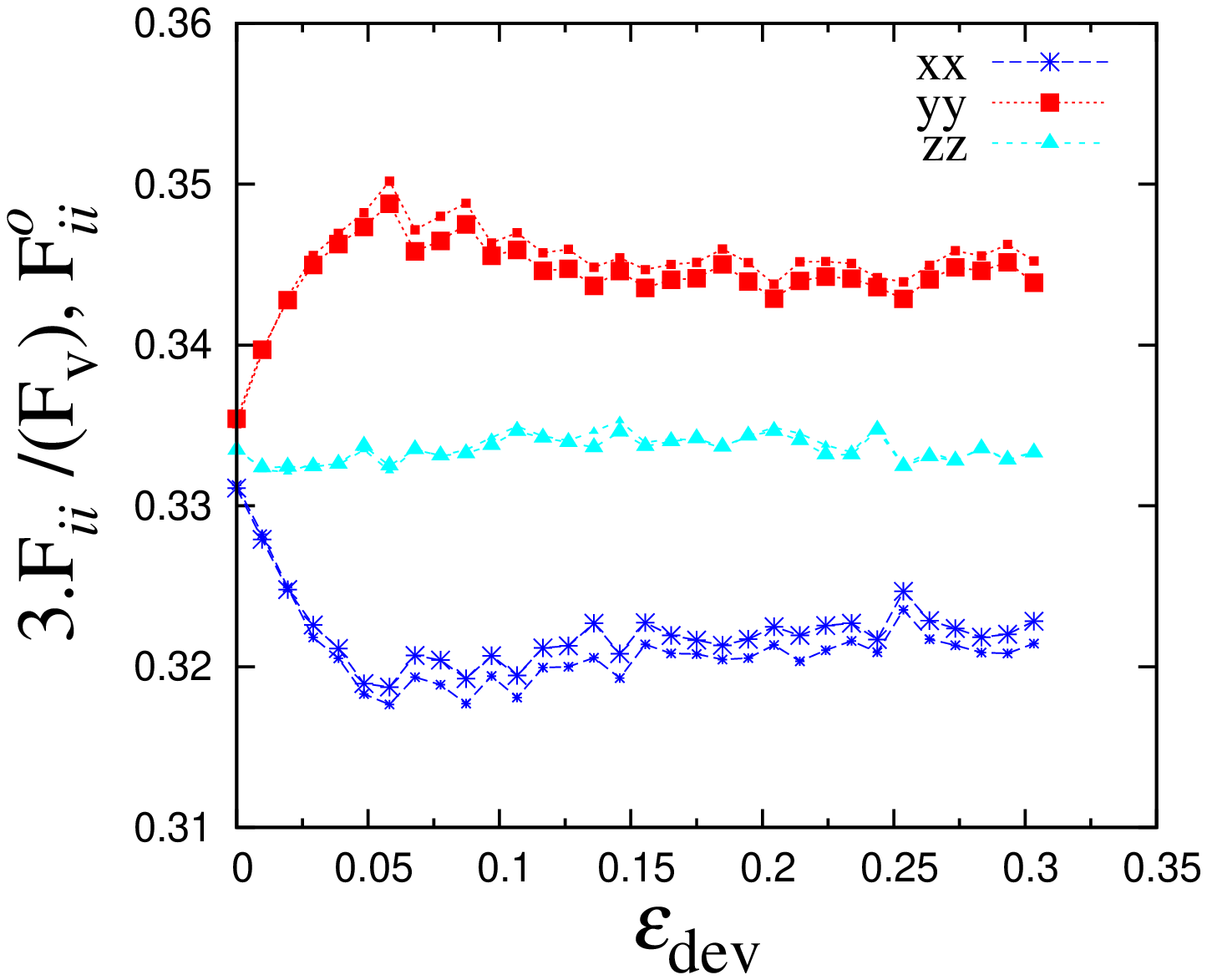}\label{FAB_collapse_w2}}
\subfigure[]{\includegraphics[scale=0.34,angle=270]{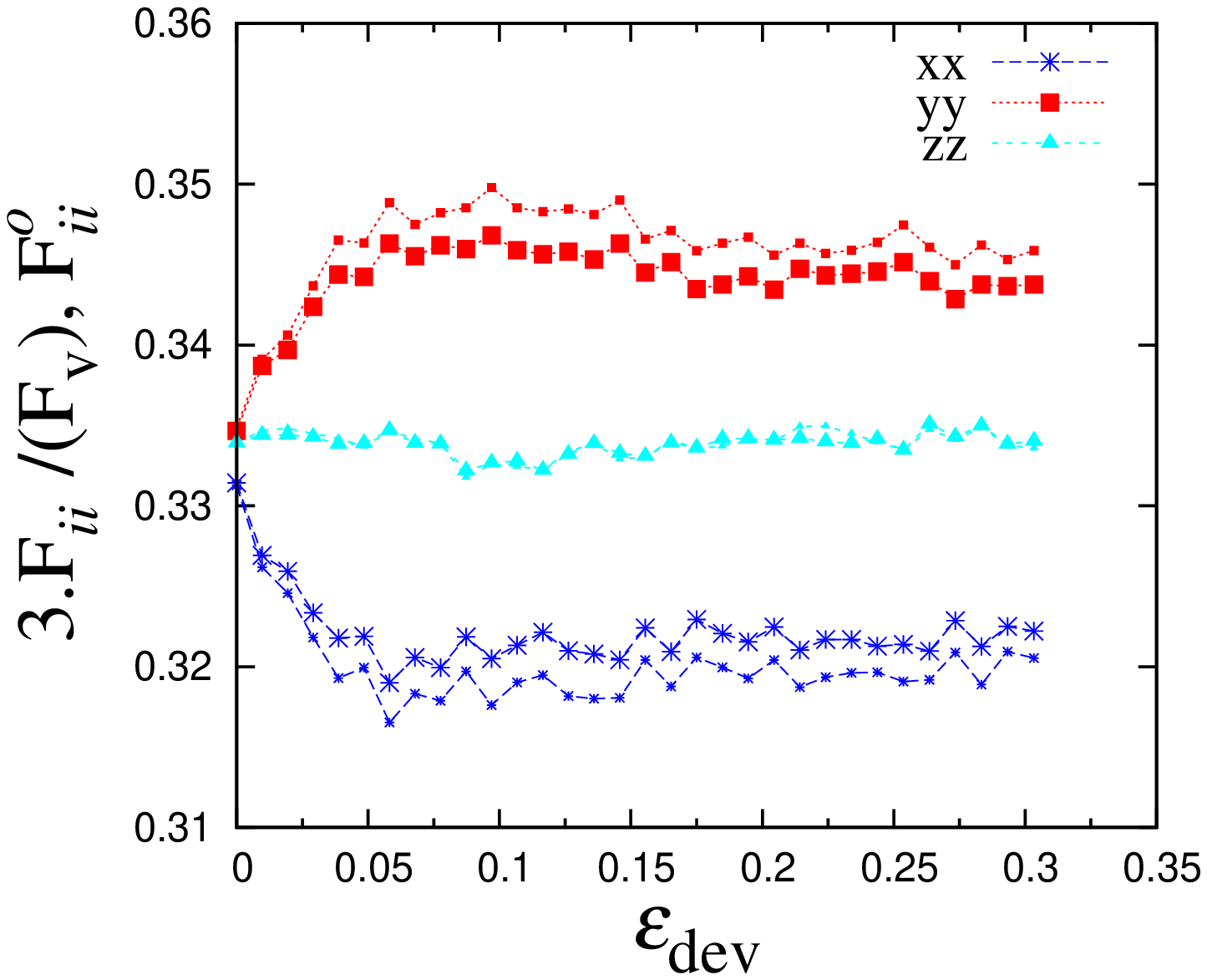}\label{FAB_collapse_w3}}
\caption{Evolution of the eigen-values of the fabric tensors (directions shown in the inset), during shear deformation at volume fraction $\nu = 0.685$, 
for the fabric definition defined in Eq.\ (\ref{eq:scaledFabricdefn}) (smaller symbols) and the relation presented in Eq.\ (\ref{eq:fabricrelation}) (large symbols), 
for three cases of polydispersity 
(a) $w=1$, i.e.,\ monodisperse (b) $w=2$ and (c) $w=3$ (present work).}
\label{fabriccomparison}
\end{figure}

The evolution of the deviatoric fabric, $\Fdev$, as function of the deviatoric 
strain is shown in Fig.\ \ref{fdev} during shear for five 
different volume fractions. It builds up from different random 
small initial values (due to the initial anisotropy in the sample that develops during preparation) and reaches different maxima.
The deviatoric fabric builds up faster at lower
volume fractions but the maximal values are higher for smaller volume fractions, qualitatively opposite to the evolution of $\sigmadkstar$ \cite{chung2013how}. 
As mentioned in section \ref{subsec:tensorial} the validity of Eq.\ (\ref{eq:fabricrelation}), that relates the two different definitions of fabric depends on polydispersity. In order to check the relation, 
in Fig.\ \ref{fabriccomparison} the evolution of the three eigenvalues of the
fabric tensor is plotted, for both definitions, Eqs.\ (\ref{eq:fabriceq}) and (\ref{eq:scaledFabricdefn}), 
during the volume conserving shear test, for three different values of polydispersity $w =$1, 2 and 3. 
For all polydispersities, the chosen volume fraction $\nu = 0.685$ is 
close to the jamming points, that slightly varies with $w$ \cite{kumar2013effect}. 
The difference between the definitions of fabric 
becomes higher for higher polydispersity $w=3$, as in Eq.\ (\ref{eq:scaledFabricdefn})
the contribution of each particle is weighted to its surface area, whereas in Eq.\ (\ref{eq:fabriceq}) 
it is weighted by the volume. 
Only for the monodisperse case, the relation is exact, as can be seen in Fig.\ \ref{FAB_collapse_w1}. 
The differences are considerable for $w=2$ and $w=3$, for both 
compressive and tensile direction, while the non-mobile direction is not affected. 
Note that the difference of the two fabrics will be smaller for denser systems.

\section{Elastic moduli}
\label {sec:perturb}
In this section, we focus on the evolution of the elastic properties of the material 
and neglect the plastic contribution to the granular behavior, 
that will be superimposed to the present analysis later in section \ref{sec:predict}.
We first describe the numerical procedure to measure the elastic moduli of the anisotropic aggregate, 
and later we analyze the data and their relation with stress and fabric.

\subsection{Numerical probes}
\label {sec:perturbprep}

\begin{figure}[!ht]
\centering
\subfigure[]{\includegraphics[scale=0.45,angle=270]{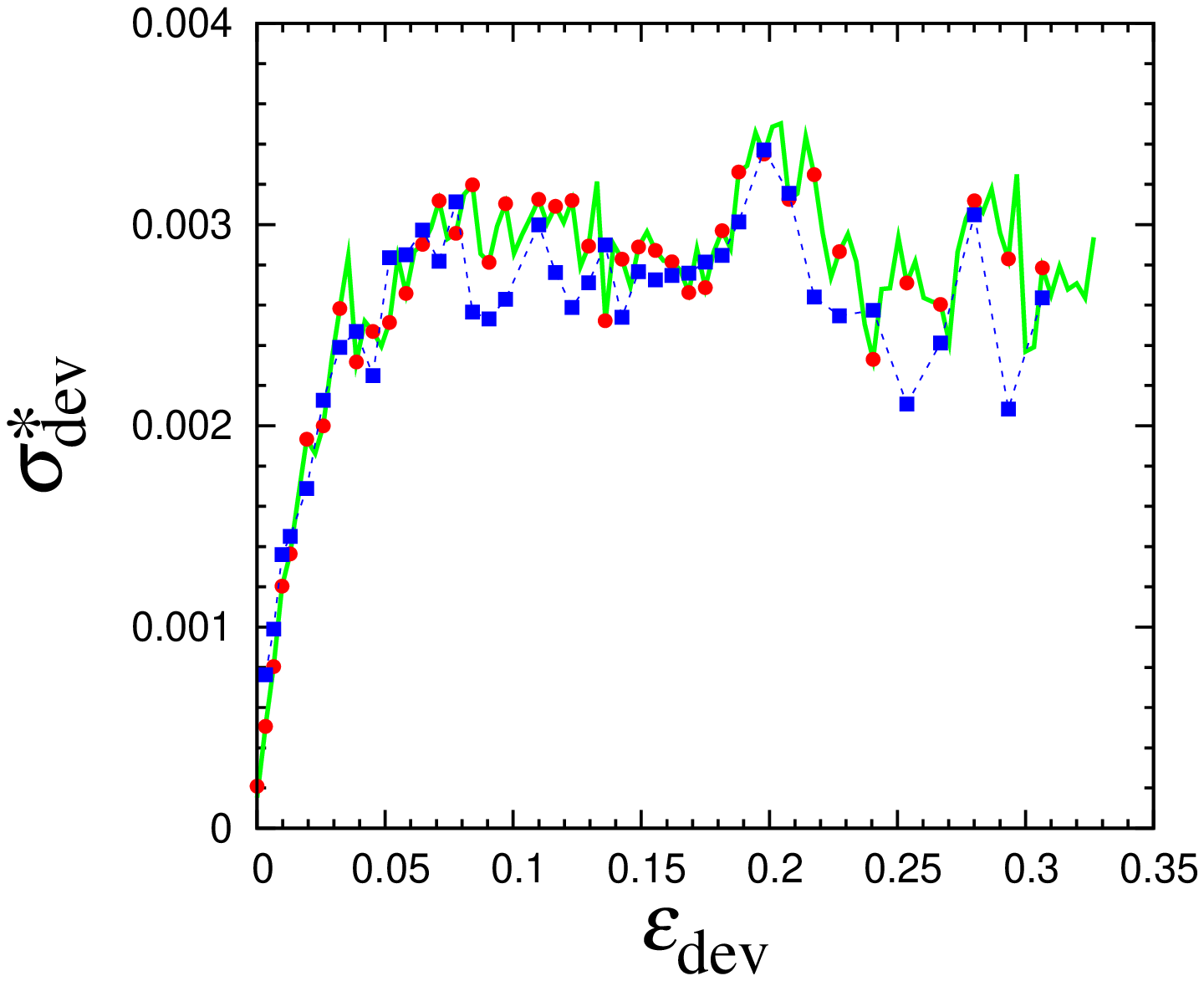}\label{scem}}
\subfigure[]{\includegraphics[scale=0.45,angle=270]{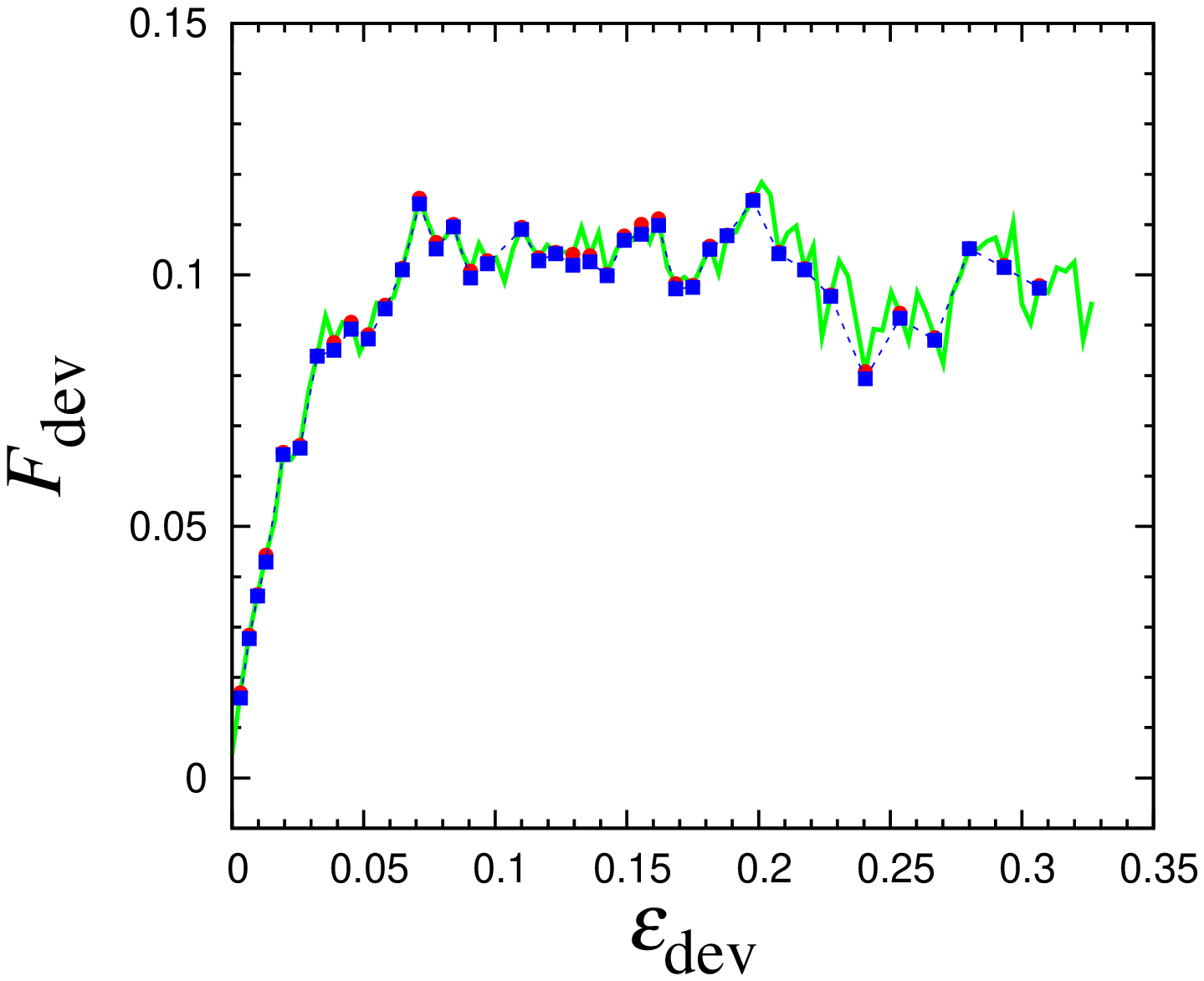}\label{scem_fab}}
\caption{Evolution of (a) non-dimensional shear stress $\sigmadkstar$ and 
(b) deviatoric fabric $\Fdev$, along the main strain path $\epsd$ for the pure shear deformation mode, for volume fraction $\nu = 0.706$. 
The red `$\bullet$' symbols in (a -- b) are the chosen states, which are 
first relaxed  (blue `$\blacksquare$' symbols in (a) and (b)) and then used as initial configurations for the 
purely isotropic ${3\delta\epsiso}$ and purely deviatoric ${\delta{\epsd}}$ perturbations.
 }
\label{schematic}
\end{figure}

In a general framework, a possible description for the incremental, elastic behavior of an anisotropic material is
\begin{eqnarray} \left[ \begin{array}{cc}
\delta \pkstar \\
\delta\sigmadkstar
\end{array} \right]
=
\left[ \begin{array}{cc}
B & A_1 \\
A_2 & \Goct \\
\end{array} \right]
\left[ \begin{array}{cc}
3\delta\epsiso \\
\delta\epsd
\end{array} \right],
\label{eqnarray:hypo}
\end{eqnarray}
where the isotropic and deviatoric components of stress have been isolated and are expressed 
as functions of $\epsiso$ and $\epsd$ via a non-dimensional stiffness matrix \cite{graham1983anisotropic}
(by multiplying the moduli with $\kstar$, the real stiffnesses can be extracted).
$B$ is the classical bulk modulus, and
$\Goct$ the octahedral shear modulus. The anisotropy moduli 
$A_1$ and $A_2$ provide a cross coupling between 
the two parts (isotropic and deviatoric) of stress and strain increments. 
Eq.(\ref{eqnarray:hypo}) provides a partial description for the evolving stress and stiffness of a sheared material, as it applies to a triaxial-box 
configuration (with eigensystem coincident with the axes of the box), 
where no shear strain/stress are measured and stress and moduli are assumed to be collinear. 
Moreover, the increase of stress and stiffness in the out-of-plane direction ($z$-direction here) due to the
non-planar (triaxial) stress state associated with a the plane deformation mode, 
is not independently accounted for. These are rather hidden in the expression for deviatoric stress as proposed in Eq.(\ref{eq:devQ}) 
and used in Eq.(\ref{eqnarray:hypo}). 
However, we have chosen this representation, since advantages are obtained by investigating the elasticity of a granular material (e.g.\ soil), not through its resistance to direct stresses expressed by 
Young's modulus and Poisson's ratio, but rather in terms of (purely volumetric and deviatoric) 
stress-response to volume and shape changes, as described by the bulk modulus $B$ and the octahedral shell modulus $G^{oct}$. 
This aspect will be further addressed in section \ref{sec:predict}, where Eq.(\ref{eqnarray:hypo}) will be included in the theoretical model.

To study the evolution of the effective moduli during shear, 
we choose different initial states (forty) as shown in Fig.\ \ref{schematic}, 
and apply sufficient relaxation, so that the granular assemblies dissipate the kinetic energy 
accumulated during the original shearing path. 
When the states along the shear path are relaxed, a much higher drop is visible in $\sigmadkstar$ rather than in $\Fdev$, see Fig.\ \ref{schematic}. 
This shows that the contact network remains almost intact and $\Fdev$ does not change;
on the other hand, the average particle overlap is more sensitive to the relaxation stage and decreases, 
leading to a finite drop in $\sigmadkstar$.
Then we perform a small strain perturbation
to these relaxed anisotropic states, i.e.,\ we probe the samples, and measure the incremental stress response
\cite{magnanimo2008characterizing, kumar2013evolution}. 
Finally, the elastic moduli are calculated as the ratio between the measured increment in stress and the applied strain.  
We can obtain all the different moduli in Eq.\ (\ref{eqnarray:hypo}), by applying an 
incremental pure volumetric or pure deviatoric strain and measuring the incremental 
volumetric or shear stress response:
\begin{eqnarray}
B = \frac{\delta{\pkstar}} {3\delta\epsiso}\bigg|_{\delta\epsd=0}  &,&  A_1 = \frac{\delta{\pkstar}}{\delta{\epsd}}\bigg|_{\delta\epsiso=0} \nonumber  , \\ 
A_2 =\frac{\delta{\sigmadkstar}} {3\delta\epsiso}\bigg|_{\delta\epsd=0} &,&  \Goct = \frac{\delta{\sigmadkstar}}{\delta{\epsd}}\bigg|_{\delta\epsiso=0}.   
\label{eqnarray:moduli}
\end{eqnarray}
Also for this part of the numerical experiment, the system is allowed to relax after the incremental strain is applied, 
that is the stress is measured after relaxation \cite{makse2000packing, magnanimo2008characterizing}
Since the numerical probe experiments are carried out with zero contact friction, 
we are measuring the resistance of the frictionless material \cite{kumar2013evolution}, where only 
normal forces are involved. 
The first big question concerns the amplitude of the applied perturbation to get the elastic response \cite{sibille2009analysis, froiio2010incremental, calvetti2003numerical}.

\subsection{How small is small?}
\label {sec:howsmall}
In this section, we discuss the amplitude of the perturbations applied to measure the elastic stress response of the granular material.
Also, we will discuss the results for larger amplitudes and the threshold between elastic and plastic regimes.

\subsubsection{Effect of isotropic perturbations ${3 \delta \epsiso}$}
\label {sec:isoperturb}
Figs.\ \ref{LONG} (column 1 and 2) show the changes in non-dimensional pressure $\delta\pkstar$, non-dimensional shear stress $\delta\sigmadkstar$, isotropic fabric $\delta\Fv$ and deviatoric fabric $\delta\Fdev$ 
for different amplitudes of the isotropic perturbation ${3\delta \epsiso}$, applied to two relaxed states that 
have been sheared until $\epsd = 0.0065$ (nearly isotropic configuration: column 1) and 
$\epsd = 0.31$ (steady state configuration: column 2) respectively. The data correspond the the shear test with $\nu=0.706$. 
The linear elastic response is also plotted (red solid curve) in the whole strain range, as derived from the 
incremental behavior at very small strain, to give an idea of the deviation form elasticity when strain increases.

$\delta \pkstar$ initially increases linearly and smoothly with ${3 \delta \epsiso}$, in agreement with the prediction of linear elasticity. 
Also the difference between the two initial states (near isotropic and steady state as shown in Figs.\ \ref{dPdepsvA_longgg} and \ref{dPdepsvB_longgg}, respectively) 
is minimal, meaning that the bulk modulus $B$ (slope of $\delta \pkstar$ with ${3 \delta \epsiso}$ in the elastic regime) 
is almost constant. This is not surprising, as we expect $B$ to be dependent on isotropic quantities that, which  
stay mostly unchanged during the shear deformation, as discussed in section \ref{sec:perturbresults}. 
$\delta \sigmadkstar$ behaves similar as $\delta \pkstar$ for small strain, but shows several sharp drops for large strain. These correspond to 
sudden changes in the coordination number $\delta C^*$ (see Fig.\ \ref{smallperlongp_C}(a--b)), due to rearrangements in the system during the probe. 
For the nearly isotropic state (Fig.\ \ref{dtaudepsvA_longgg}), the ratio of $\delta \sigmadkstar$ with ${3 \delta \epsiso}$ in the linear elasticity regime, 
i.e.\ $A_2$, is small when compared with the steady state (Fig.\ \ref{dtaudepsvB_longgg}).
This clearly tells that $A_2$ evolves during the shear deformation for a given volume fraction, 
and must be linked with deviatoric quantities.

$\delta\Fv$ increases with ${3\delta \epsiso}$, with more fluctuations compared to $\delta\pkstar$, for both states considered here, 
$\epsd = 0.0065$ (nearly isotropic state, Fig.\ \ref{dFvdepsvA_longgg}) and $\epsd = 0.31$ (steady state, Fig.\ \ref{dFvdepsvB_longgg}).
Moreover, the prediction using Eq.\ (\ref{eq:Fveqn}) for $\Fv$, matches the dataset very well.
$\Fdev$ does not change ($\delta \Fdev = 0$) with increasing ${3\delta \epsiso}$, until the first rearrangement in structure occurs (see Figs.\ \ref{smallperlongp_C}(c--d)).
After this $\delta \Fdev$ starts to decrease with increasing amplitude ${3\delta \epsiso}$, faster in the steady state (Fig.\ \ref{dFdevdepsvA_longgg}) than in the near isotropic state, see Fig.\ \ref{dFdevdepsvB_longgg}.
We note here that, when a non-incremental volumetric strain (${3\delta \epsiso} > 10^{-4}$) is applied, 
the system moves from a volume-conserving  to a new non-volume-conserving deformation path.
As this system is already anisotropic, this leads to a decrease ($\delta \Fdev<0$) in deviatoric fabric $\Fdev$, 
opposite to the increase ($\delta \sigma_{dev}>0$) in deviatoric stress, see Figs.\ \ref{dtaudepsvA_longgg} and \ref{dtaudepsvB_longgg},
higher in the steady state (Fig.\ \ref{dFdevdepsvB_longgg}) than in the nearly isotropic state (Fig.\ \ref{dFdevdepsvA_longgg}). 
The last observation suggests that the distance between the volume conserving and non-volume conserving 
configurations increases with $\epsd$. 

Hence, during isotropic compression (increasing ${3\delta \epsiso}$) of a pre-sheared (anisotropic) state, 
both the pressure $\pkstar$ and shear stress $\sigmadkstar$ increase, with pressure increasing much faster leading to 
a decrease in deviatoric stress ratio $\sd=\sigma^{*}_{dev}/P^{*}$. 
The deviatoric fabric $\Fdev$ also decreases with isotropic compression of a pre-sheared state, 
and the decrease is initially faster than the exponential decay of $\Fdev$ (see section\ \ref{sec:predict} below) with volume fraction $\nu$, as seen in Fig.\ \ref{dFdevdepsvB_longgg}. 
This decrease in $\Fdev$ becomes slower for large strain, as also seen in Fig.\ \ref{dFdevdepsvA_longgg}. 
These observations are consistent with the findings of Imole \textit{et al.\ }\cite{imole2013hydrostatic}, 
where the authors noticed a decreasing steady state deviatoric fabric and deviatoric stress ratio with the increasing volume fraction, or $\epsiso$. \\

\subsubsection{Effect of deviatoric perturbations ${\delta \epsd}$}
\label {sec:devperturb}
Figs.\ \ref{LONG}(column 3 and 4) show the changes in the same quantities as before
for different amplitudes of the deviatoric perturbation ${\delta \epsd}$, applied to a relaxed state with volume fraction 
$\nu=0.706$ that has been sheared until $\epsd = 0.0065$ (nearly isotropic configuration: column 3) and 
$\epsd = 0.31$ (steady state configuration: column 4). 

$\delta \pkstar$ increase linearly with $\delta \epsd$ (the slope in the elastic regime is $A_1$), 
with $A_1$ much smaller for the nearly isotropic state (Fig.\ \ref{dPdepsdevA_longgg}) than for the steady state (Fig.\ \ref{dPdepsdevB_longgg}). 
This shows that $A_1$ evolves during the shear deformations, like $A_2$, for a given volume fraction, and must 
be linked with the deviatoric state of the system.
Moreover, after large deformation, both states show drops in $\delta \pkstar$, which can be linked to the particle rearrangements at large deformation (see Fig.\ \ref{smallperlongp_C}(c--d)).
A non-linear, irregular behavior shows up for $\delta \epsd > 10^{-4}$, with $\delta \pkstar$ positive in case of loose sample 
(present sample) and negative for dense samples (data not shown), in agreement with the observations in Fig.\ \ref{dpwithnu}. 
$\delta \sigmadkstar$ also increases linearly with $\delta \epsd$, 
with $\Goct$ (slope of the line) slightly higher for the near isotropic state (Fig.\ \ref{dtaudepsdevA_longgg}) than for the steady state (Fig.\ \ref{dtaudepsdevB_longgg}).
Again, similar to $\delta \pkstar$, $\delta \sigmadkstar$ shows drops after large deformations, 
which can be linked to the particle rearrangements at large deformation (see Fig.\ \ref{smallperlongp_C}(c--d)).
In the steady state, the incremental stresses ($\delta \pkstar$ and $\delta \sigmadkstar$) increase linearly for very small strain, 
as the relaxed configuration, starting point for the probes, has lower stress than the main deviatoric path (see Fig.\ \ref{scem}) 
and the system tends  to regain the "missed" stress, when the shear restarts.  
After the first elastic response, $\delta \pkstar$ and $\delta \sigmadkstar$ fluctuate around zero for larger amplitudes (Figs.\ \ref{dPdepsdevB_longgg} and \ref{dtaudepsdevB_longgg}),
as no change in stress is expected with increasing deviatoric strain in the steady state.

$\delta\Fv$ stays mostly zero when small $\delta \epsd$ is applied for both near isotropic and steady state configurations 
(Figs.\ \ref{dFvdepsdevA_longgg} and  \ref{dFvdepsdevB_longgg}).
With increasing strain amplitude, $\delta\Fv$ increases in the case of a loose sample close to the isotropic state (Fig.\ \ref{dFvdepsdevA_longgg}), and decreases for denser samples (data not shown), 
in agreement with the behavior in Fig.\ \ref{dfvwithnu}. 
In Fig.\ \ref{dFdevdepsdevA_longgg}, $\delta\Fdev$ for the nearly isotropic state, stays zero for $\delta \epsd < 10^{-4}$, when no rearrangements happen and the behavior is elastic, while it reaches a 
positive finite value for larger amplitude (that coincides with the slope of the curve in Fig.\ \ref{fdev}). 
This finite value increases with increasing anisotropy (or deviatoric strain state) until it reaches zero in the steady state, 
where no variation of deviatoric fabric is expected with further applied deviatoric strain (see Fig.\ \ref{dFdevdepsdevB_longgg}).
When compared to the model predictions in Ref.\ \cite{imole2013hydrostatic}, 
the simulation data for $\Fdev$ well match with the theoretical line, where $\Fdev$ increases due to shear for the near isotropic state, and does not change for the steady state simulation.

\subsubsection{Discussion and comparison}
\label {sec:isodevperturb}
\begin{figure}[!ht]
\hspace{8mm} {nearly isotropic} \hspace{14mm}  {steady state} \hspace{16mm} {nearly isotropic} \hspace{16mm} {steady state} \\
\centering
\subfigure[]{\includegraphics[scale=0.25,angle=270]{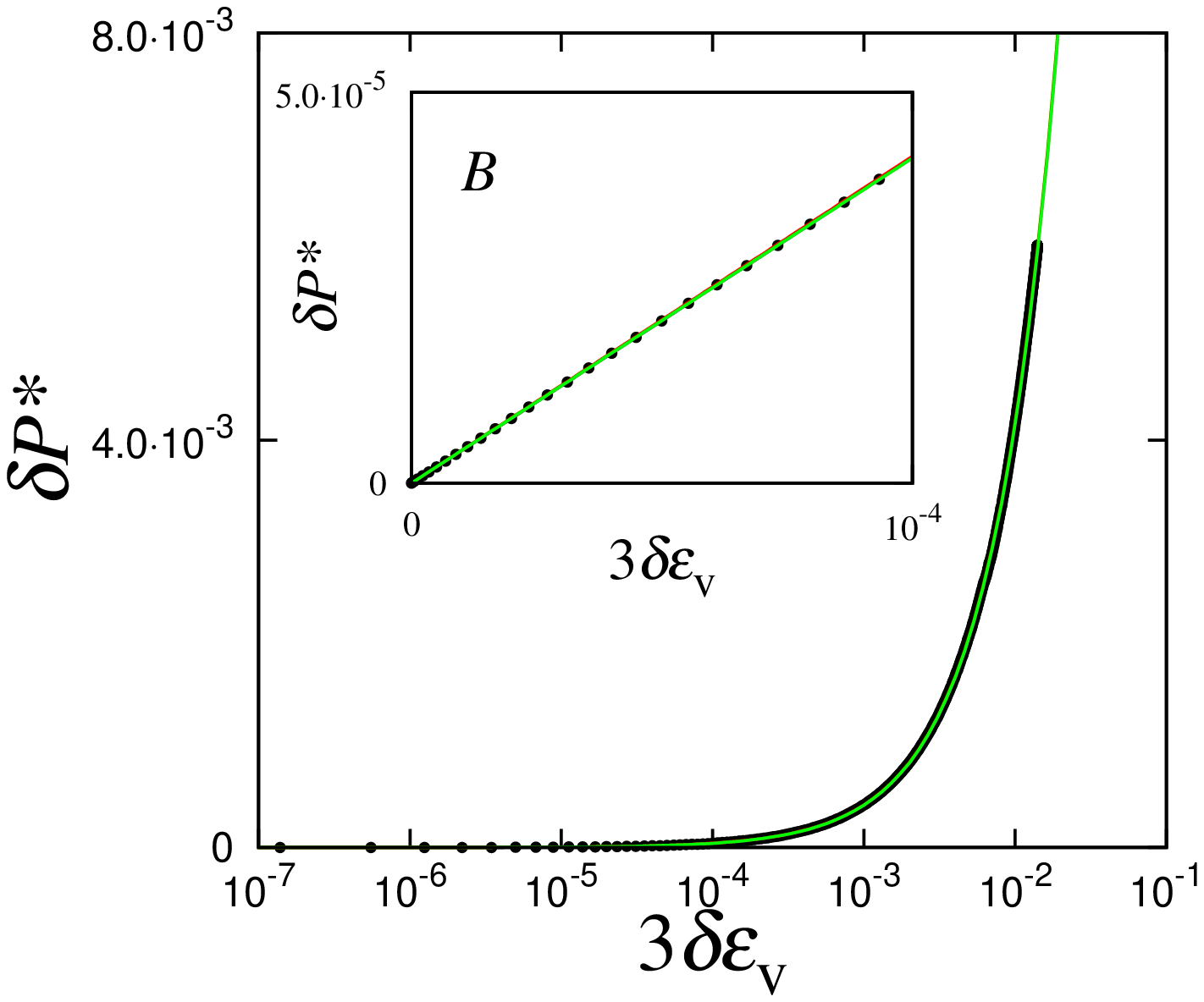}\label{dPdepsvA_longgg}}
\subfigure[]{\includegraphics[scale=0.25,angle=270]{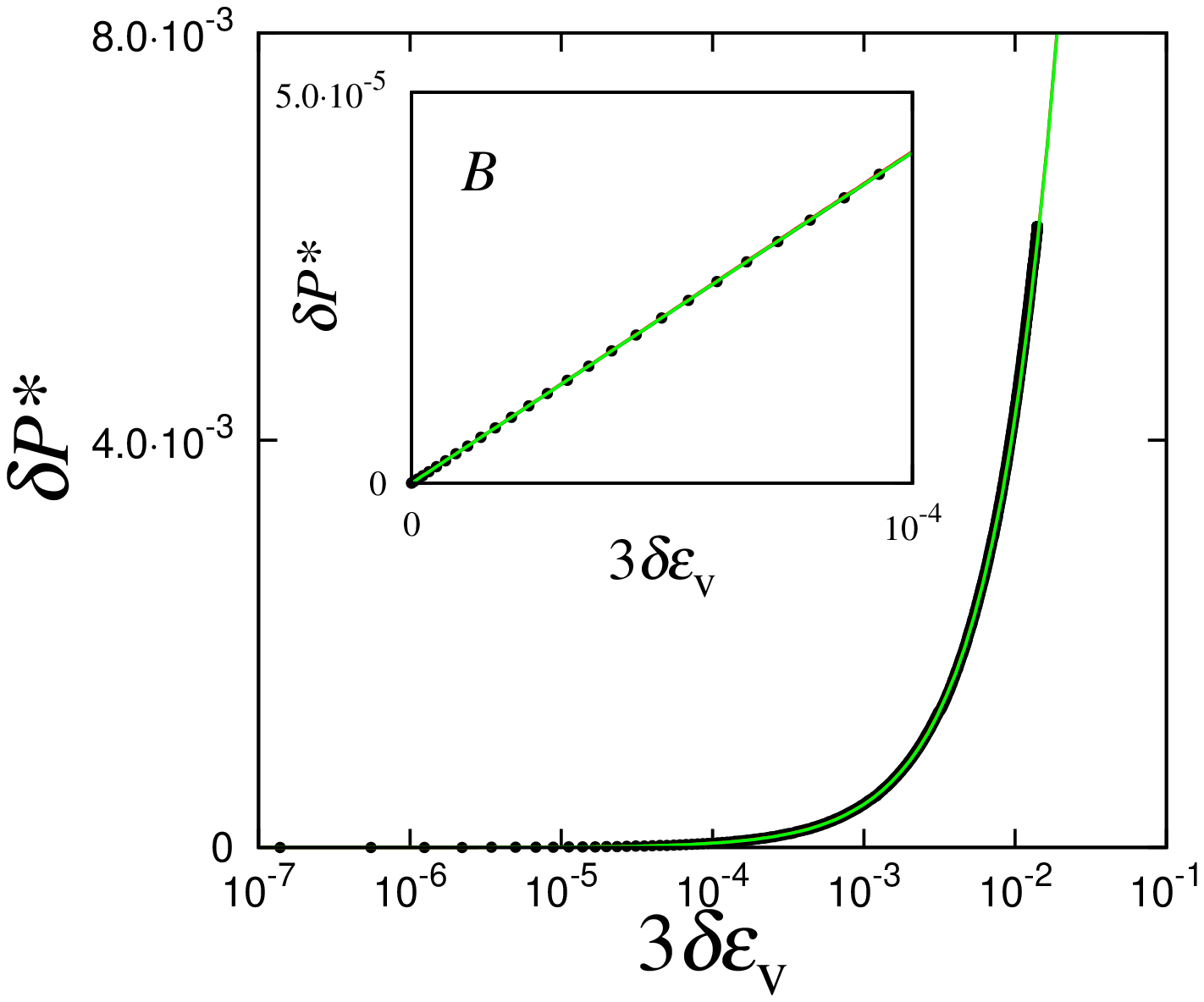}\label{dPdepsvB_longgg}}
\subfigure[]{\includegraphics[scale=0.25,angle=270]{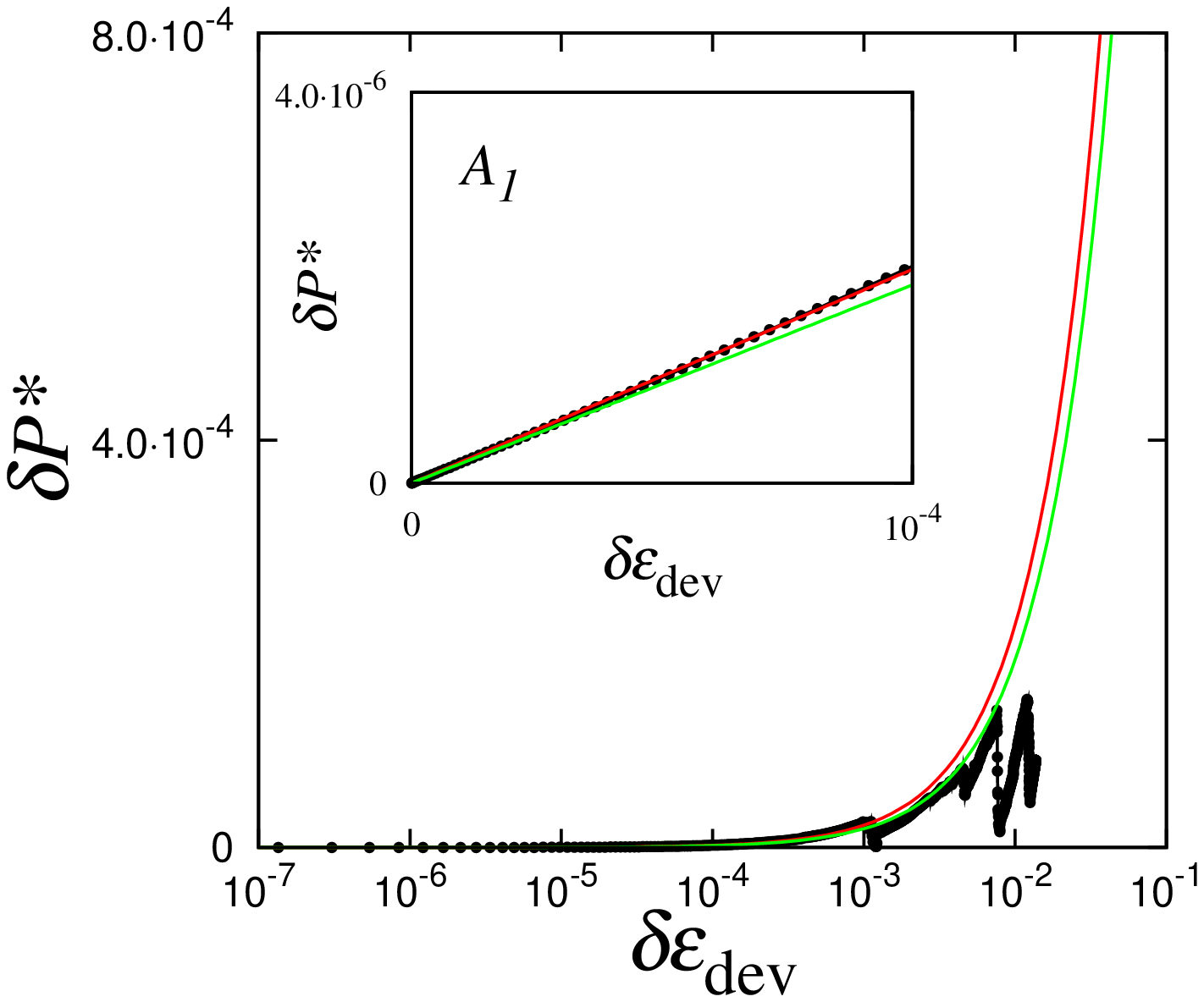}\label{dPdepsdevA_longgg}}
\subfigure[]{\includegraphics[scale=0.25,angle=270]{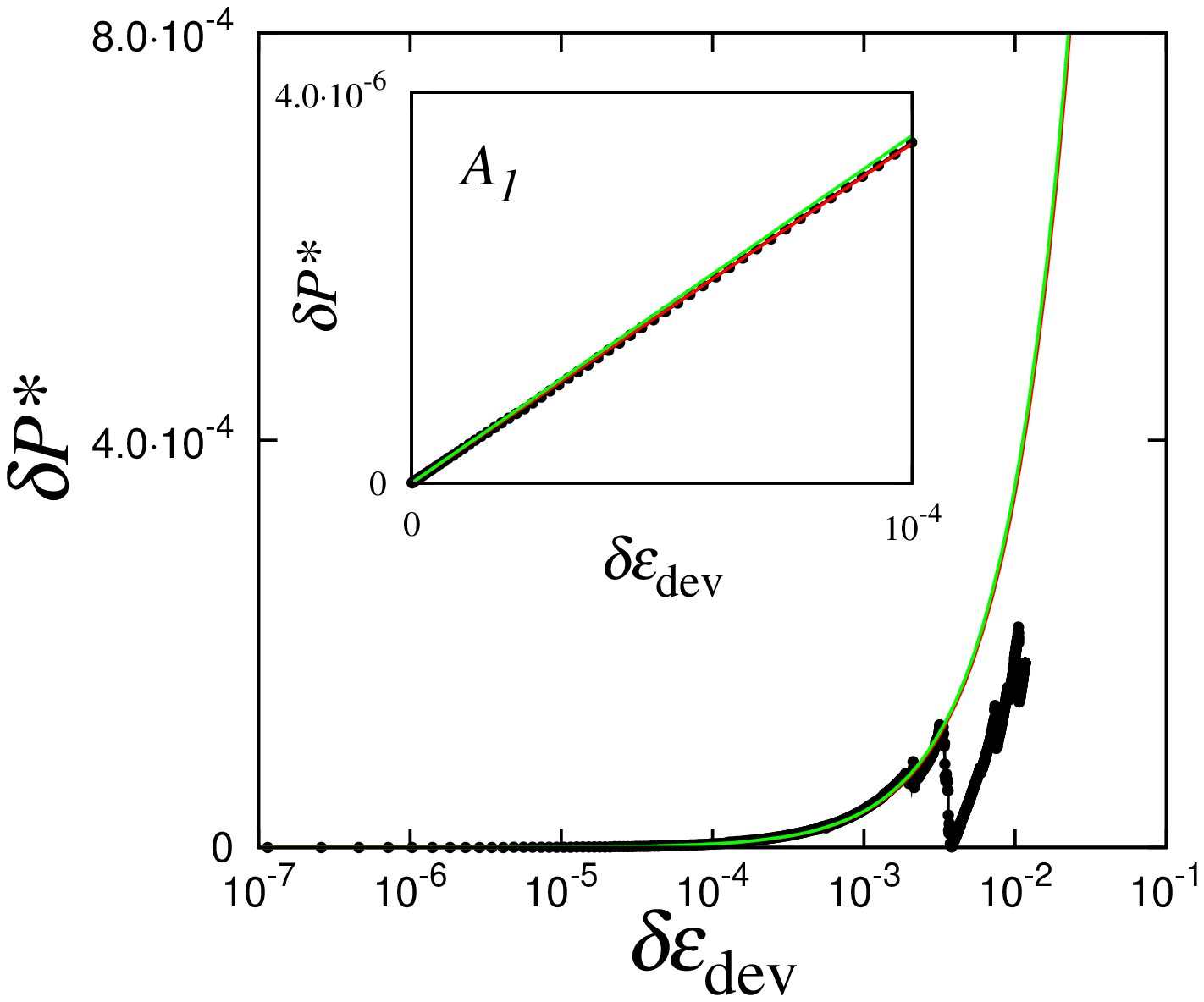}\label{dPdepsdevB_longgg}}\\
\hrule
\subfigure[]{\includegraphics[scale=0.25,angle=270]{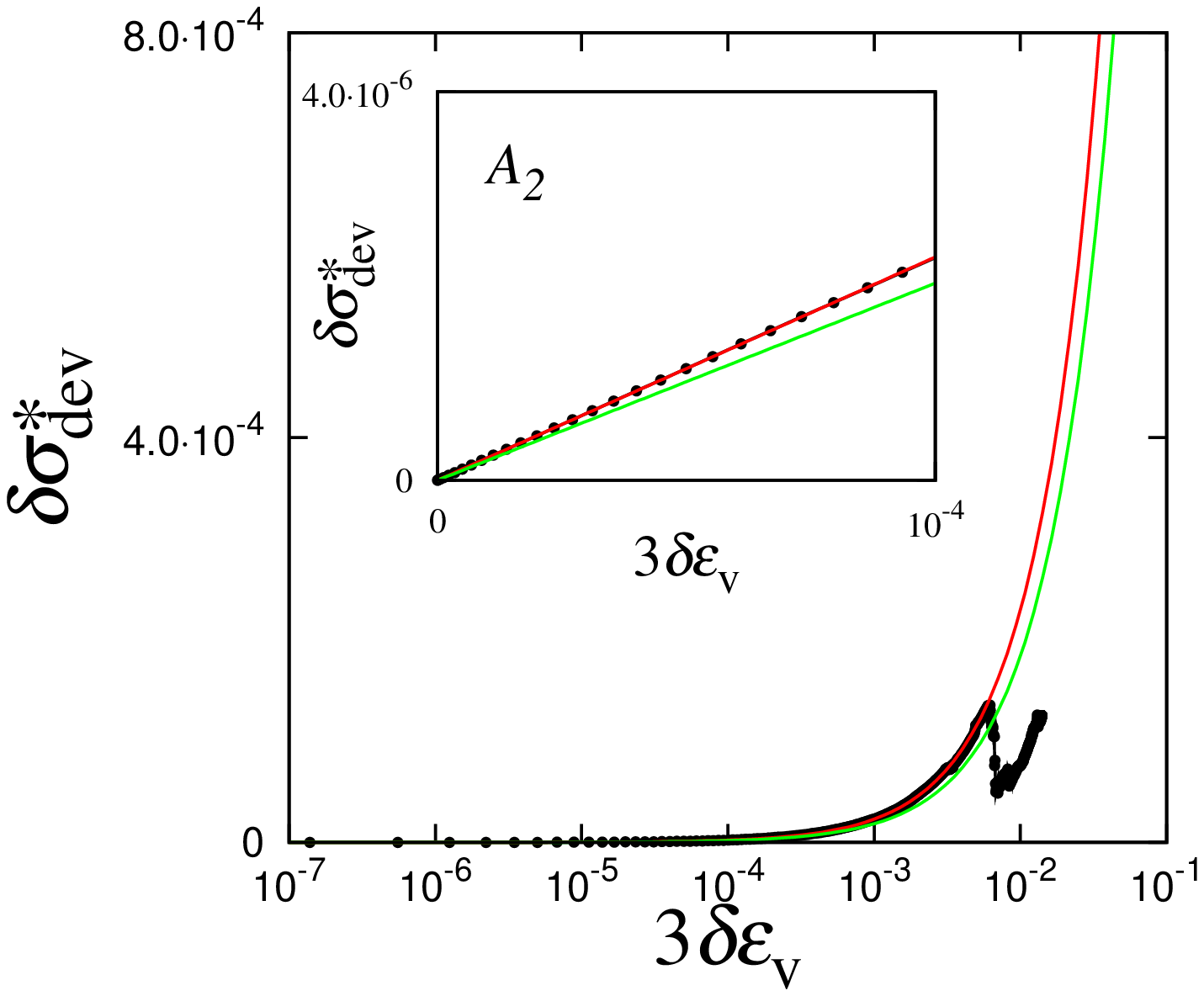}\label{dtaudepsvA_longgg}}
\subfigure[]{\includegraphics[scale=0.25,angle=270]{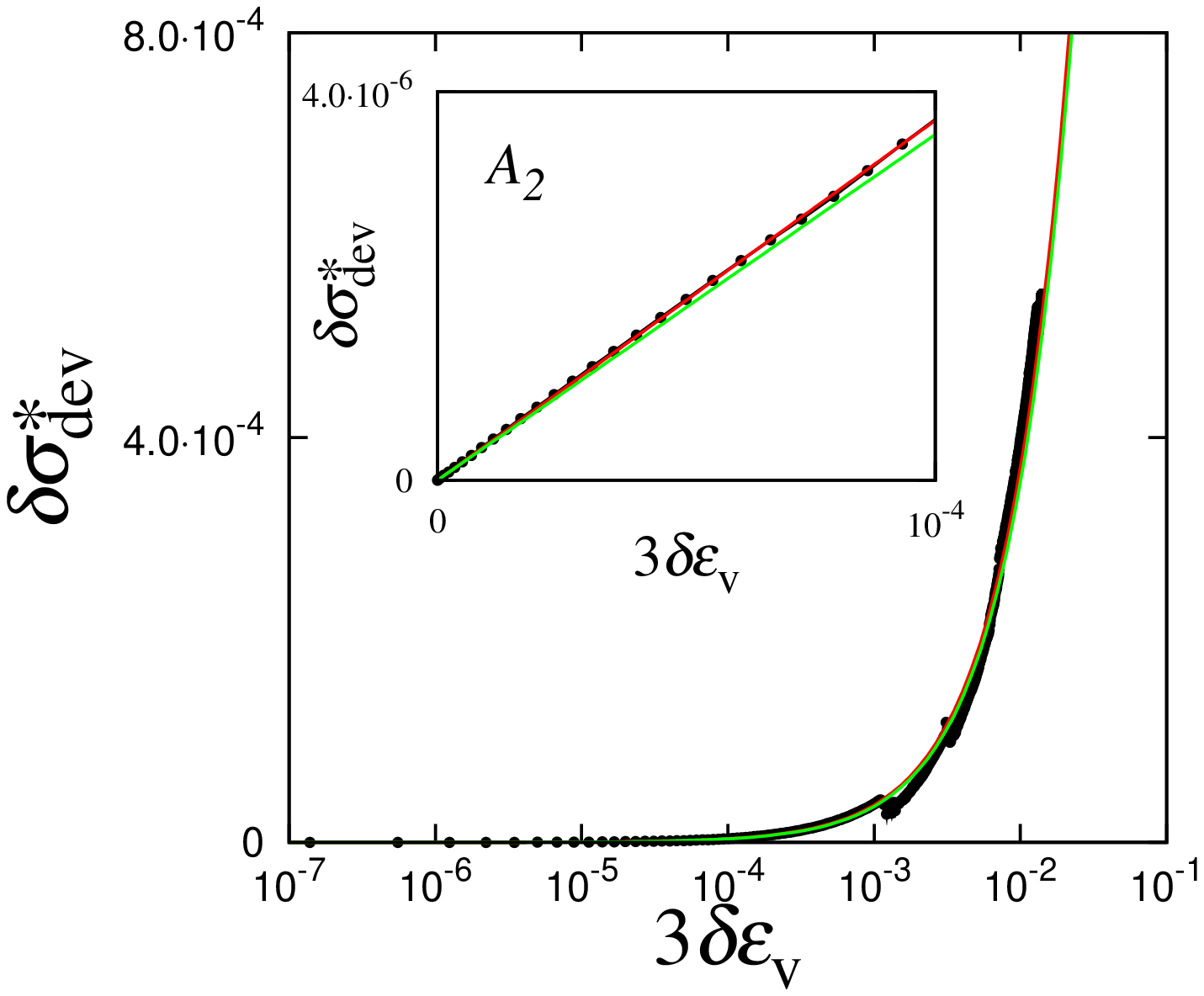}\label{dtaudepsvB_longgg}}
\subfigure[]{\includegraphics[scale=0.25,angle=270]{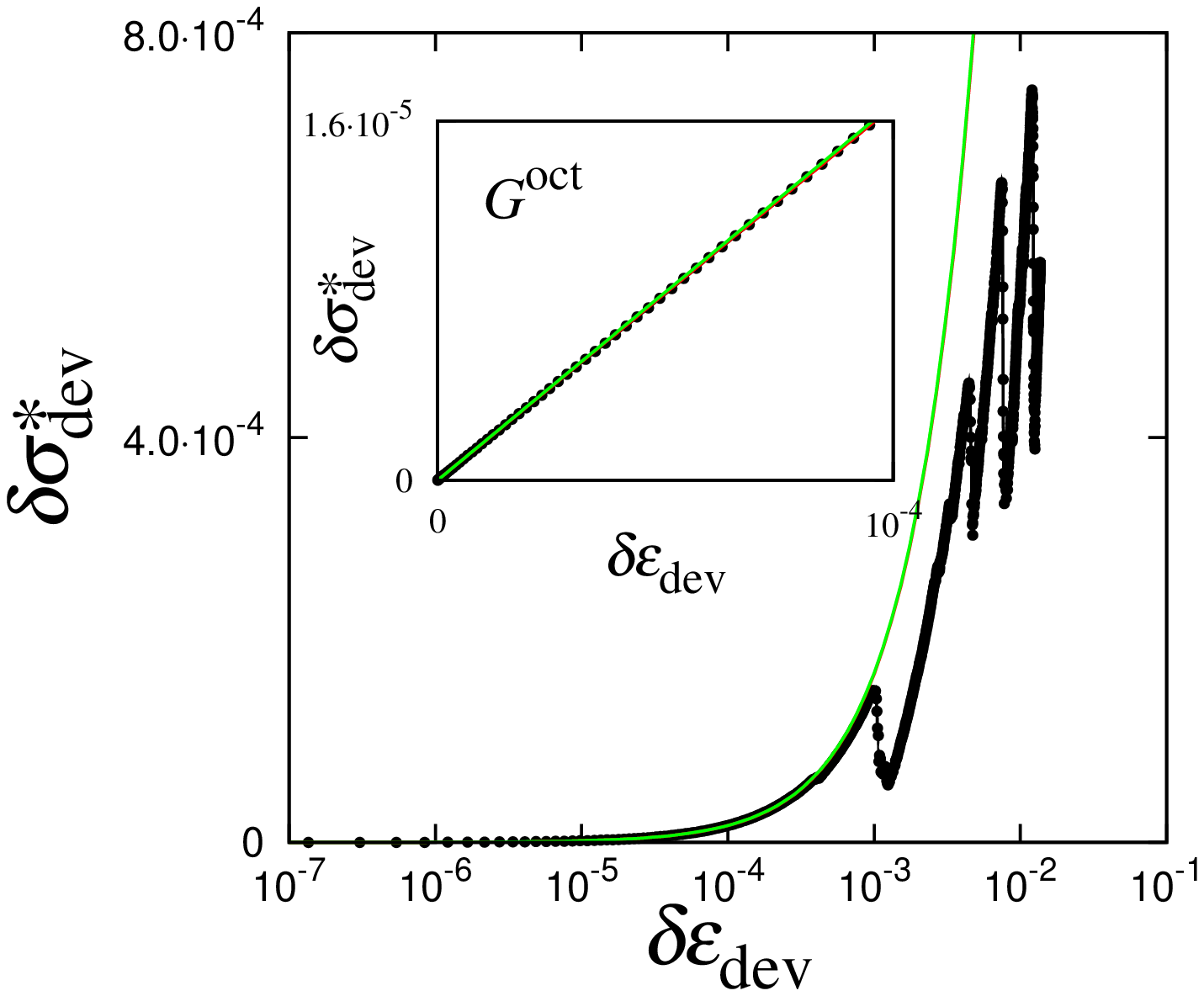}\label{dtaudepsdevA_longgg}}
\subfigure[]{\includegraphics[scale=0.25,angle=270]{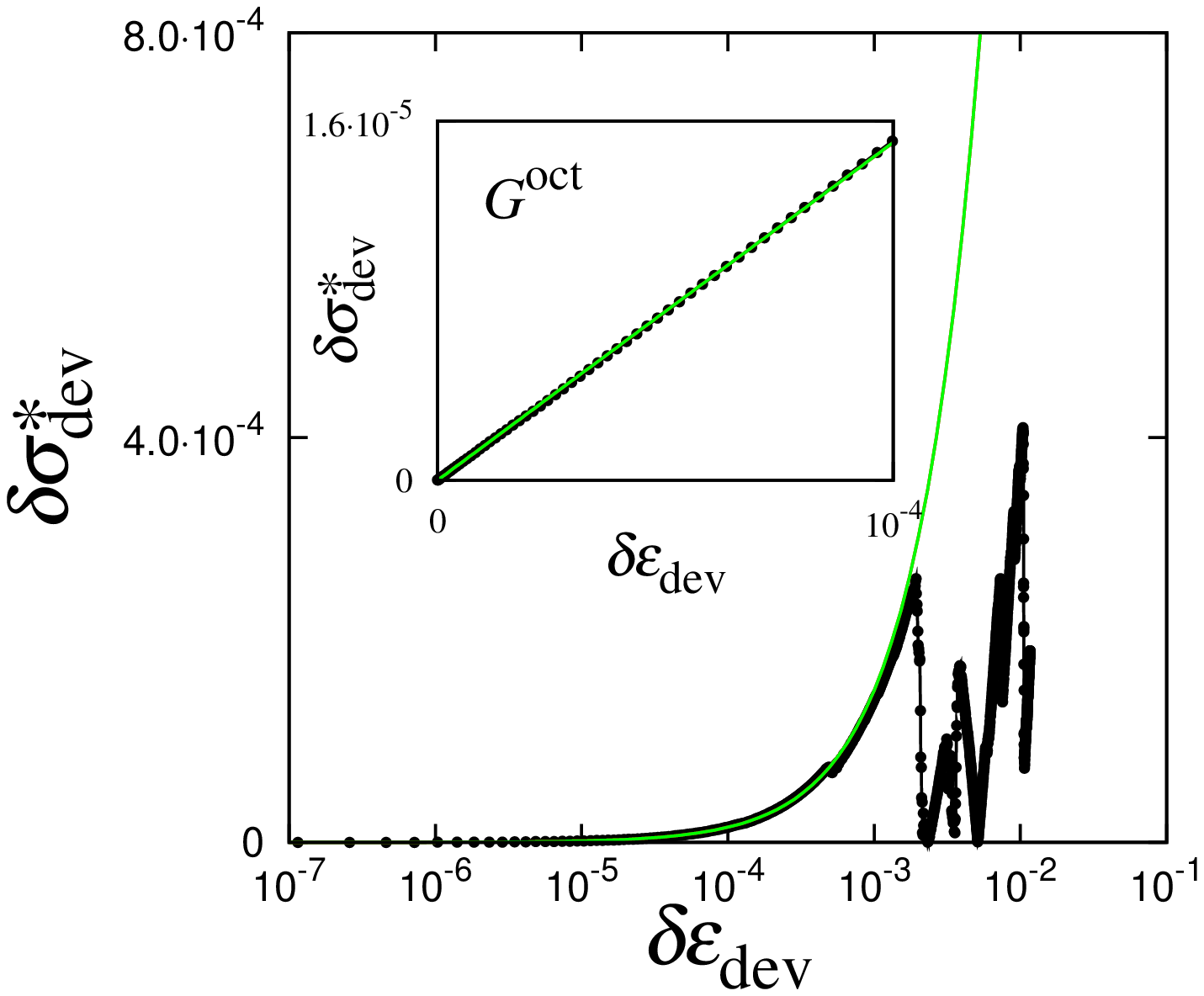}\label{dtaudepsdevB_longgg}}\\
\hrule
\subfigure[]{\includegraphics[scale=0.25,angle=270]{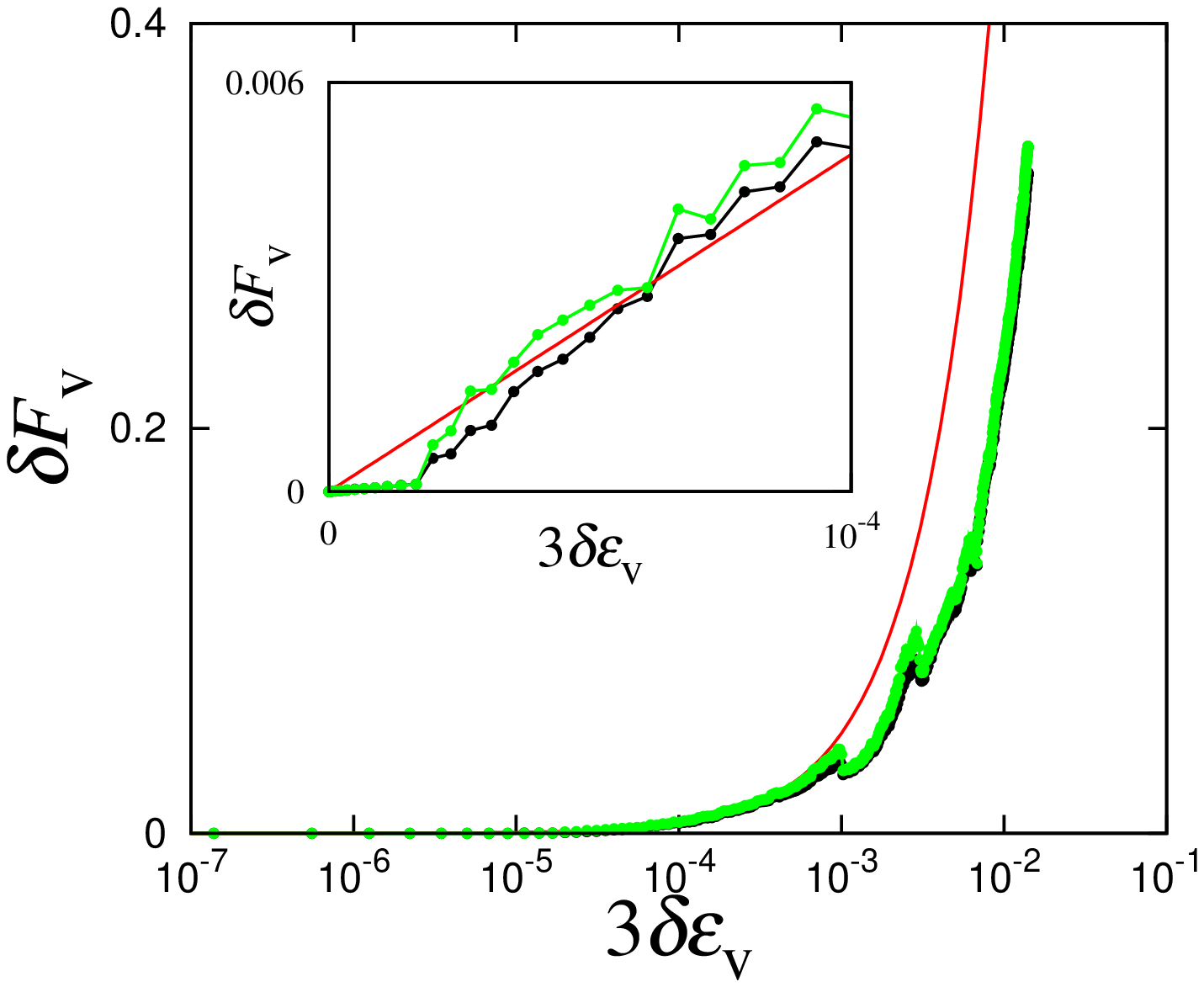}\label{dFvdepsvA_longgg}}
\subfigure[]{\includegraphics[scale=0.25,angle=270]{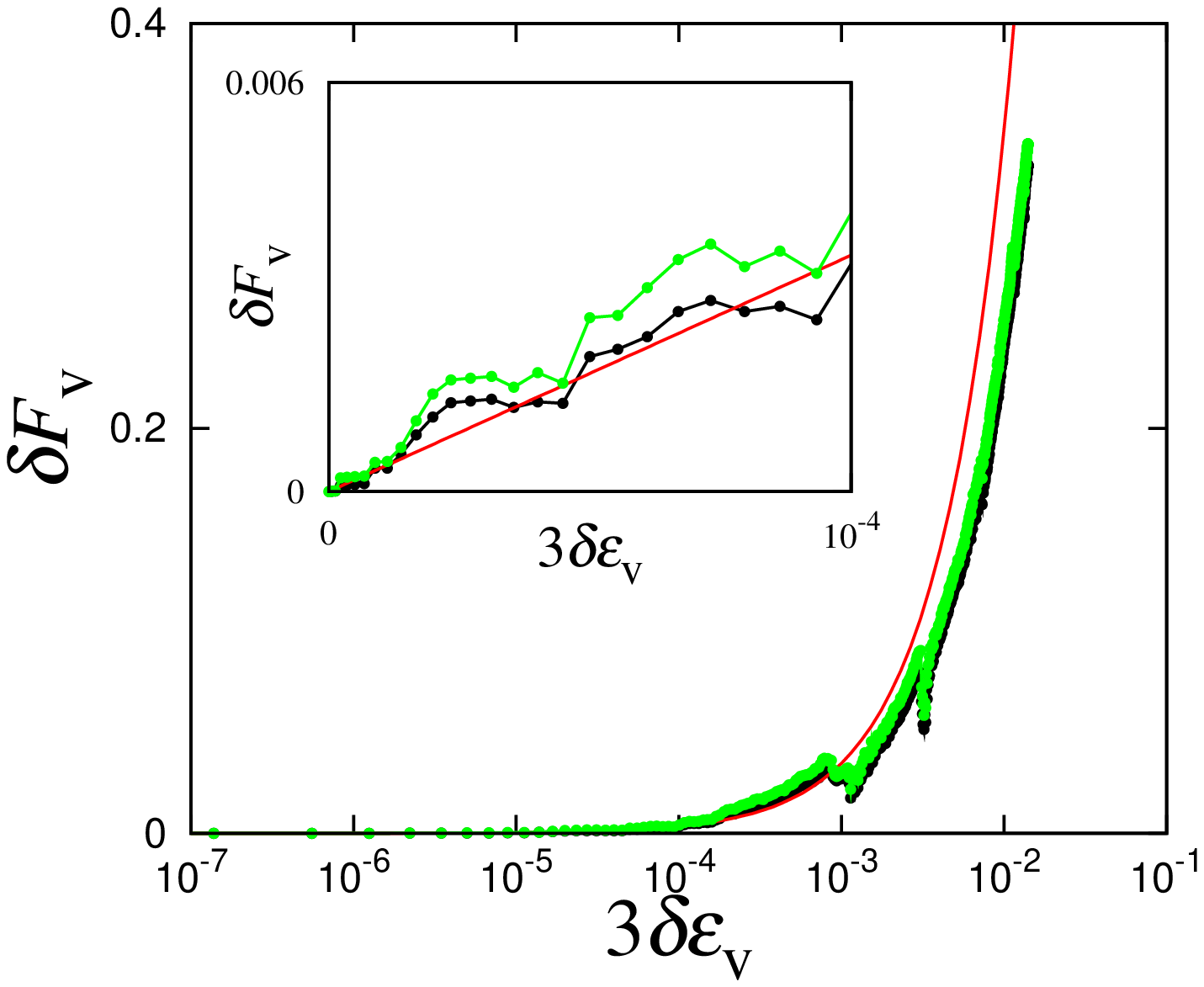}\label{dFvdepsvB_longgg}}
\subfigure[]{\includegraphics[scale=0.25,angle=270]{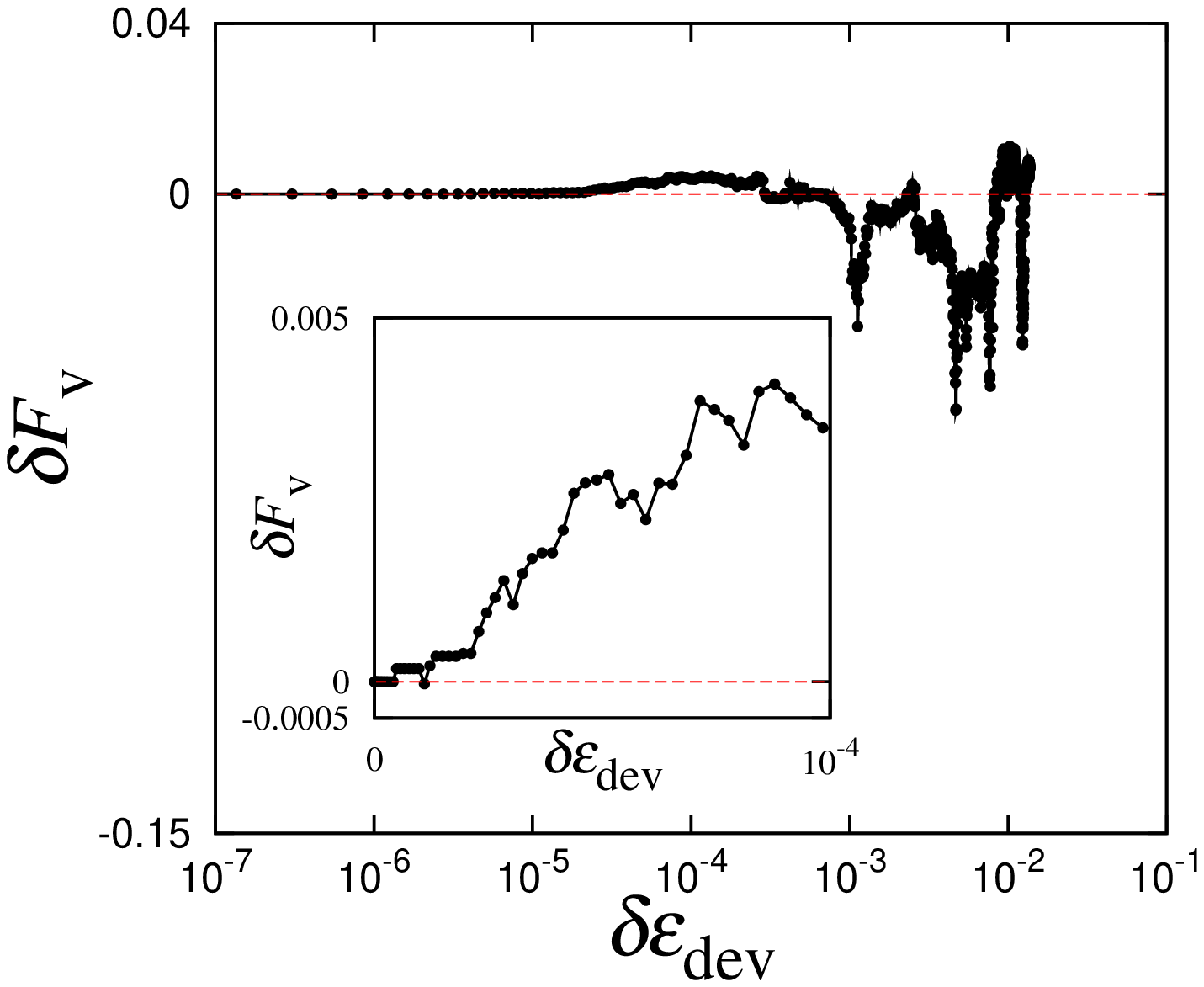}\label{dFvdepsdevA_longgg}}
\subfigure[]{\includegraphics[scale=0.25,angle=270]{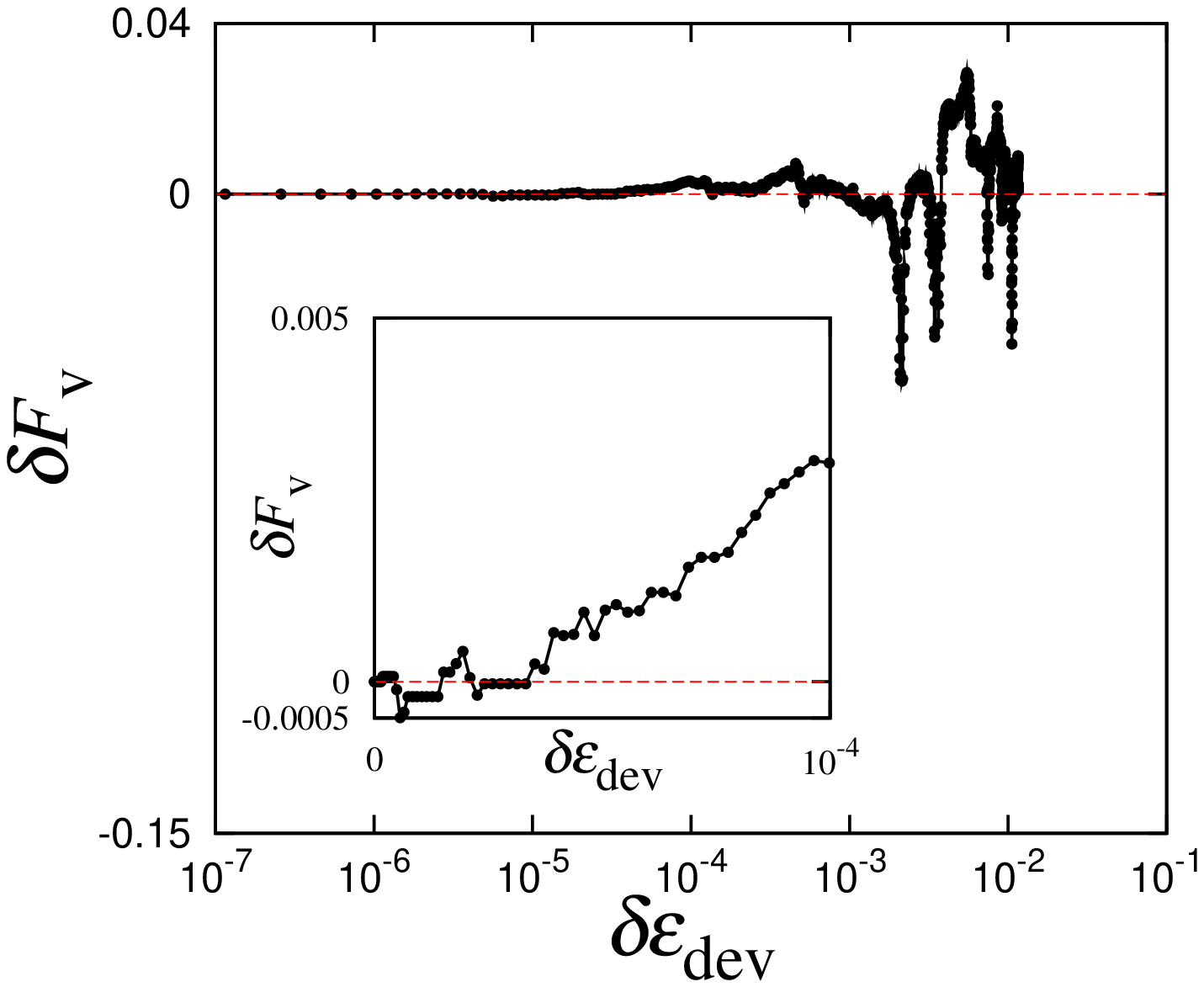}\label{dFvdepsdevB_longgg}}\\
\hrule
\subfigure[]{\includegraphics[scale=0.25,angle=270]{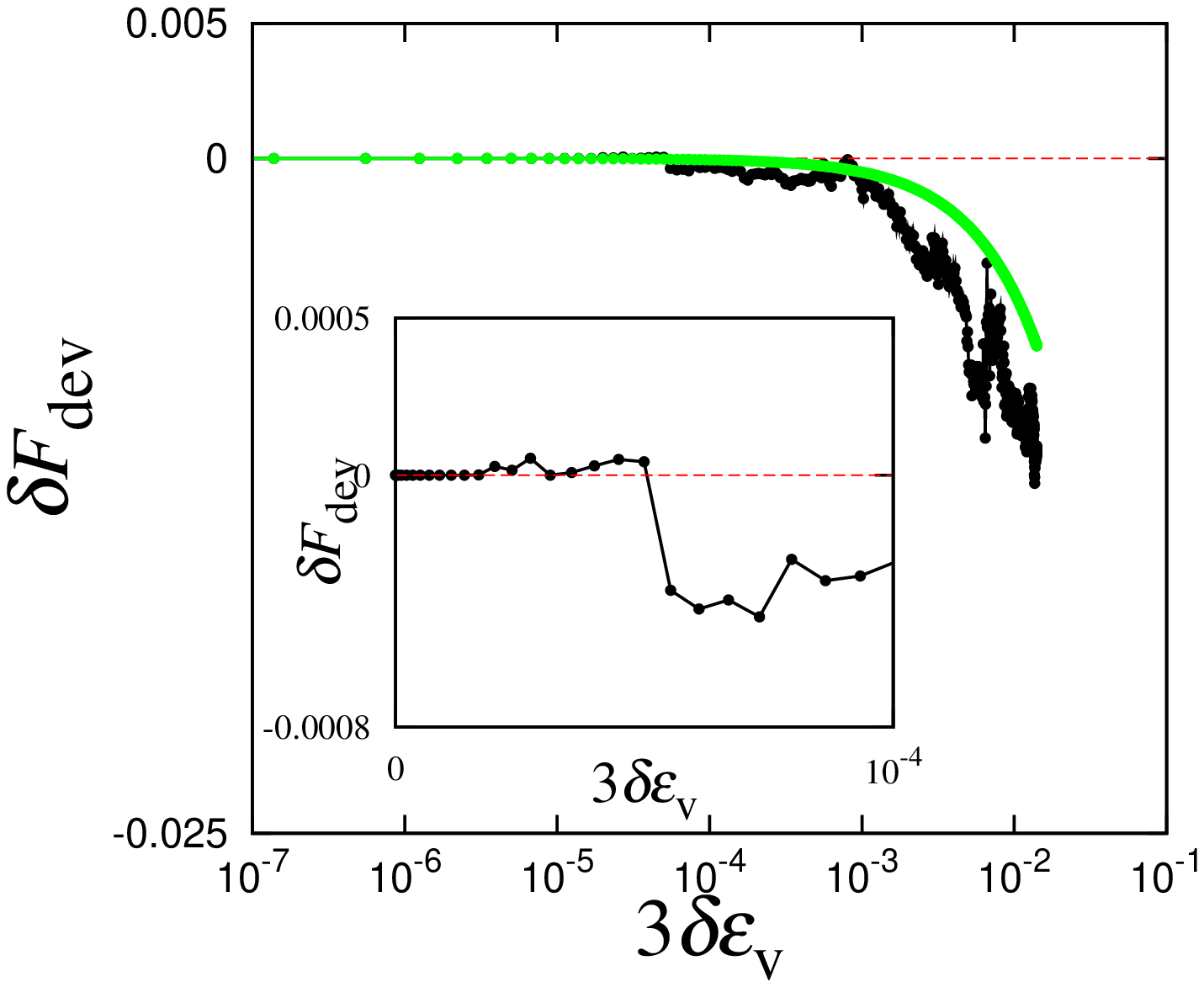}\label{dFdevdepsvA_longgg}}
\subfigure[]{\includegraphics[scale=0.25,angle=270]{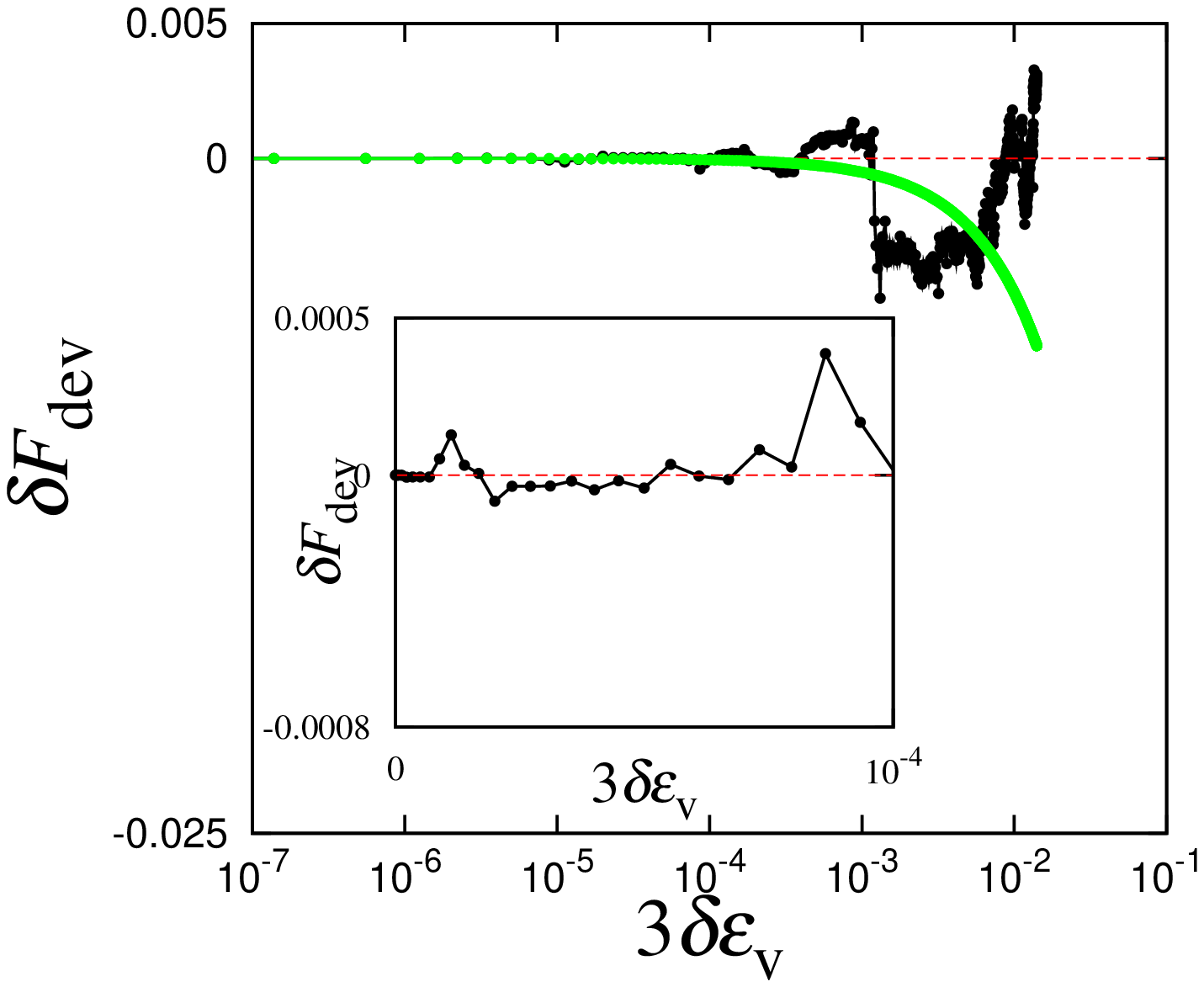}\label{dFdevdepsvB_longgg}}
\subfigure[]{\includegraphics[scale=0.25,angle=270]{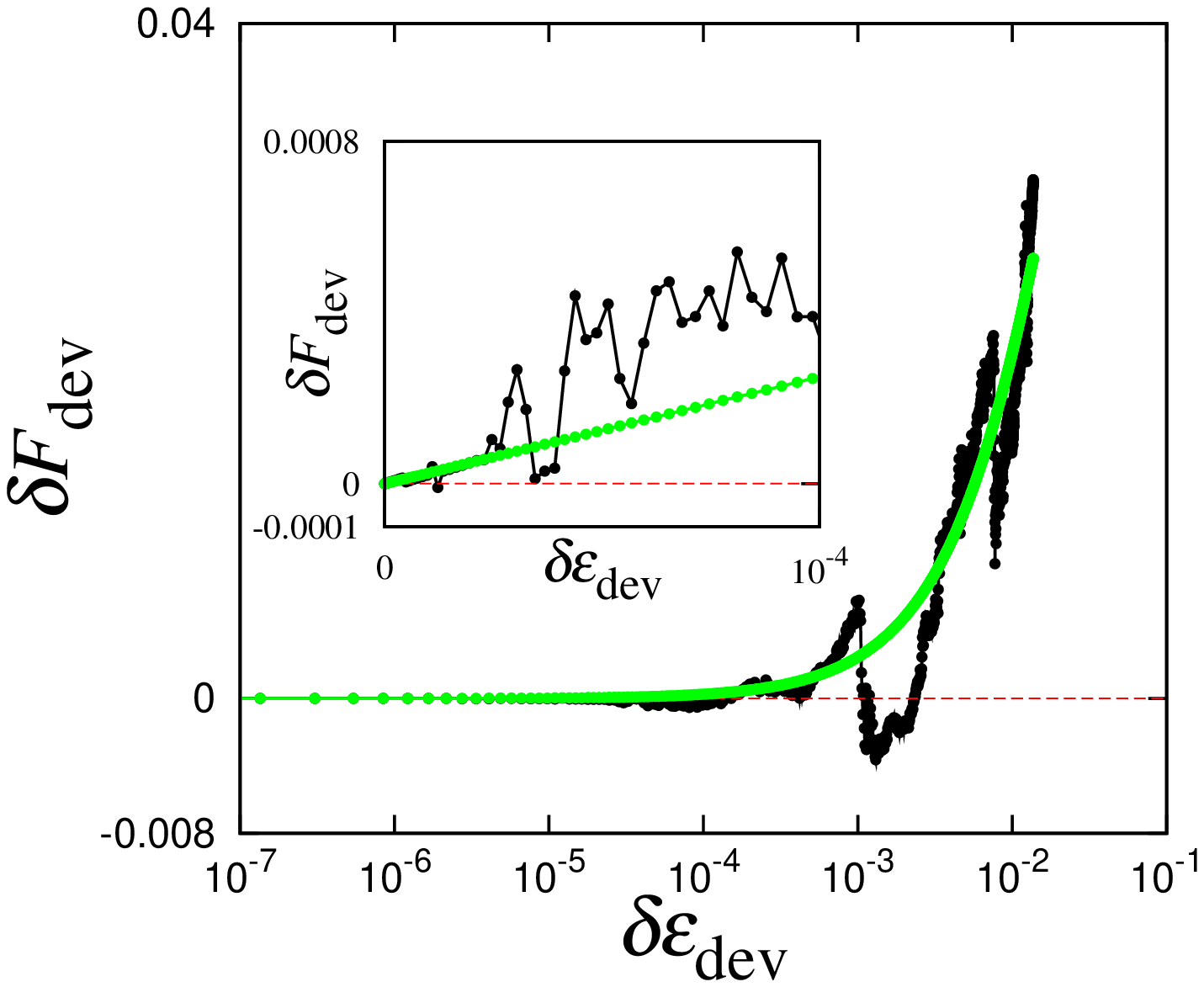}\label{dFdevdepsdevA_longgg}}
\subfigure[]{\includegraphics[scale=0.25,angle=270]{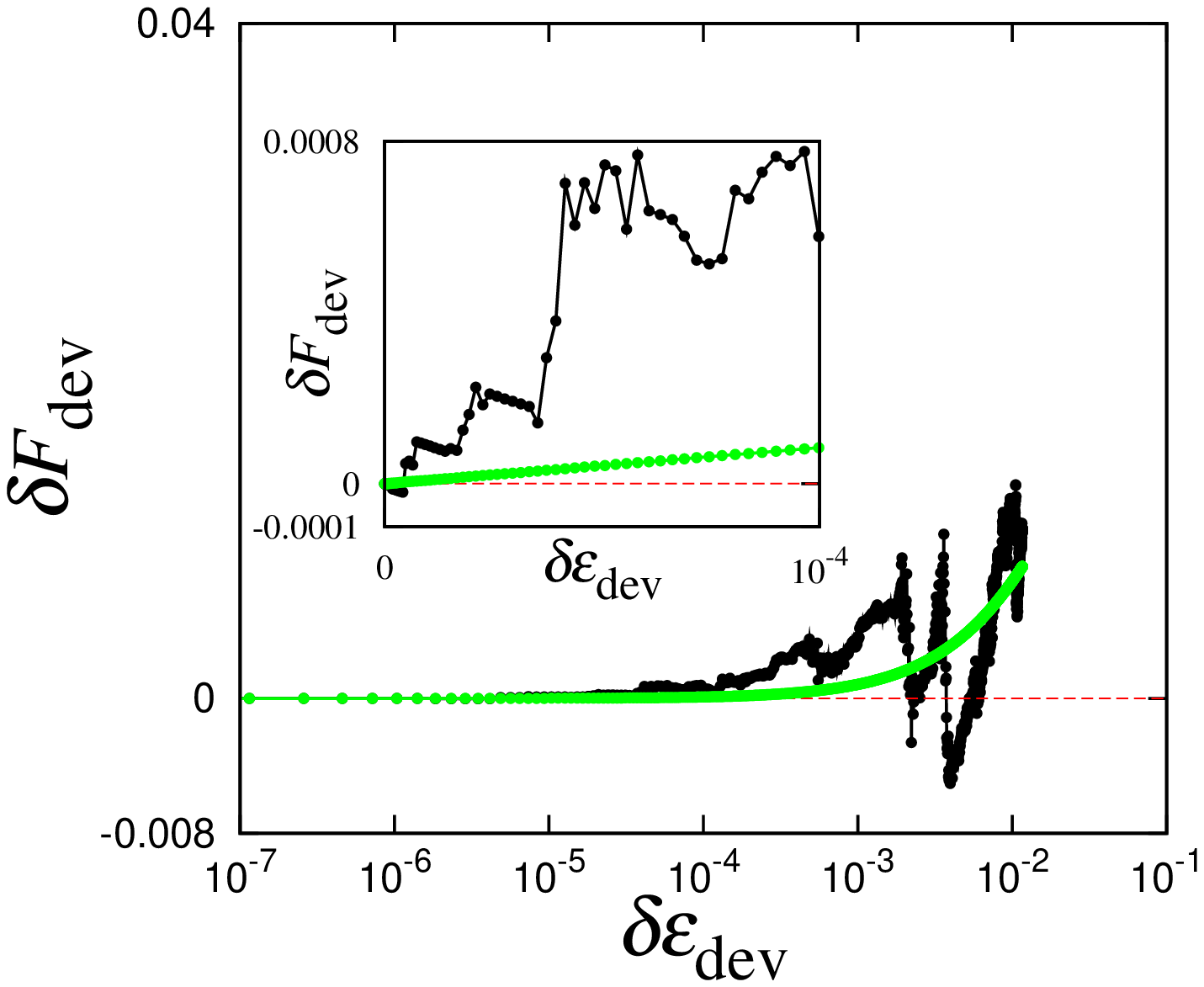}\label{dFdevdepsdevB_longgg}}
\caption{(Rows) Change of non-dimensional pressure $\delta\pkstar$, non-dimensional shear stress $\delta\sigmadkstar$, isotropic fabric $\delta\Fv$ and deviatoric fabric $\delta\Fdev$
versus strain amplitude. Column 1 and 2 represent purely isotropic while column 3 and 4 represent deviatoric perturbation experiments.  
The perturbation is applied to the state corresponding to $\epsd = 0.0065$ (nearly isotropic configuration: column 1 and 3) and 
$\epsd = 0.31$ (steady state configuration: column 2 and 4) of the main deviatoric experiment with volume fraction $\nu = 0.706$. 
Note that the $x$-axis is log-scale, with inset plots in linear scale. 
The red line passing through the dataset in (a-j) represents a linear fit in the elastic regime for $3\delta \epsiso; \delta \epsd < 10^{-4}$. 
The analytical predictions for the elastic range from our results section\ \ref{sec:perturbresults} 
in Eqs.\ (\ref{eq:B})--(\ref{eq:G}) are plotted as green line in (a--h). 
The green line in (i) and (j) represents $\Fv=g_3 \nu C$ calculated using Eq.\ (\ref{eq:Fveqn}), when subtracted from its initial value.
The dashed horizontal line in (k)--(p) represents zero. 
The green line in (m) and (n) represent the evolution of change in deviatoric fabric $\delta\Fdev$ in critical state using parameters from Table 3 of Ref.\ \cite{imole2013hydrostatic}, 
with the assumption that the new state after volumetric deformation is also in critical state.
The green line in (o) and (p) represents Eq.\ (18) from Ref.\ \cite{imole2013hydrostatic} when subtracted from its 
initial value $\Fdev^\mathrm{0}=0.03$ for (o) and $\Fdev^\mathrm{0}=0.113$ for (p), with the growth rate $\betaF=39$ and $\Fdev^\mathrm{max}=0.12$.  }
\label{LONG}
\end{figure}

Since we are interested in measuring the pure elastic response of the material, we take care that no rearrangements happen in the system
during the numerical probe, that is ${3\delta\epsiso}$ and ${\delta\epsd}$ are 
applied only up to $ 10^{-4}$ (with very slow wall movement rate $\sim 10^{-6}$,i.e.,\ smaller than for the main large shear strain preparation experiment). 
Looking at Fig.\ \ref{LONG}, we note that much bigger drops appear in the deviatoric response when the isotropic perturbation is applied.
Vice-versa, the fluctuations/drops are much larger in pressure rather than in shear stress, when we deal with deviatoric perturbations. 
It is worthwhile to mention here that we have tested our method by applying strain perturbations in opposite directions 
i.e.,\ ${3\delta\epsiso}$ and $-{3\delta\epsiso}$, ${\delta{\epsd}}$ and $-{\delta{\epsd}}$.
This does not lead to any difference in the elastic response, as long as we stay in the limit of elastic perturbations.

\begin{figure}[!ht]
\centering
\subfigure[]{\includegraphics[scale=0.25,angle=270]{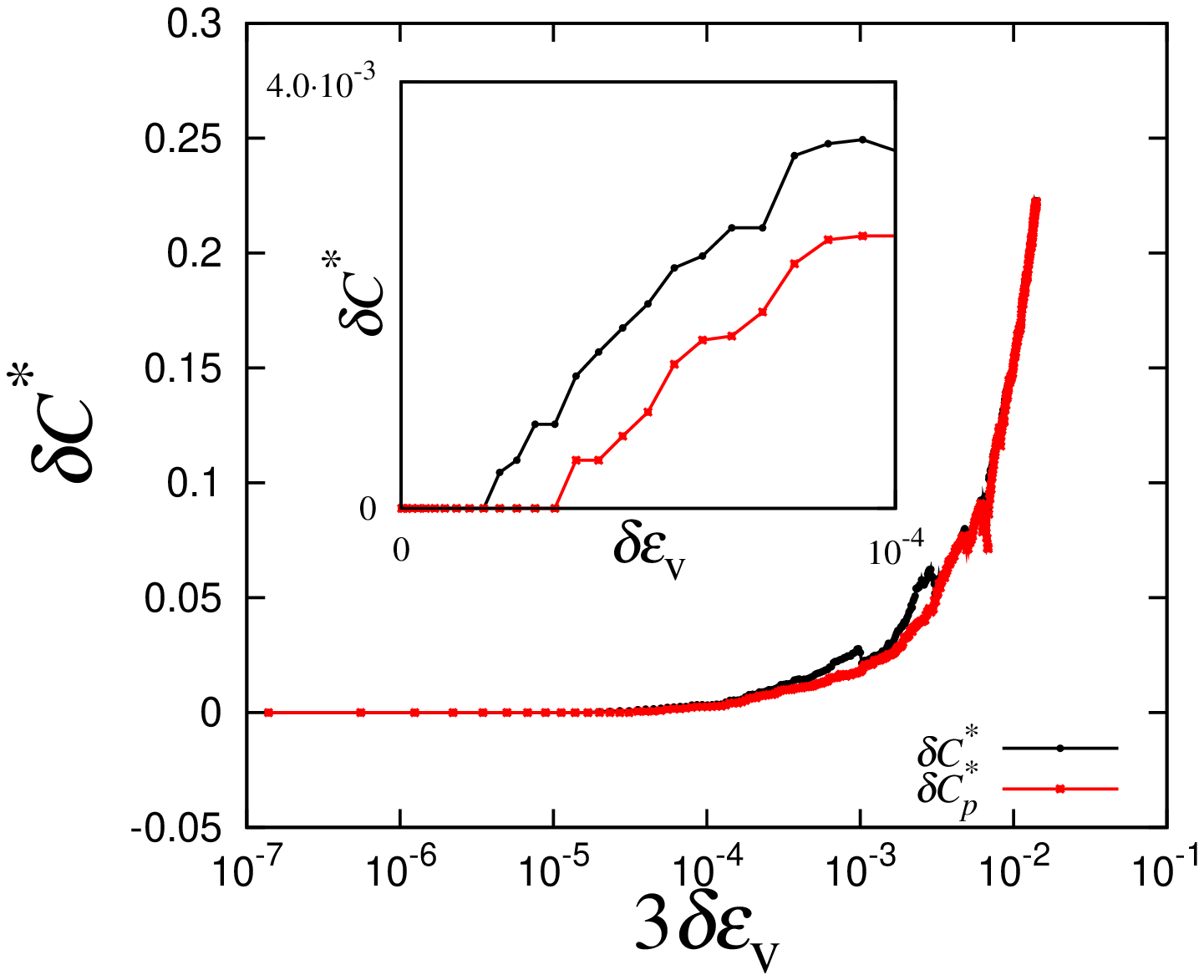}\label{dCstararvA_longgg}}
\subfigure[]{\includegraphics[scale=0.25,angle=270]{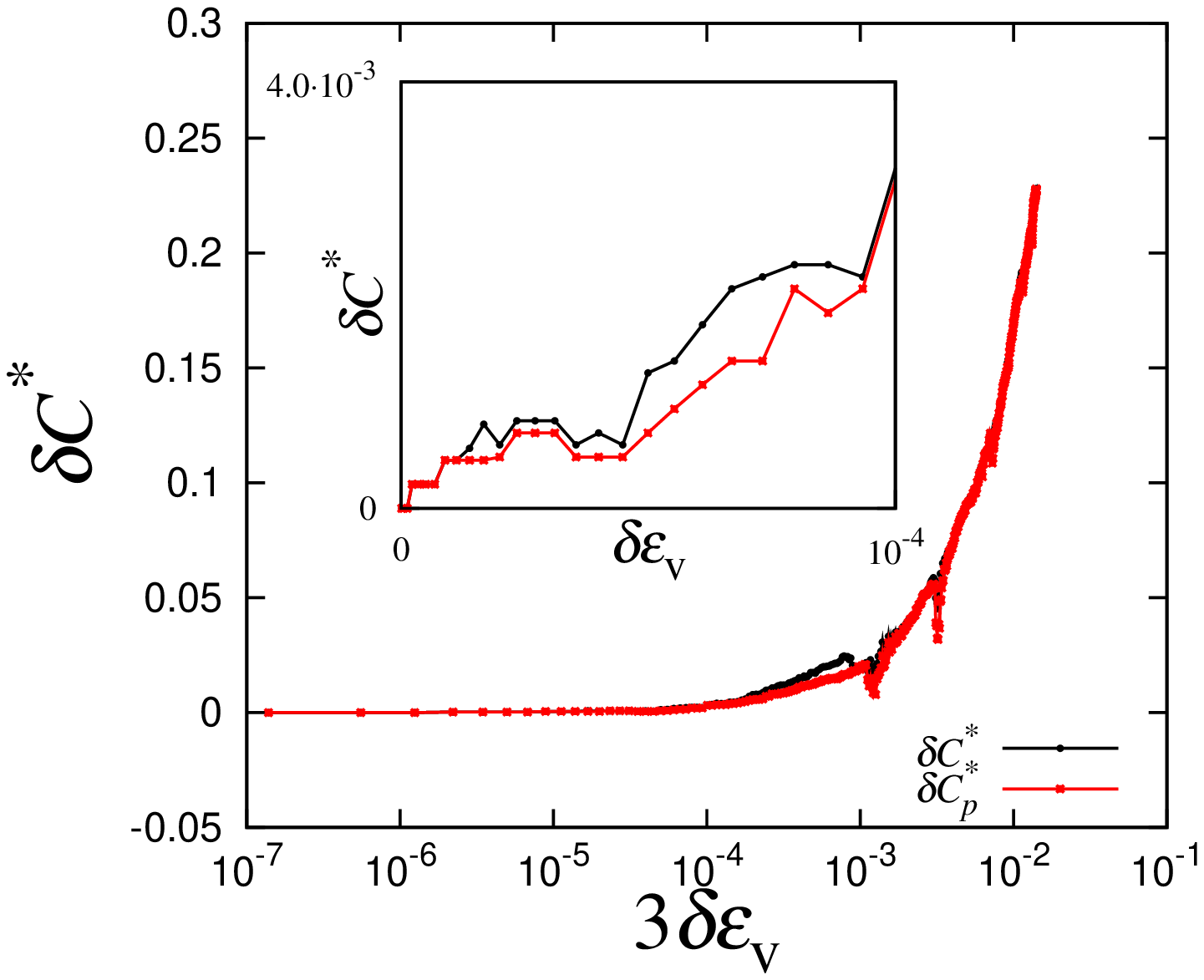}\label{dCstararvB_longgg}}
\subfigure[]{\includegraphics[scale=0.25,angle=270]{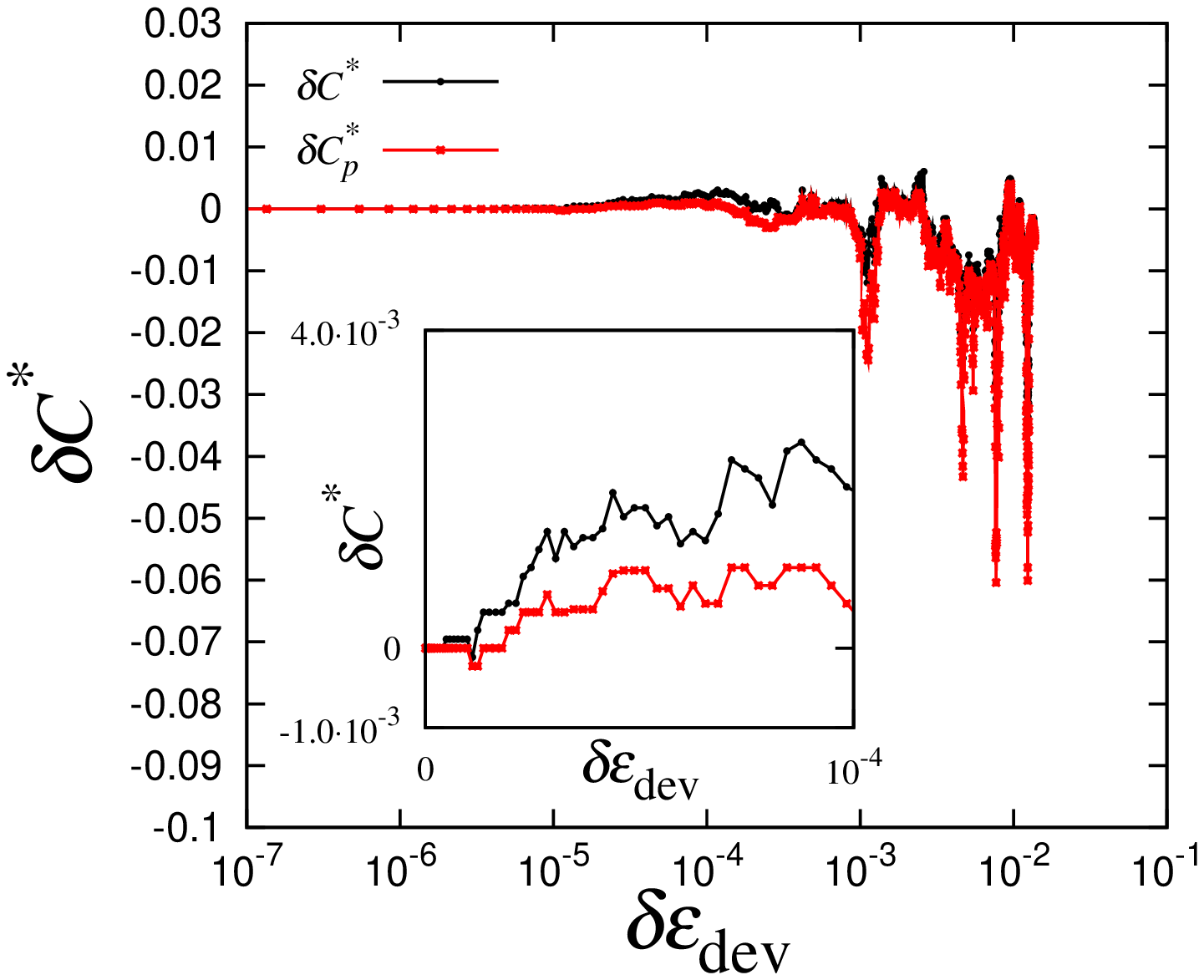}\label{dCstarardevA_longgg}}
\subfigure[]{\includegraphics[scale=0.25,angle=270]{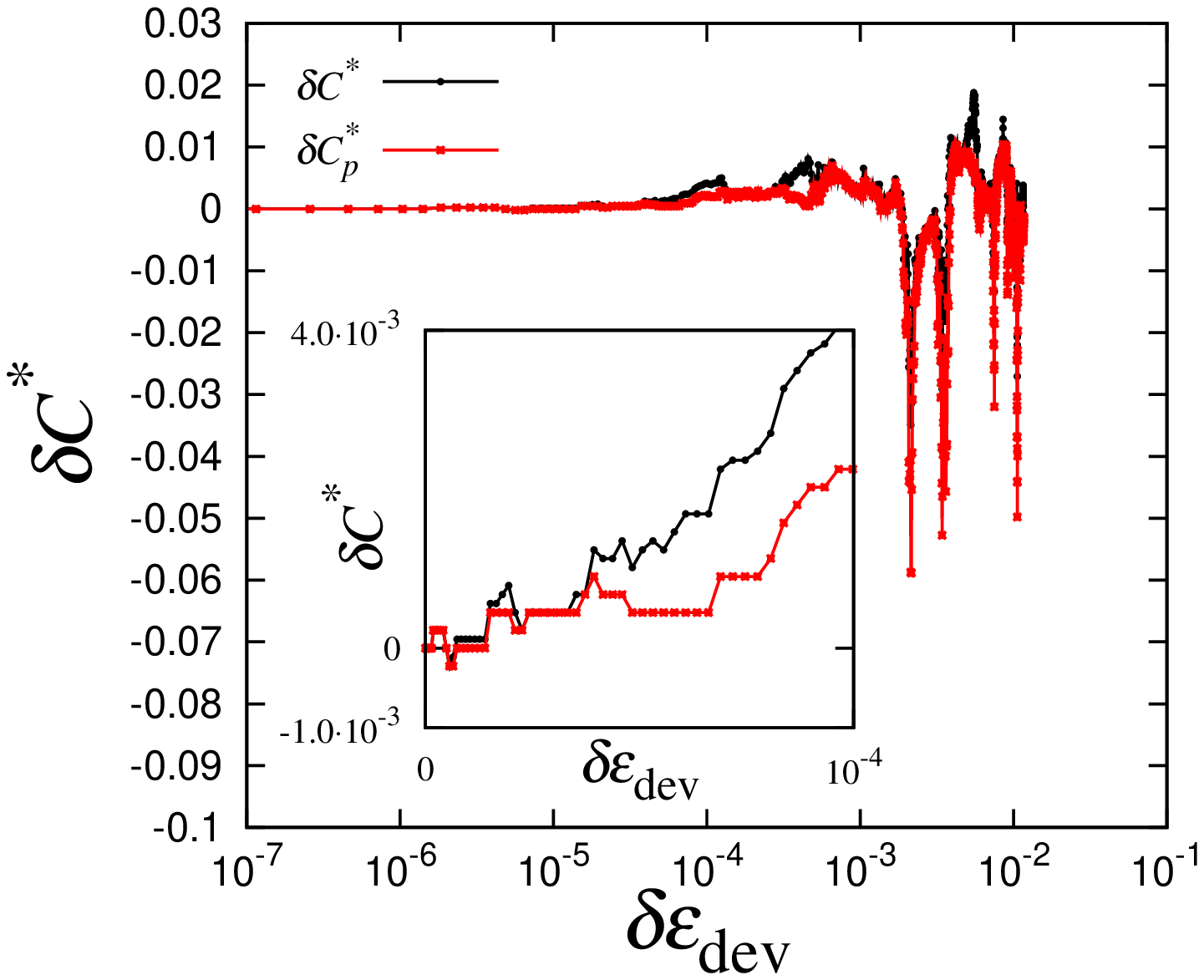}\label{dCstarardevB_longgg}}
\caption{Change of the coordination number $\delta C^* = \delta \left(M_4/N_4\right)$ (black `$\bullet$' curve) and 
the modified coordination number $\delta C^*_p = \delta \left(M_4^p/N_4\right)$ (red `*' curve), defined in section\ \ref{sec:simmeth}, 
versus strain amplitude during purely (a--b) isotropic and (c--d) deviatoric perturbation experiments (corresponding plots as in Fig.\ \ref{LONG}). 
The perturbation is applied to the state corresponding to $\epsd = 0.0065$ (nearly isotropic configuration: (a) and (c)) and 
$\epsd = 0.31$ (steady state configuration: (b) and (d)) of the main deviatoric experiment with volume fraction $\nu = 0.706$. 
Note that the $x$-axis is on log-scale, with inset plots in linear scale.}
\label{smallperlongp_C}
\end{figure}

\begin{figure}[!ht]
\centering
\subfigure[]{\includegraphics[scale=0.44,angle=270]{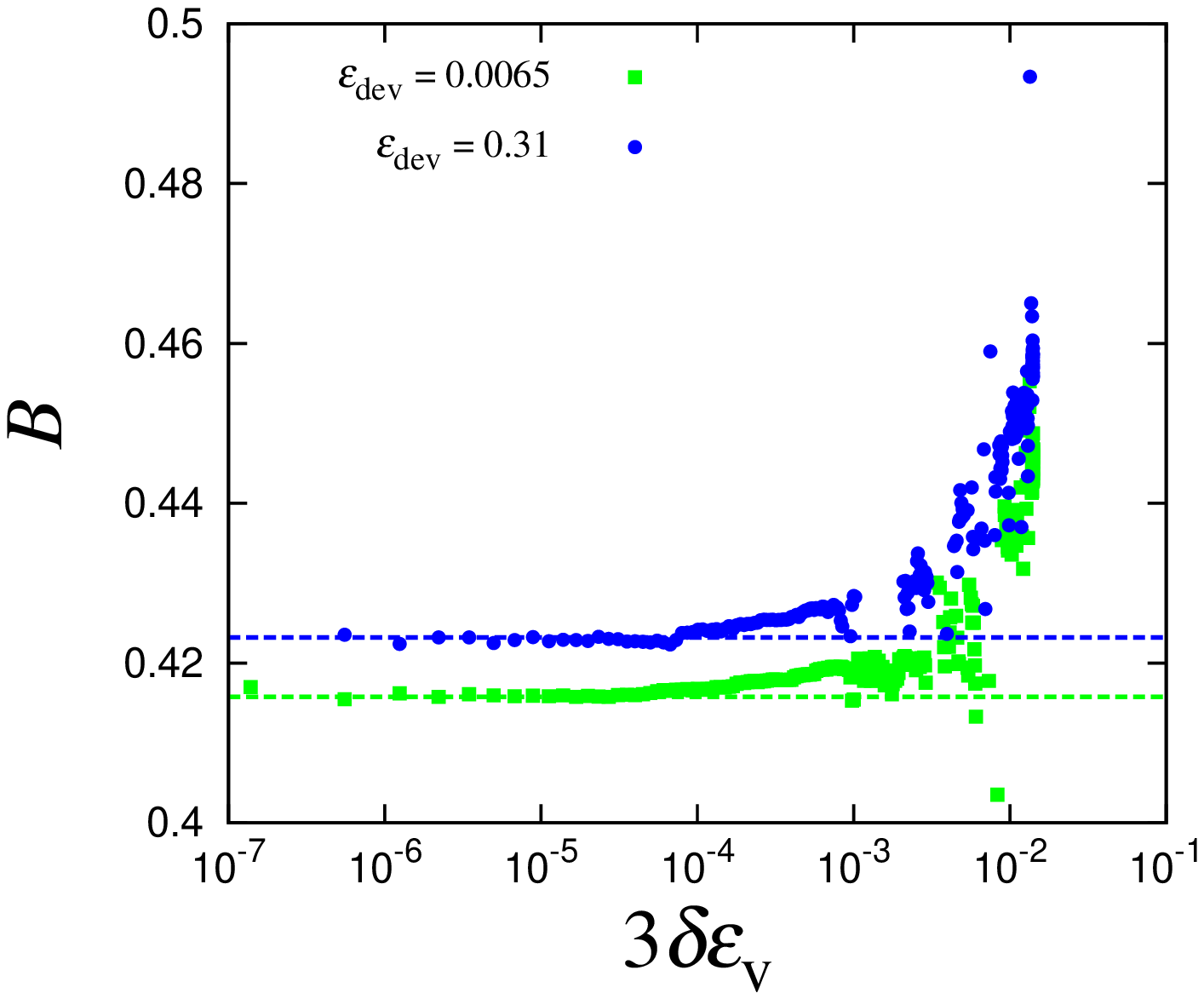}}\label{Bampli}
\subfigure[]{\includegraphics[scale=0.44,angle=270]{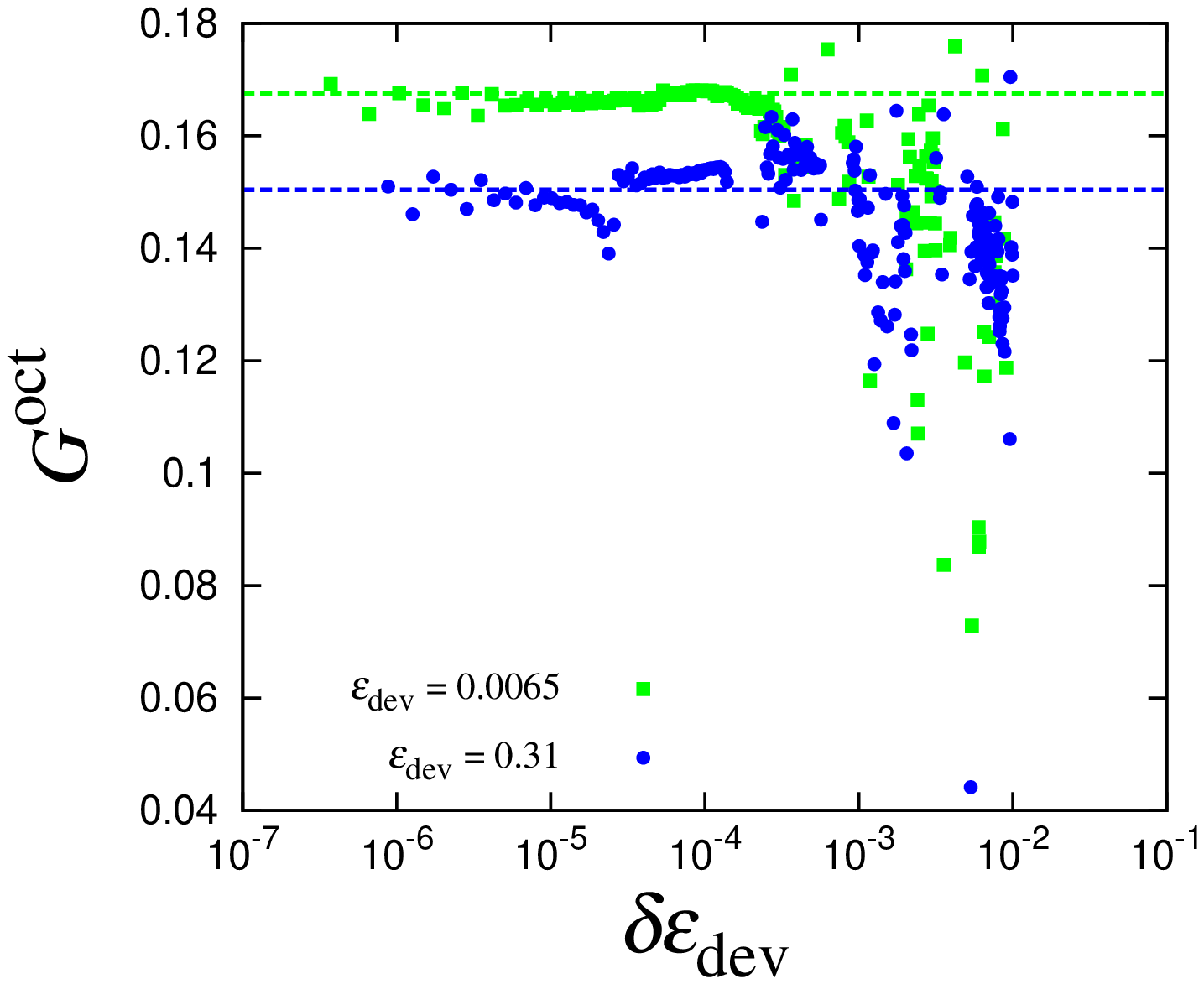}}\label{Gampli}
\caption{Evolution of (a) bulk modulus $B$ and (b) octahedral shear modulus $\Goct$ with the respective applied isotropic $3\delta\epsiso$ and deviatoric $\delta\epsd$ strain amplitudes for a 
state corresponding to $\epsd = 0.0065$ (nearly isotropic configuration: green `$\blacksquare$') and 
$\epsd = 0.31$ (steady state configuration: blue `$\bullet$') of the main deviatoric experiment with volume fraction $\nu = 0.706$. 
Corresponding dashed horizontal lines represents the initial values of $B$ and $\Goct$.}
\label{BGampli}
\end{figure}

We test the rearrangements argument in Fig.\ \ref{BGampli}, by plotting the calculated bulk modulus $B$ and octahedral shear modulus $\Goct$ against 
the amplitude of the applied isotropic $3\delta \epsiso$ and deviatoric $\delta \epsd$ strain, respectively, for 
states at $\epsd = 0.0065$ and $0.31$ (nearly isotropic and steady state configurations, respectively) of the main deviatoric experiment.
Both $B$ and $\Goct$ stay practically constant for small amplitudes and we can assume the regime to be linear elastic \cite{chung2013how}. 
At $3\delta \epsiso\simeq10^{-4}$, the first change in the number of contacts
happens (Fig.\ \ref{smallperlongp_C}(a--b)) and $B$ starts to increases non-linearly.
Similarly, when $\epsd\simeq10^{-4}$, the first change in the number of contacts
happens (Fig.\ \ref{smallperlongp_C}(c--d)) and $\Goct$ starts to decay. 
It is interesting to notice that for both $B$ and $\Goct$, the elastic regime shrinks when the main 
deviatoric strain $\epsd$ increases (Fig.\ \ref{BGampli}) 
and, also, when the volume fraction reduces, going towards the jamming volume fraction (data not shown).
A similar modulus may be plotted for fabric as $\delta\Fdev/\delta\epsd$ that, due to the finite size of the system, 
would be identically zero, until the first rearrangement occurs (see Fig.\ \ref{LONG}).
%

\begin{figure}[!ht]
\centering
\subfigure[]{\includegraphics[scale=0.25,angle=270]{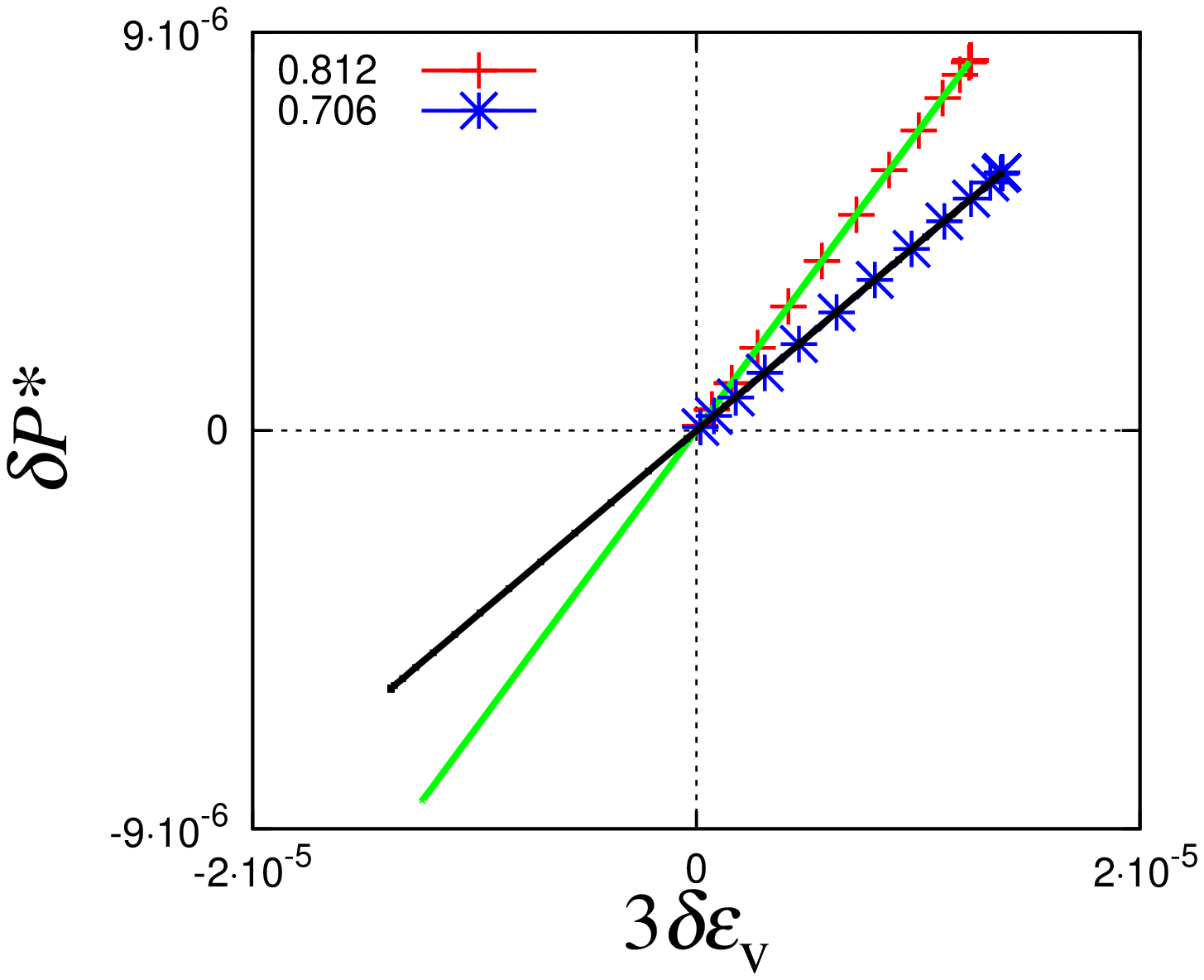}\label{P_V_small}}
\subfigure[]{\includegraphics[scale=0.25,angle=270]{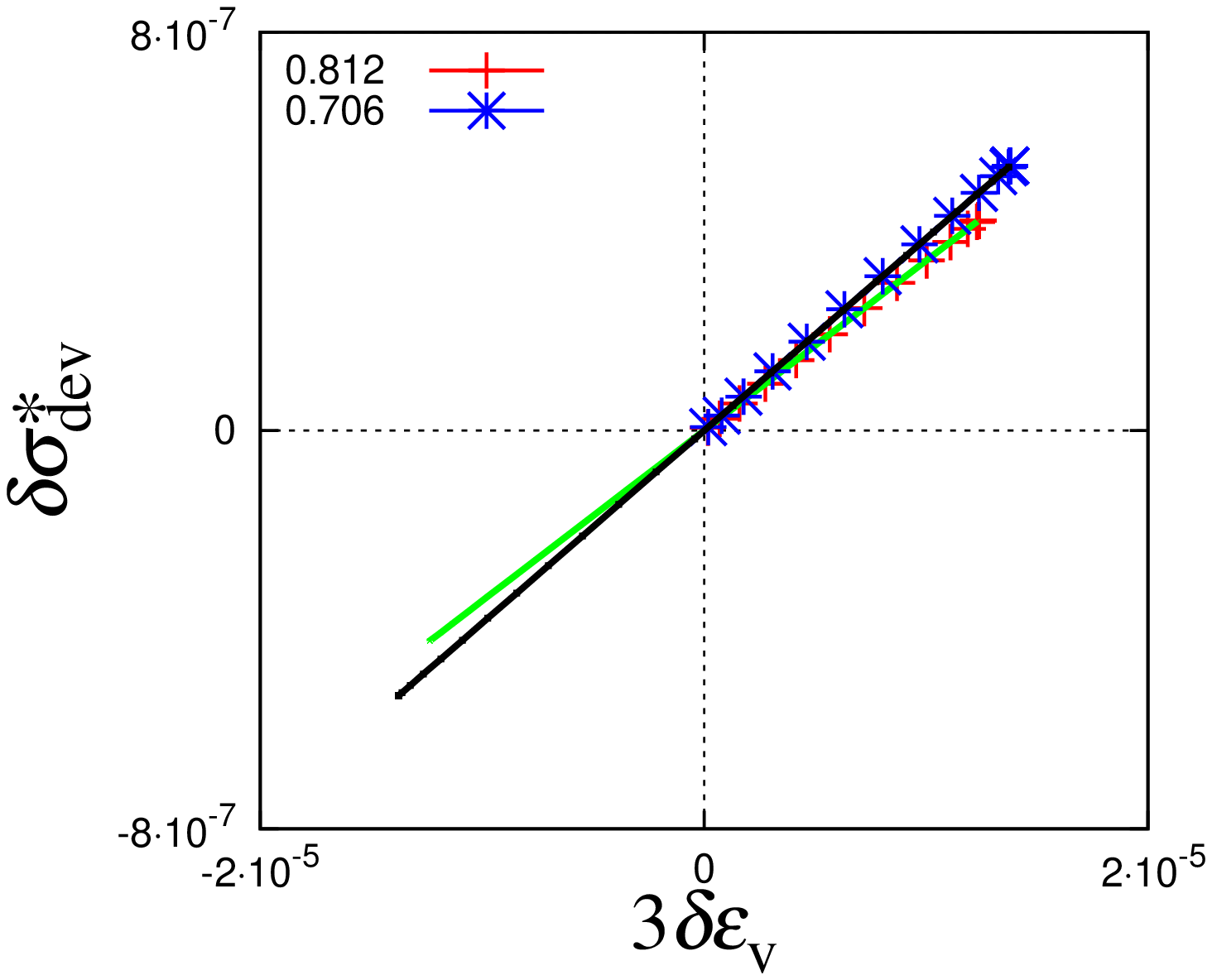}\label{tau_V_small}}
\subfigure[]{\includegraphics[scale=0.25,angle=270]{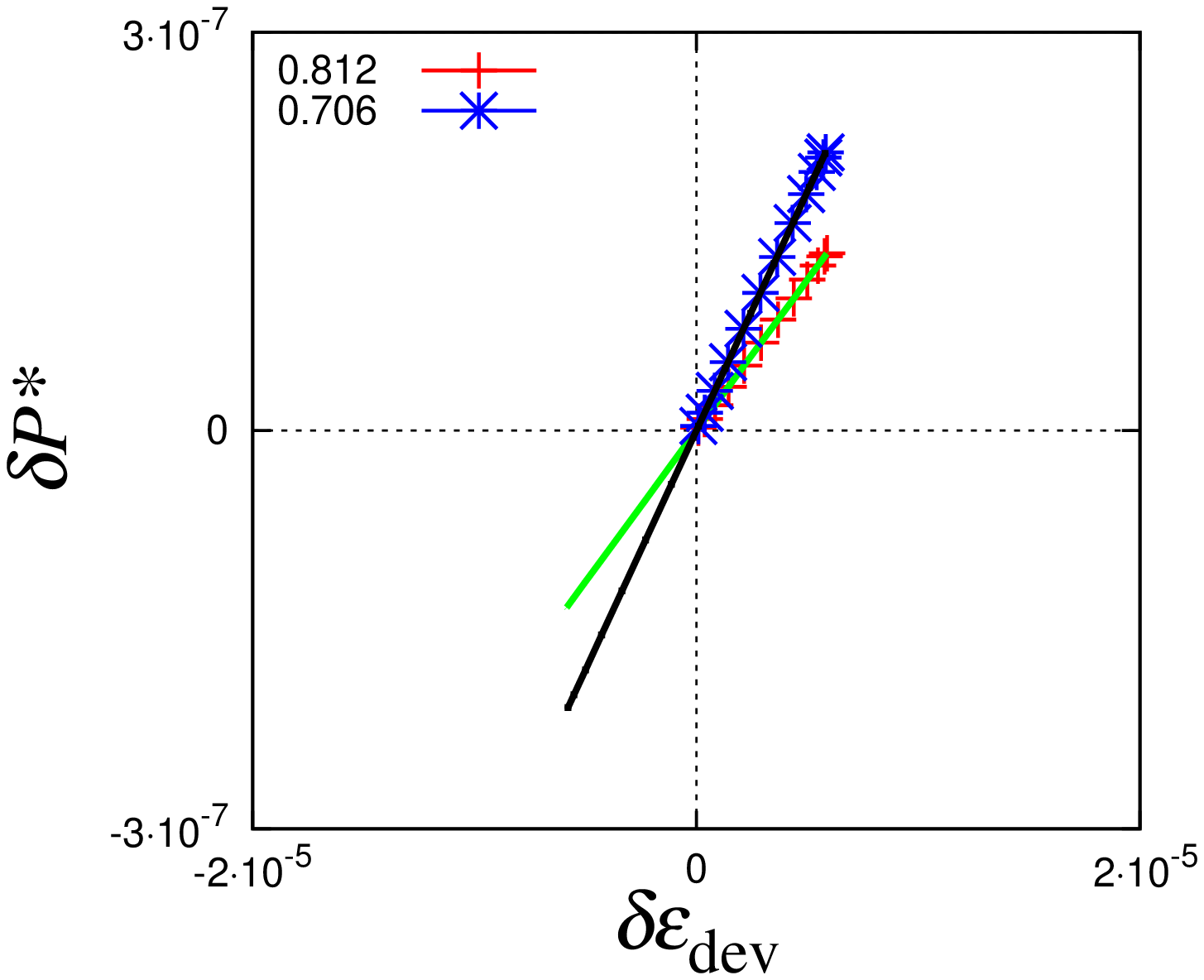}\label{P_DEV_small}}
\subfigure[]{\includegraphics[scale=0.25,angle=270]{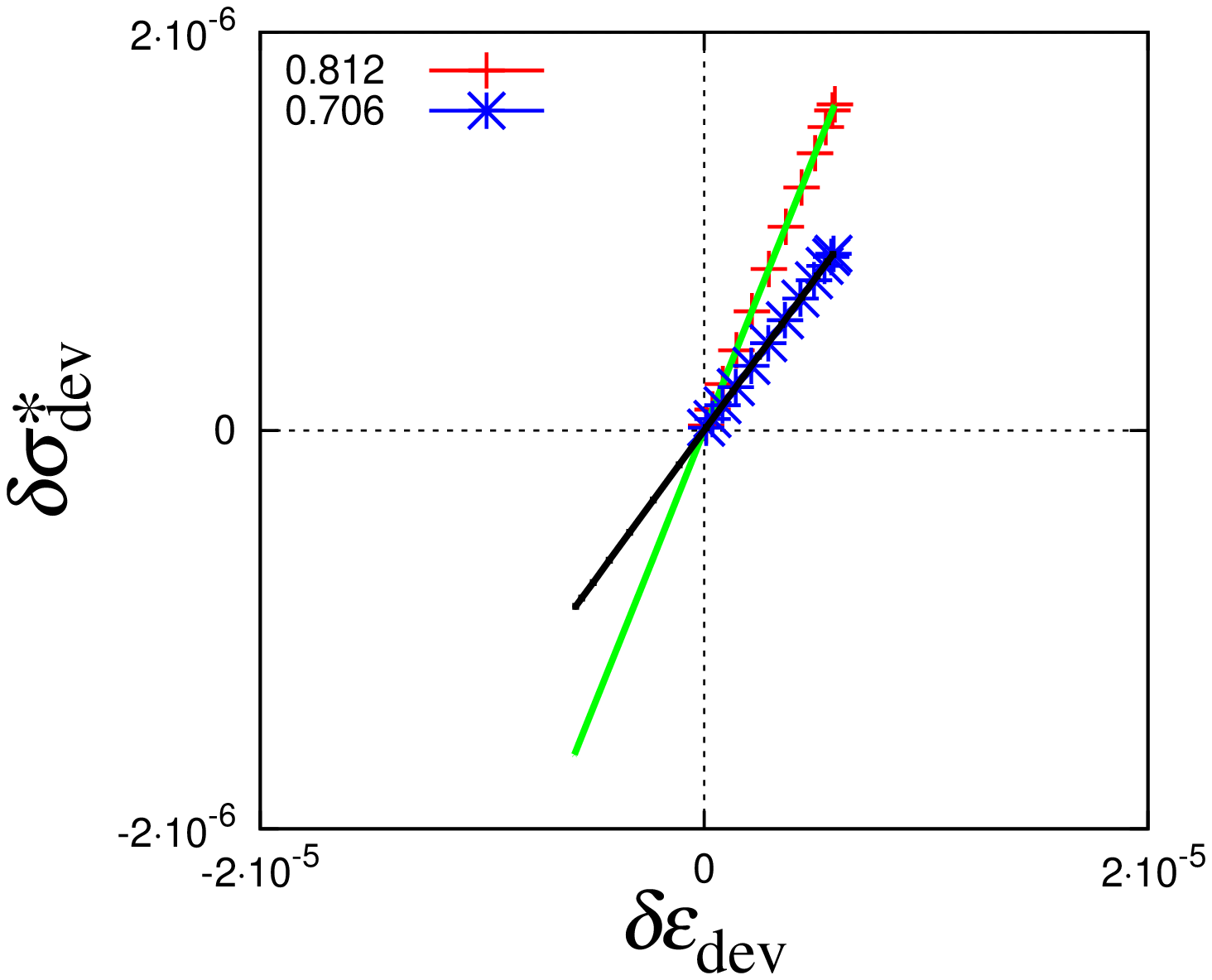}\label{tau_DEV_small}}\\
\subfigure[]{\includegraphics[scale=0.25,angle=270]{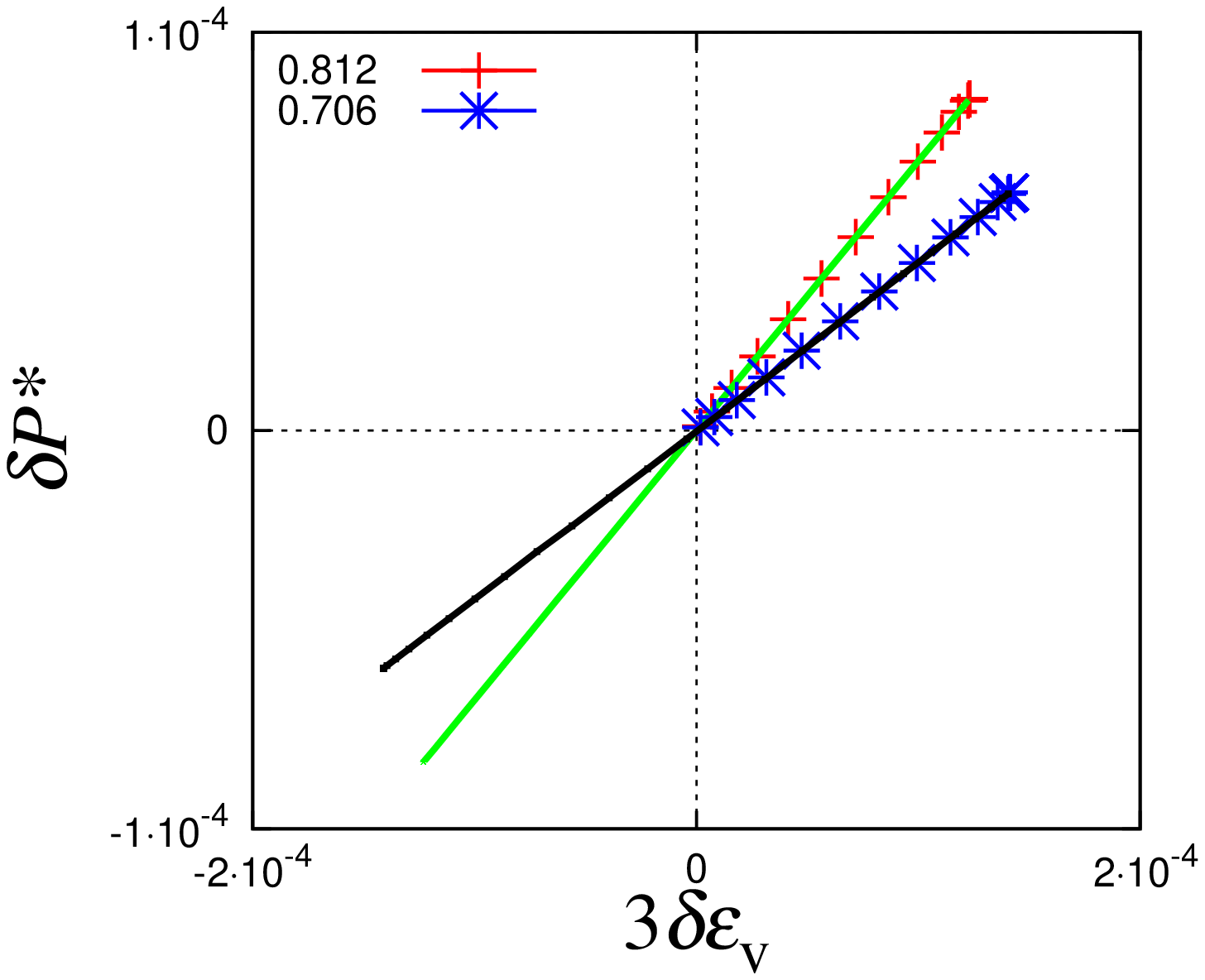}\label{P_V_medium}}
\subfigure[]{\includegraphics[scale=0.25,angle=270]{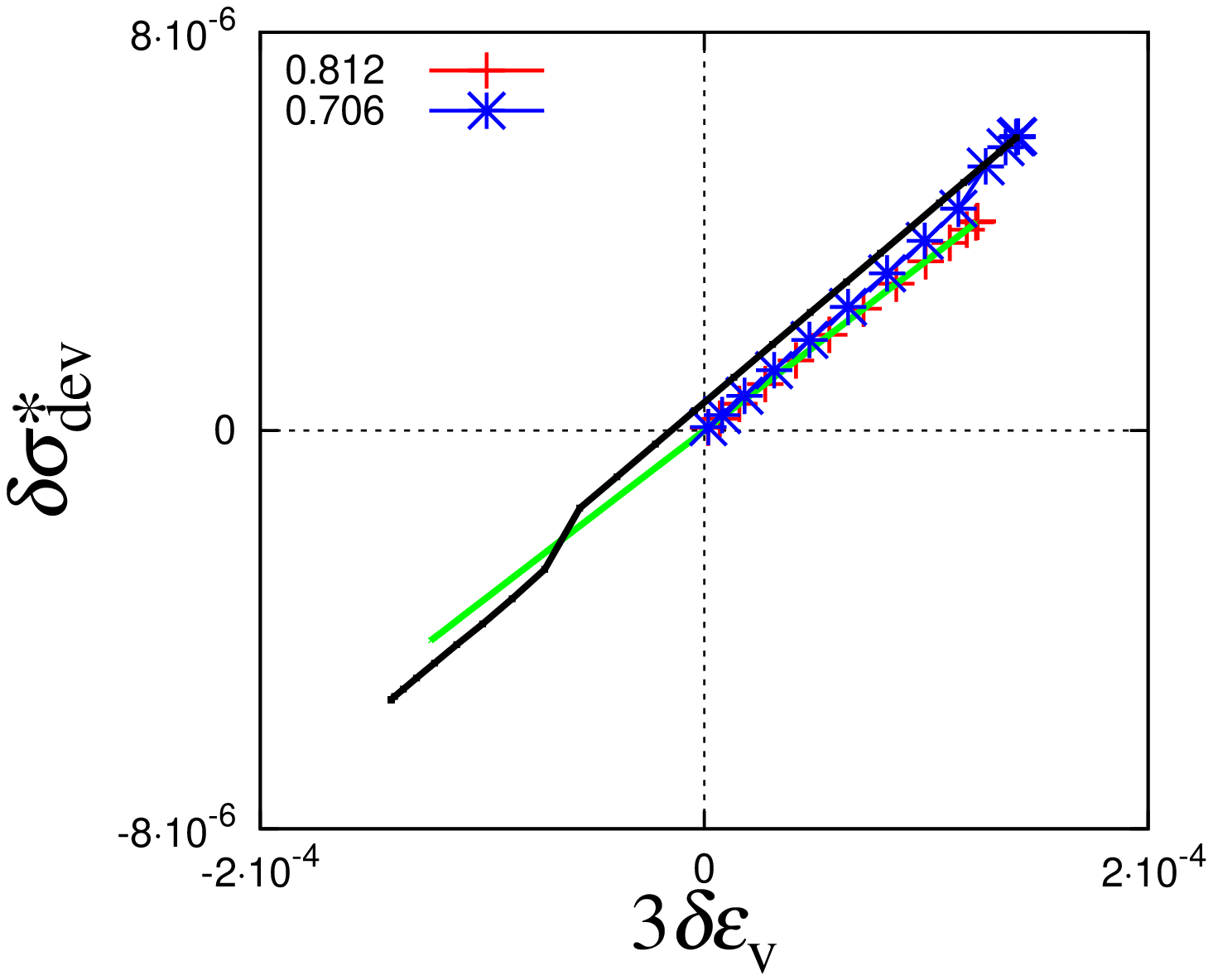}\label{tau_V_medium}}
\subfigure[]{\includegraphics[scale=0.25,angle=270]{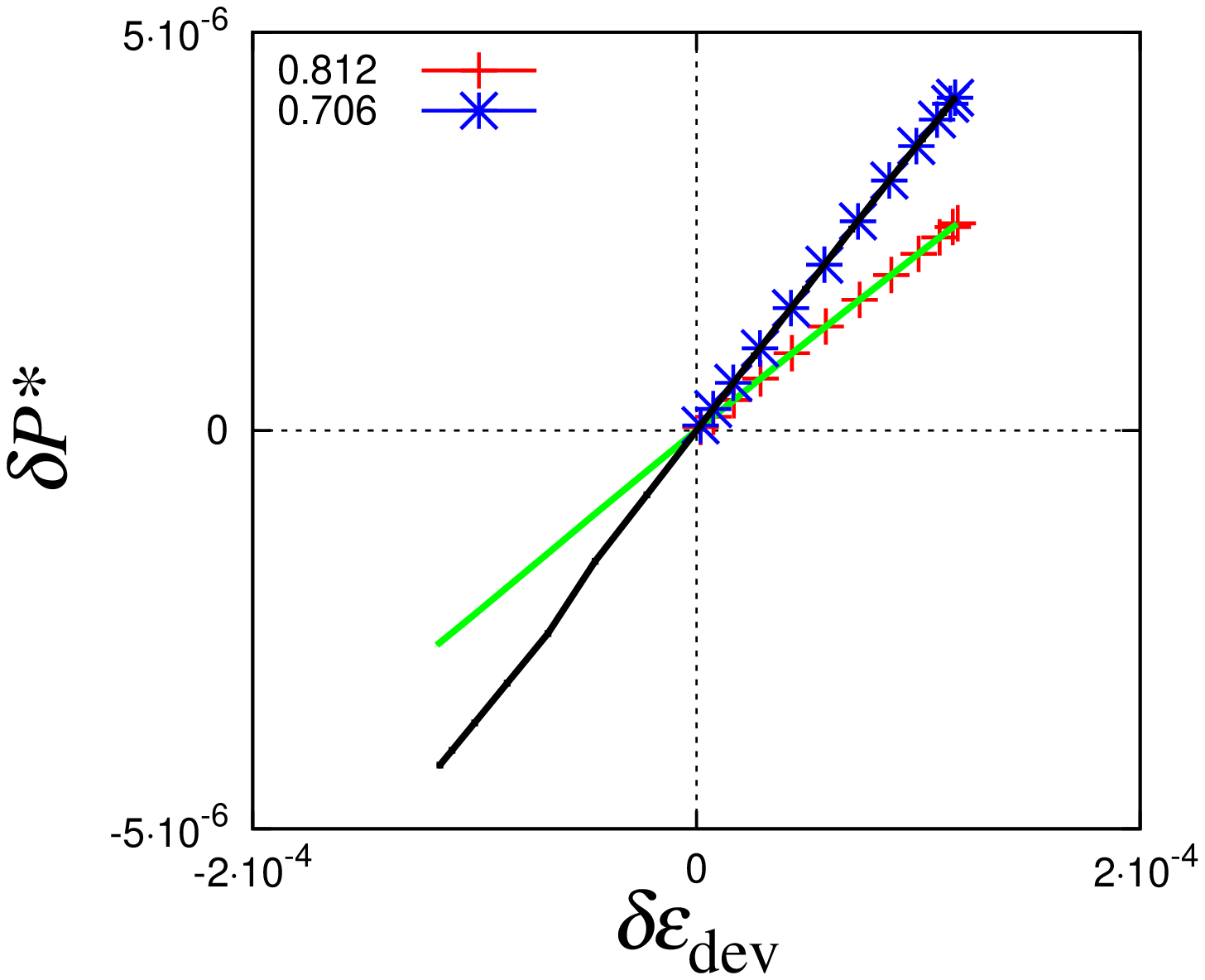}\label{P_DEV_medium}}
\subfigure[]{\includegraphics[scale=0.25,angle=270]{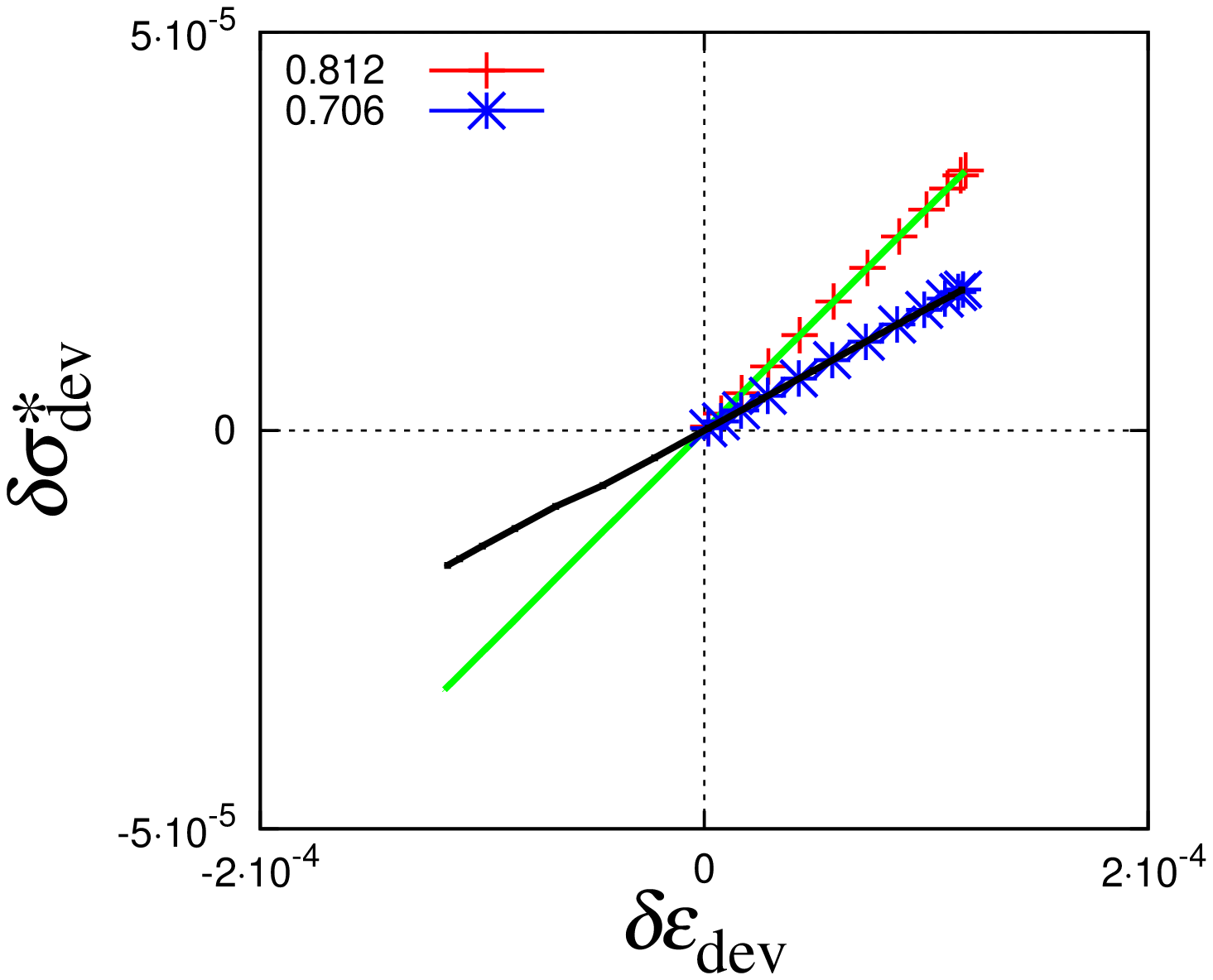}\label{tau_DEV_medium}}\\
\subfigure[]{\includegraphics[scale=0.25,angle=270]{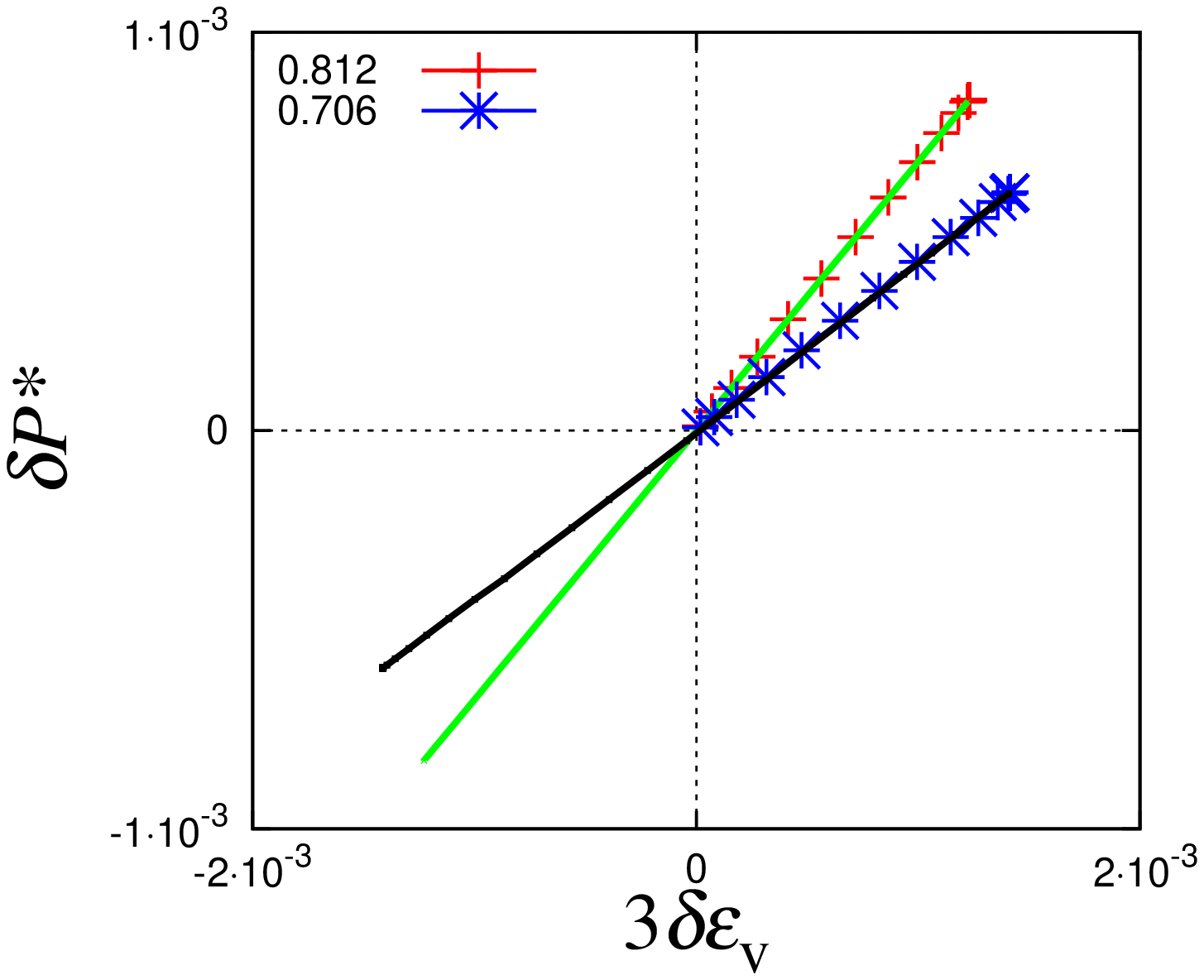}\label{P_V_large}}
\subfigure[]{\includegraphics[scale=0.25,angle=270]{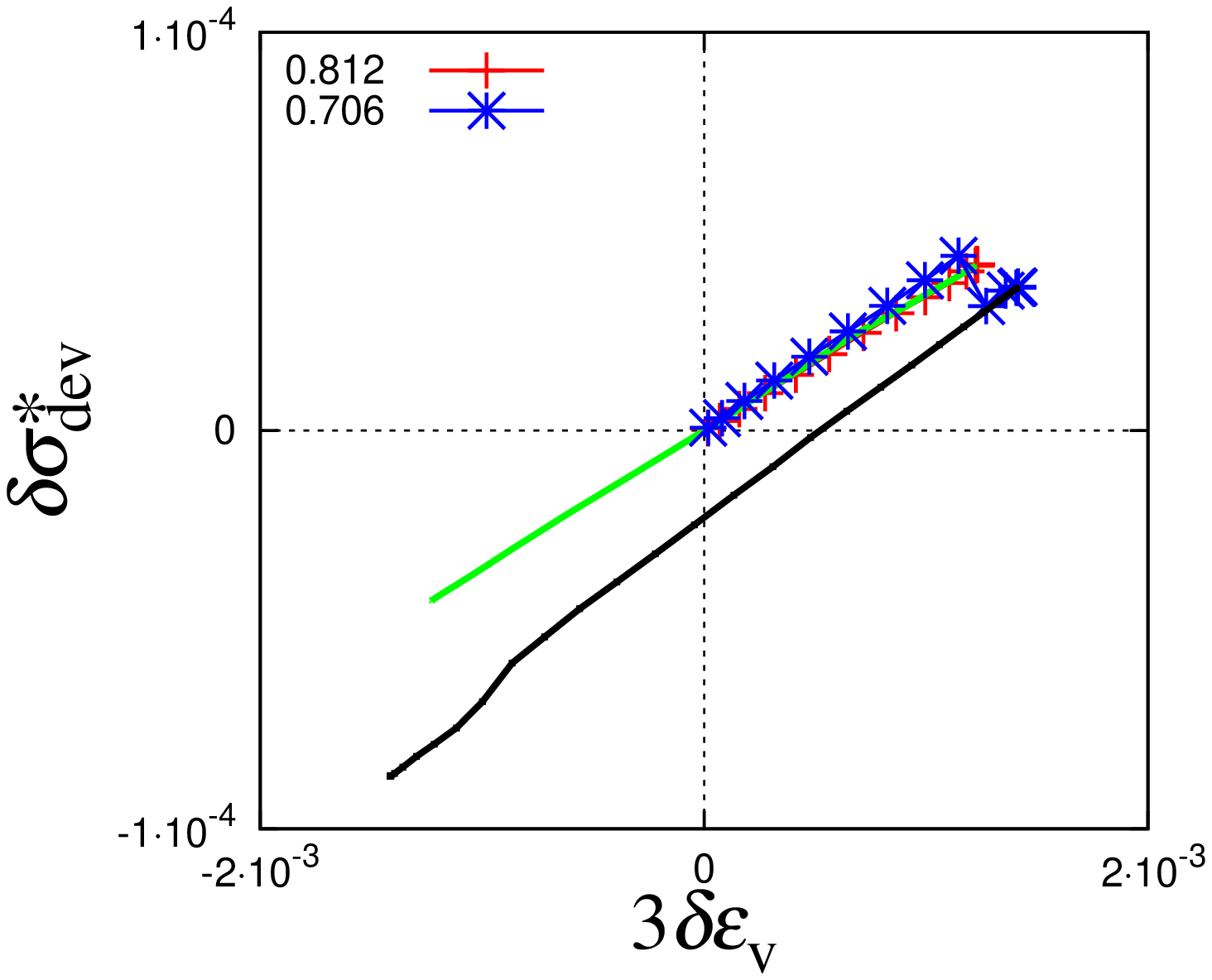}\label{tau_V_large}}
\subfigure[]{\includegraphics[scale=0.25,angle=270]{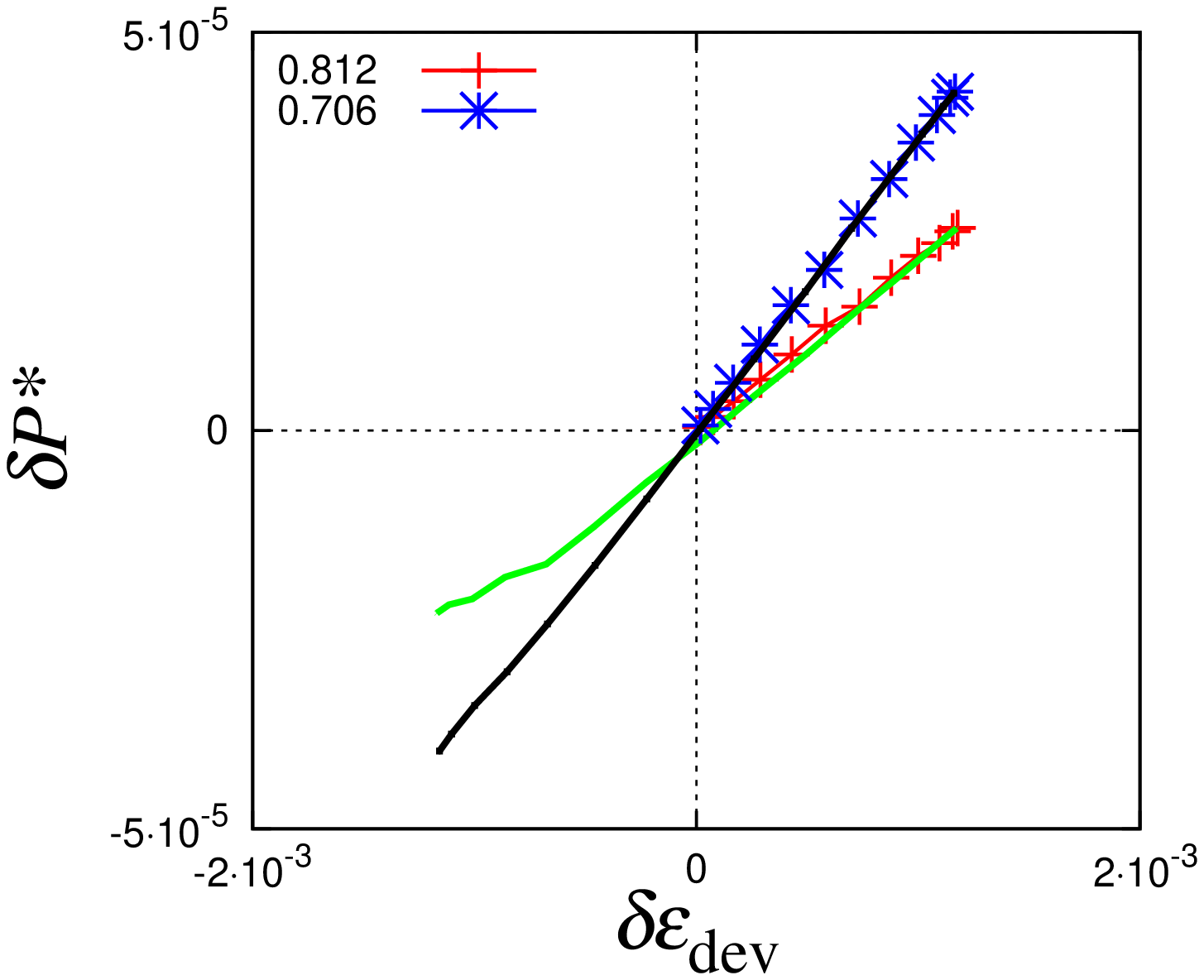}\label{P_DEV_large}}
\subfigure[]{\includegraphics[scale=0.25,angle=270]{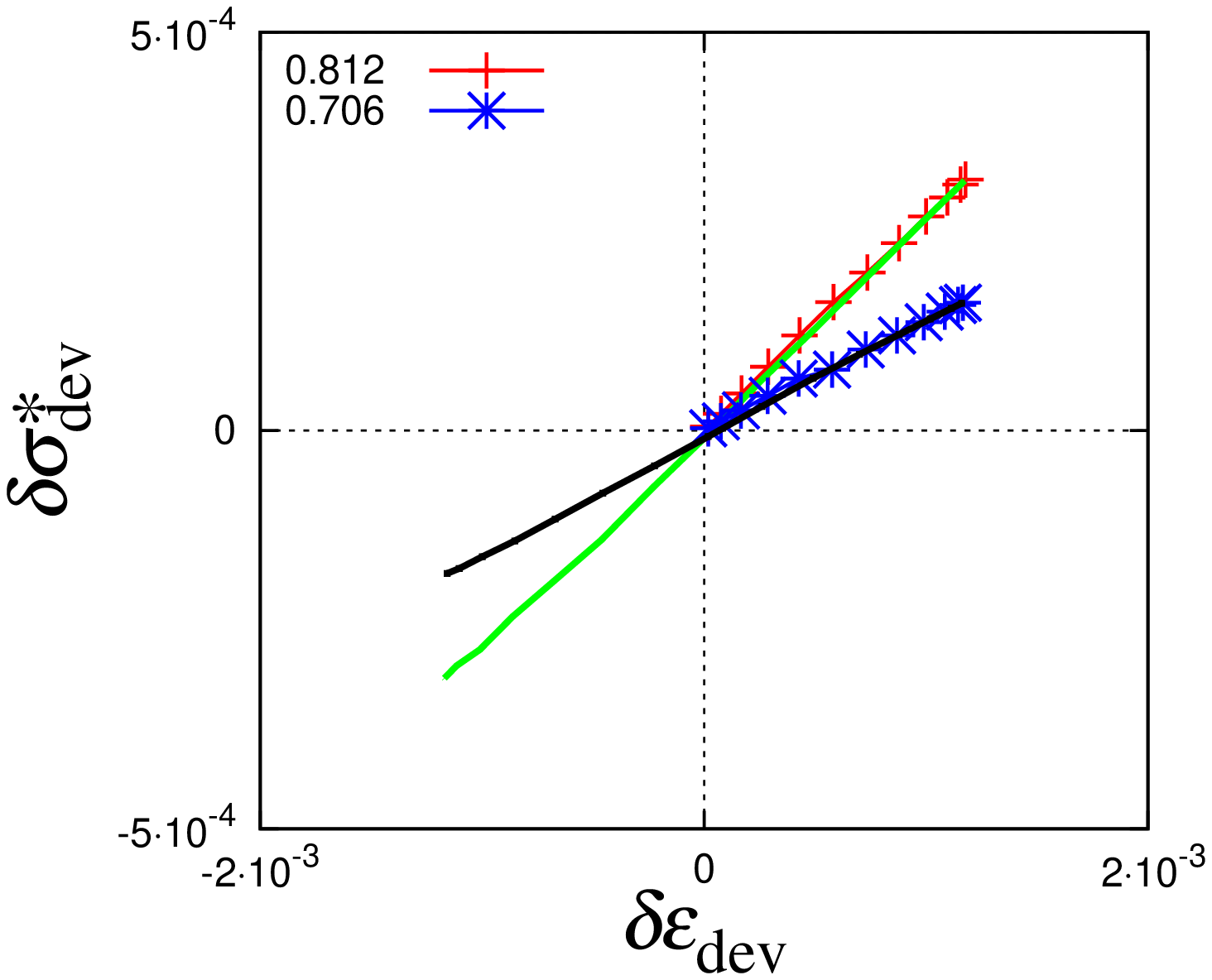}\label{tau_DEV_large}}
\caption{Evolution of non-dimensional pressure $\pkstar$, non-dimensional shear stress $\sigmadkstar$ during small (a -- d), medium (e--h), and large (i -- l) perturbations 
in the loading (symbols) and then unloading (solid lines) direction. 
Red `+' represents loading and the green line represents unloading for $\nu = 0.812$. 
Similarly, blue `*' represents loading and the black line represents unloading for $\nu = 0.706$. 
The deformation is applied to the state corresponding to $\epsd = 0.31$ (steady state configuration) of the main deviatoric experiment.}
\label{elas}
\end{figure}

We further check the elasticity of the probe by reversing the incremental strain. 
We plot the stress responses to volumetric/deviatoric strain in Fig.\ \ref{elas} and compare loading and unloading probes 
for different volume fractions ($\nu=0.706$ and $0.812$) and amplitudes. 
Looking at Figs.\ \ref{LONG}, \ref{smallperlongp_C} and \ref{elas} together, three regimes seem to appear.
The first one for very small strain ($<5.10^{-6}$),
due to the finite size of the system, is characterized by no opening and closing of contacts, and shows perfect reversibility of the data, i.e.,\ elasticity in Figs.\ \ref{elas}(a--d).
The second regime in Figs.\ \ref{elas}(e--h) shows some weakly irreversible behavior, but only for the smallest volume fraction and a 
mixed perturbation mode, see the sample at $\nu=0.706$ in Fig.\ \ref{tau_V_medium}; we associate this behavior to minor contact changes, as visible in 
Figs.\ \ref{LONG} and \ref{smallperlongp_C}, but no large scale rearrangements occur.
Finally, the third regime, for perturbations two orders of magnitude higher ($>10^{-4}$), 
a residual strain after reversal shows up for both volume fractions and all types of perturbations, see Figs.\ \ref{elas}(i--l), 
proving also that plasticity is much more pronounced in the deviatoric modes than in isotropic ones. 
We claim that small drops are related to local (weak, almost reversible) re-structuring, while in the last case, the whole system (or big portion of it) is involved in the collapse of the structure,
with a more pronounced effect for samples close to the jamming volume fraction \cite{magnanimo2014irreversible, keim2014mechanical}. 

For granular materials, the strain can not be split in elastic and plastic contributions
by ``trivially" referring to the residual deformation like in classical solids: 
as soon as we are out of the elastic range, rearrangements happen during loading and (even though less probably) during unloading, 
and most likely no original particle position is recovered.
Finally, we note that the results shown here are valid for finite-size systems; for much larger (real) samples of much smaller particles, we expect the first elastic
regime to reduce to much smaller strains. The boundary between the second and third regime is an issue for further research \cite{saitoh2014master}.

\subsection{Evolution of the moduli}
\label{sec:perturbresults}

Using the four packings at different $\nu_{i}$, we next 
determine which variables affect the incremental response 
of the aggregates at different deviatoric strains along the main path.
In order to understand the role of the microstructure, i.e.,\ the fabric tensor $\mathbf{F}$, the volumetric 
and deviatoric components, $\Fv$ and $\Fdev$, are considered.
We postulate that the incremental response of the granular material can be uniquely predicted, 
once its fabric state (along with the stress state) is known, 
irrespective of the path that the system experienced to reach that state. 
In this sense the fabric tensor can be referred to as a state variable.

\subsubsection{Bulk modulus $B$}
\label{sec:bulksection}

\begin{figure}[!ht]
\centering
\subfigure[]{\includegraphics[scale=0.5,angle=270]{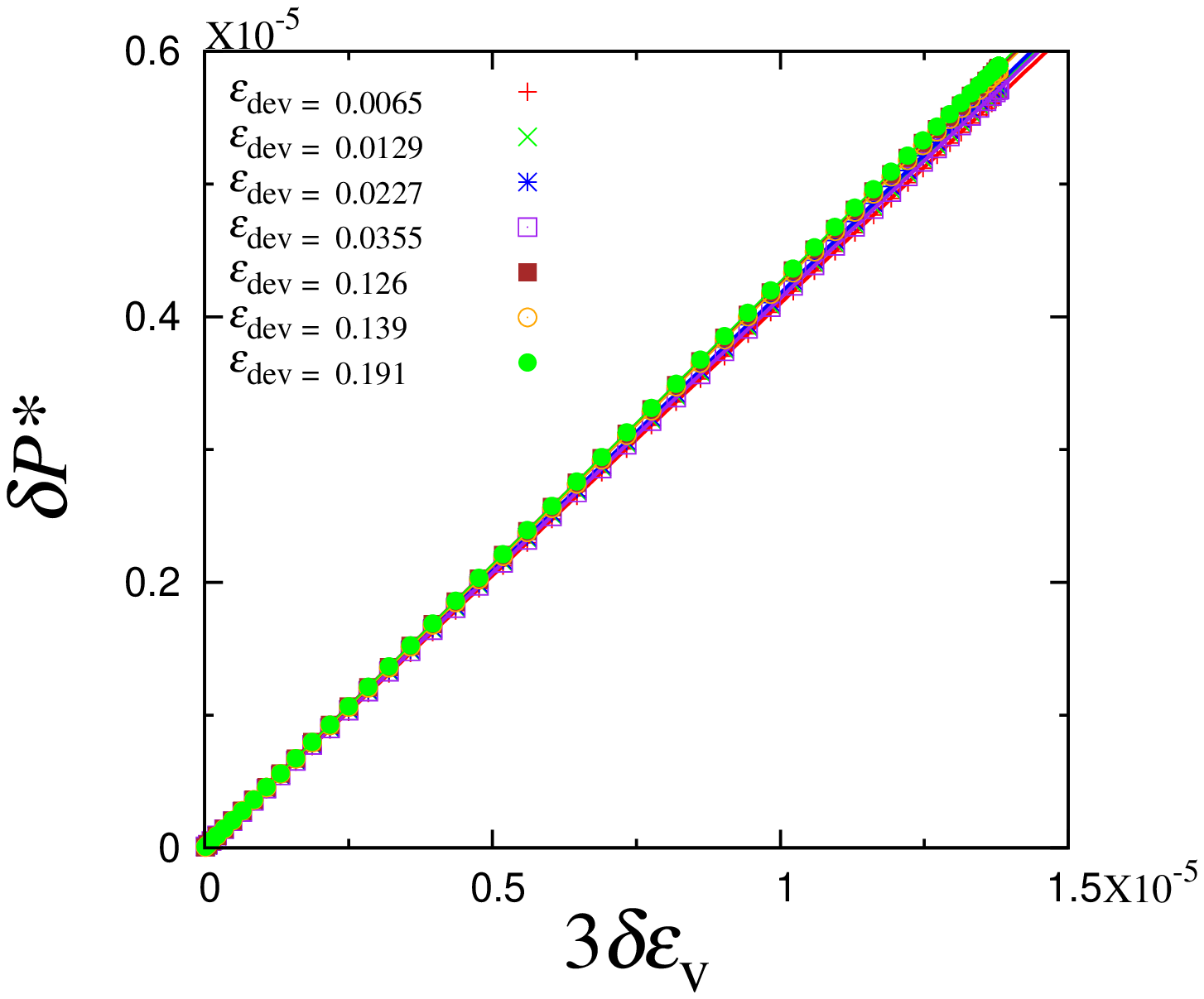}\label{dPdepsv}}
\subfigure[]{\includegraphics[scale=0.5,angle=270]{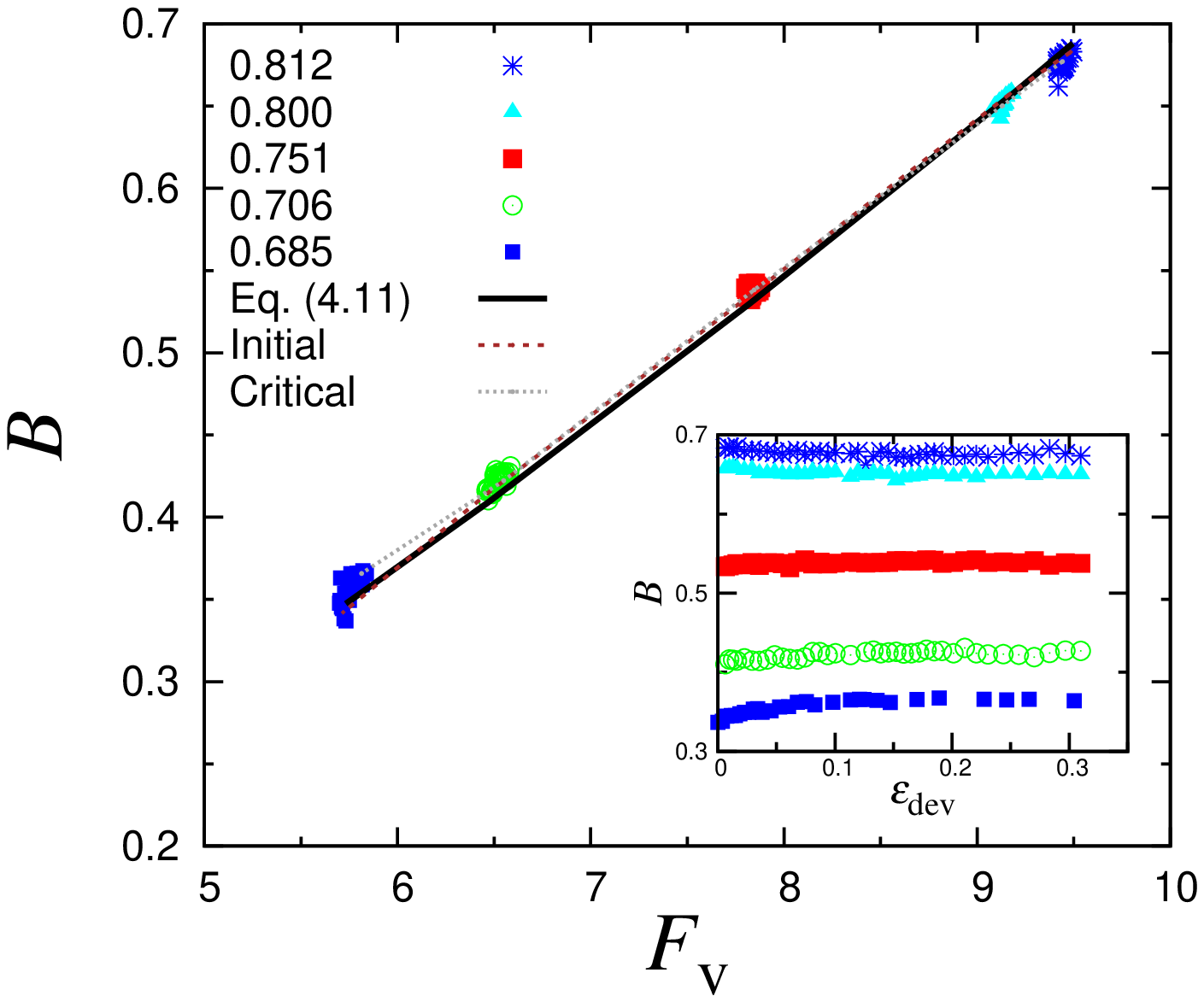}\label{bulkplot}}
\caption{ (a) Evolution of change in non-dimensional pressure $\delta \pkstar$ during purely isotropic perturbations $3\delta \epsiso$ 
for different states for volume fraction $\nu = 0.706$ along the main path as shown in the inset. 
(b) Evolution of the bulk modulus $B$ as scaled with isotropic fabric $\Fv$ for five different volume fraction as shown in the inset. 
The solid line passing through the data represents Eq.\ (\ref{eq:B}). 
The dashed lines represent the initial and steady state data, as given in the legend. }
\end{figure}

In Fig.\ \ref{dPdepsv}, we plot the incremental non-dimensional pressure $\delta \pkstar$
against the amplitude of the applied isotropic perturbation $3\delta \epsiso$
for one volume fraction, $\nu = 0.706$, and various initial anisotropic configurations.
The slope of each line is the bulk modulus of that state. 
It practically remains unchanged for different states and 
suggests that $B$ is constant for a given volume fraction. 

In Fig.\ \ref{bulkplot}, we plot the variation of the bulk modulus $B$, with the 
isotropic fabric $\Fv$ for packings with different volume fractions $\nu_i$. 
$B$ increases systematically when the five different reference configurations are 
compared, and it is related to the value of $\Fv$ constant at a given $\nu_{i}$ \cite{goncu2010constitutive,kumar2013evolution, shaebani2012unilateral}.
As expected $B$ is a purely volumetric quantity and varies with changes 
in the isotropic contact network. 
The inset in Fig.\ \ref{bulkplot} shows that the bulk modulus remains almost constant with 
applied shear during a single deviatoric experiment \cite{kumar2013evolution}, 
behaving qualitatively similar to pressure $\pkstar$ and isotropic fabric $\Fv$, see Figs.\ \ref{pres} and \ref{fv} respectively. 
That is, the contact orientation anisotropy, $\Fdev$, which changes 
during the main deviatoric deformation path (see Fig.\ \ref{fdev}) does not affect it. 
In agreement with observations on the volumetric fabric in section \ref{sec:fabricresults}, 
also $B$ shows a slight increase/decrease in the first part of the deviatoric path, more pronounced for 
loose samples, as clearly seen in Fig.\ \ref{dfvwithnu}. The trend of $B$ slightly deviates from $\Fv$ in the low strain regime, 
while the dependence is well captured in the steady state, after large strain.
The relation between bulk modulus and fabric was given in Ref.\ \cite{goncu2010constitutive} as:
\begin{equation}
\label{eq:B}
B =  \frac{\delta{\pkstar}} {3\delta\epsiso}\bigg|_{\delta\epsd=0} = \frac{p_0 \Fv}{g_3 \nu_c} \left[ 1 - 2\gamma_p \left(-\epsiso\right) + \left(-\epsiso\right)\left( 1 - \gamma_p \left(-\epsiso\right) \right)\frac{\partial \mathrm{ln}\Fv}{\partial\left(-\epsiso\right)} \right], 
\end{equation}
where $p_0$, $\gamma_p$ and the jamming volume fraction $\nu_c$ are fit parameters presented in Table\ \ref{moduli_param}.\footnote[5]{Note that $\nu_c$ for the same particulate system was reported as 0.66 for isotropic deformation in Ref.\ \cite{goncu2010constitutive}, 
as $0.6646$ for isotropic and $0.658$ for shear deformation in Ref. \cite{imole2013hydrostatic}. 
We use a similar $\nu_c=0.658$ here, which, however, is dependent on history of the sample and on the deformation mode.
The small deviations of $B$ from Eq.\ (\ref{eq:B}) can be attributed to a (small) variation of $\nu_c$, however, this is beyond the focus of this paper.}
$g_3\approx1.22$ is dependent on the particle size distribution as presented in Refs.\ \cite{goncu2010constitutive, kumar2013effect, imole2013hydrostatic}, see section\ \ref{sec:simmeth}. 
For a given volume fraction, the above relation only requires the knowledge of the isotropic fabric $\Fv=g_3\nu C = g_3 \nu C^* \left(1- \phi_r\right)$, 
where the empirical relations for $C^*(\nu)$ and $\phi_r(\nu)$ are taken from Refs.\ \cite {imole2013hydrostatic, kumar2013effect}, see section\ \ref{sec:simmeth}. 
The numerical data show good agreement with the theoretical prediction 
presented in \cite{goncu2010constitutive} and reported in Fig.\ \ref{bulkplot}. 
The minimum $\Fv$ is obtained at the jamming volume fraction, with $\nu_c=0.658$, $C^*=6$, and $\phi_r=\phi_c=0.13$, leading to $\Fv^\mathrm{min} = 4.2$. 
At the jamming transition, we can extrapolate a finite value of the bulk modulus $B^\mathrm{min} = 0.22$, 
while it suddenly drops to zero below $\nu_{c}$
\cite{pica2009jamming, ohern2003jamming, zhang2005jamming, otsuki2011critical, bohy2012soft, parisi2010hard,inagaki2011protocol, metayer2011shearing}. The discontinuity of $B$ is related to the discontinuity in $\Fv$, that jumps form zero to a finite value in $\nu_{c}$
due to equilibrium requirements.

\begin{table}
  \centering
\begin{tabular}{c@{\hskip 0.3in}||@{\hskip 0.3in}c}
\hline\hline \\[-1.5ex]
Modulus & Fit parameter  \\ [1ex] 
\hline     \\[-1.5ex]
Bulk modulus $B$ & $p_0=0.0425$, $\gamma_p \approx 0.2$, $\nu_c=0.658$ \\ [1ex]
First anisotropy modulus $A_1$ & $\CAo = 1 \pm 0.01$ \\ [1ex]
Second anisotropy modulus $A_2$ & $\CAt = 1 \pm 0.02$ \\ [1ex]
Octahedral shear modulus $\Goct$ & $\CGo = 130 \pm 3$ \\ [1ex]
\end{tabular}
\caption{Summary of fit parameters extracted from the small perturbation results in Eqs.\ (\ref{eq:B}), (\ref{eq:A_1}), (\ref{eq:A_2}), and (\ref{eq:G}). }
\label{moduli_param}
\end{table}

\subsubsection{Anisotropy moduli $A_1$ and $A_2$}

\begin{figure}[!ht]
\centering
\subfigure[]{\includegraphics[scale=0.5,angle=270]{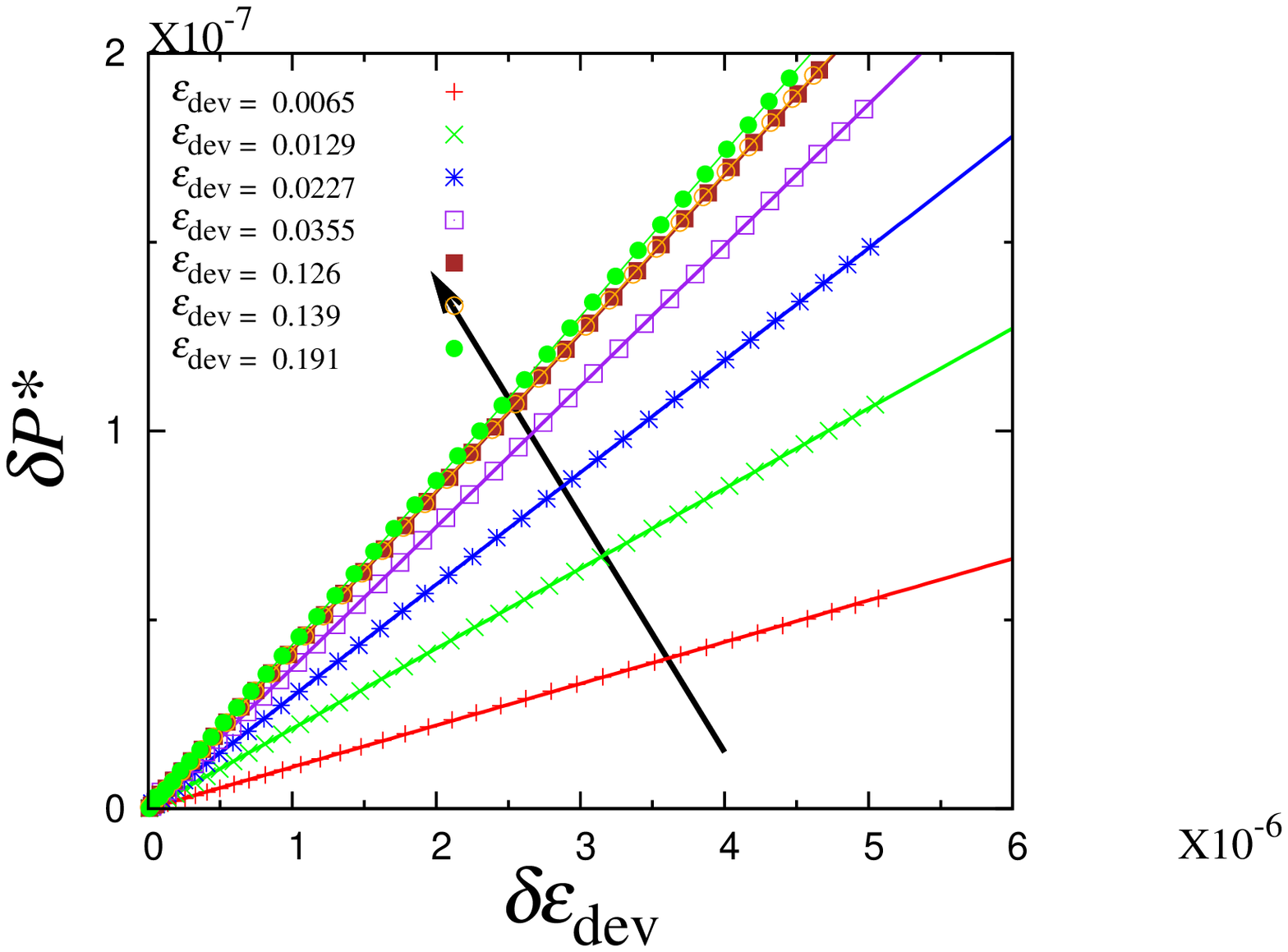}\label{dPdepsdev}}
\subfigure[]{\includegraphics[scale=0.5,angle=270]{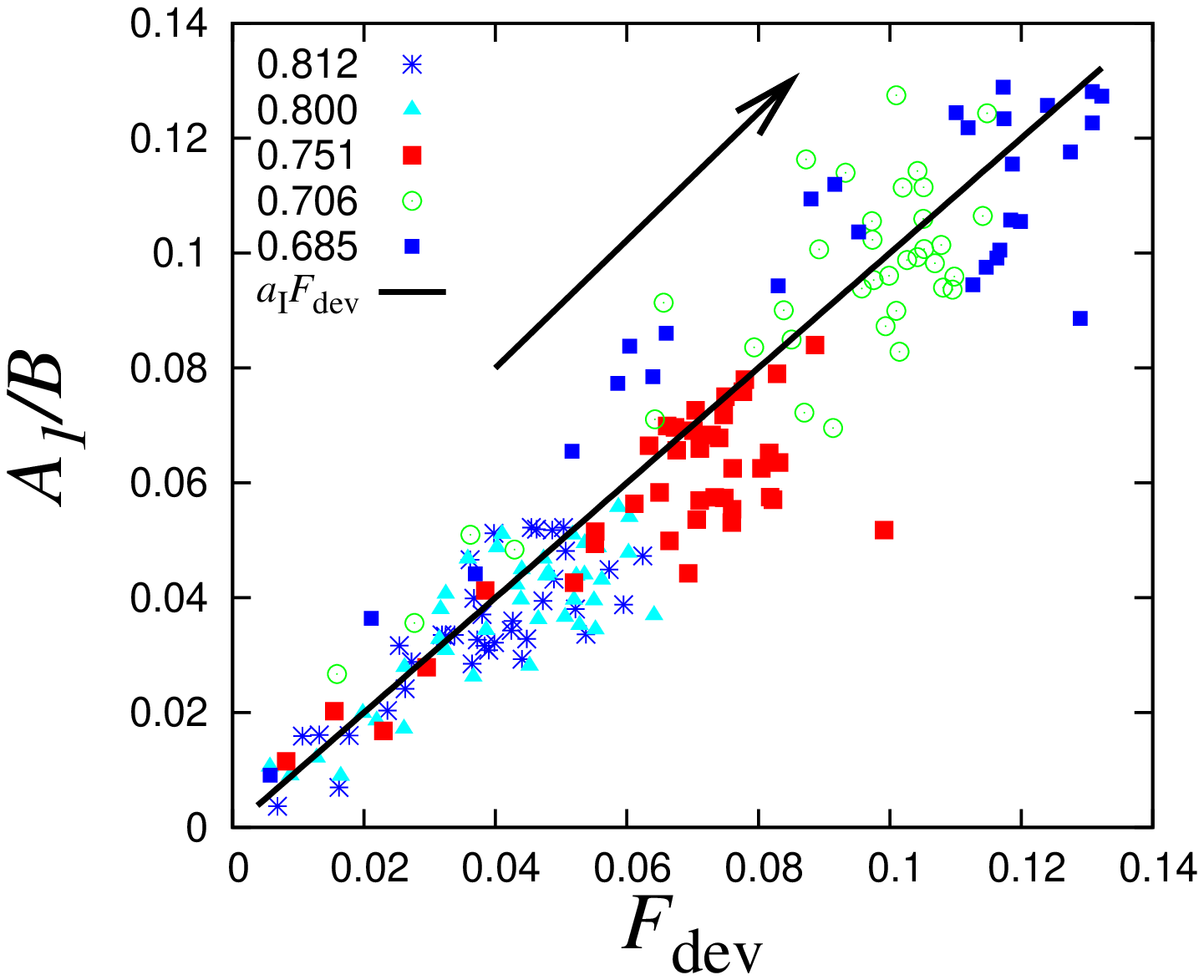}\label{A1}}
\caption{ (a) Evolution of change in non-dimensional pressure $\delta \pkstar$ during purely deviatoric perturbations $\delta \epsd$ 
for different states for volume fraction $\nu = 0.706$ along the main path as shown in the inset. The arrow indicates the direction of increasing strain states during main deviatoric experiments. 
(b) Evolution of the ratio of first anisotropy modulus with bulk modulus $A_1/B$ as function of the deviatoric fabric $\Fdev$ for five different volume fractions as shown in the inset. 
The solid line passing through the data represents Eq.\ (\ref{eq:A_1}) divided by $B$.}
\end{figure}

\begin{figure}[!ht]
\centering
\subfigure[]{\includegraphics[scale=0.5,angle=270]{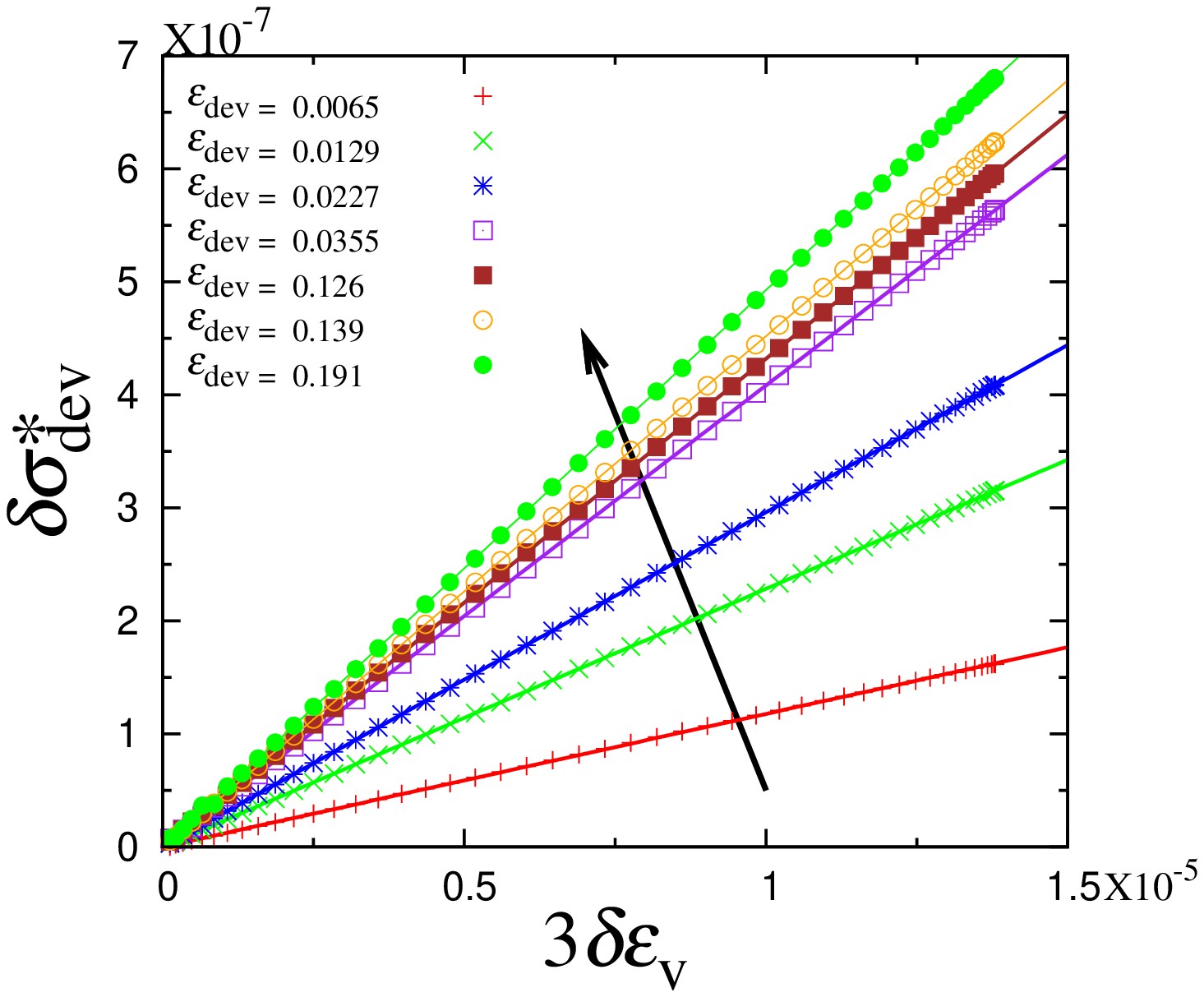}\label{dtaudepsv}}
\subfigure[]{\includegraphics[scale=0.5,angle=270]{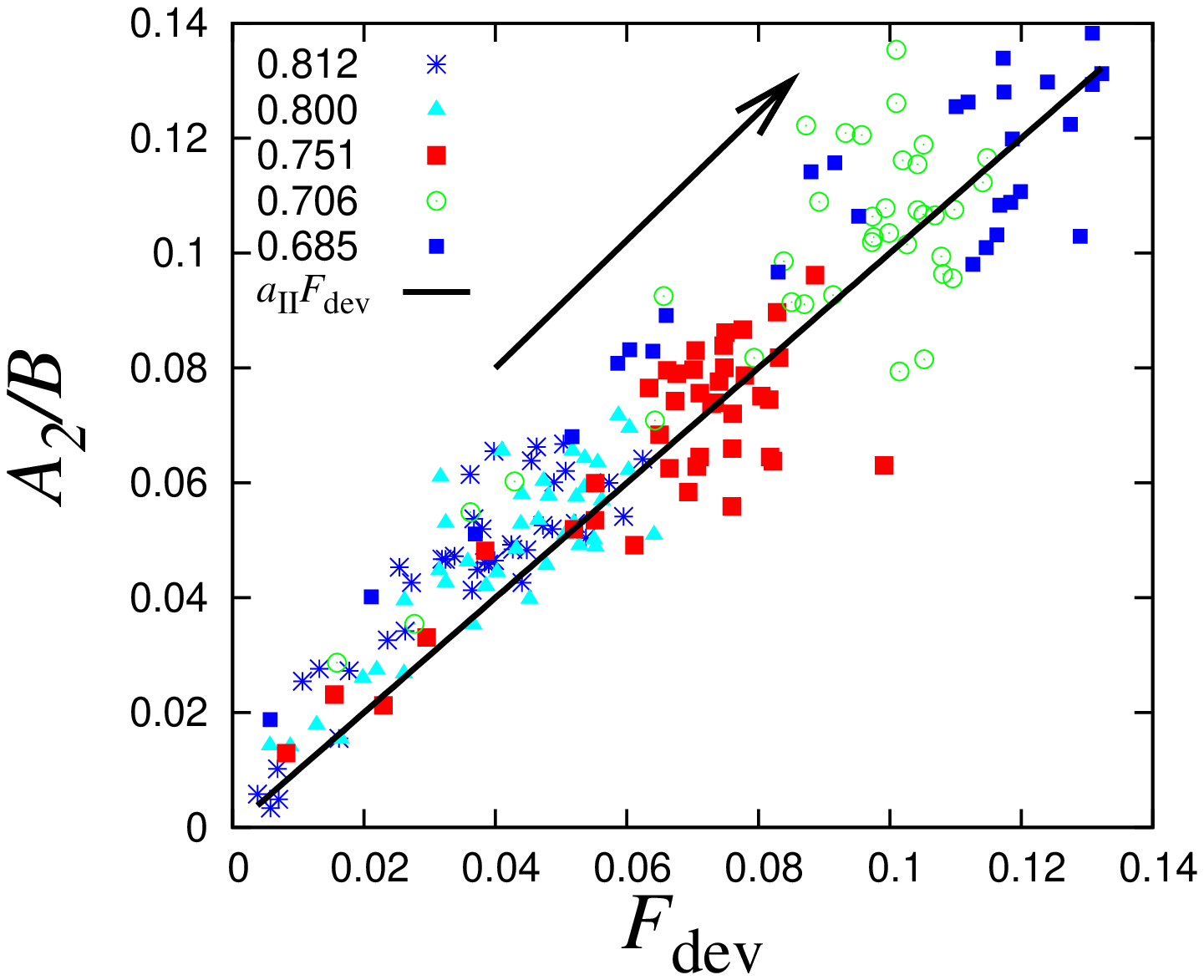}\label{A2}}
\caption{(a) Evolution of change in non-dimensional shear stress $\delta \sigmadkstar$ during purely isotropic perturbations $3\delta \epsiso$ 
for different states for volume fraction $\nu = 0.706$ along the main path as shown in the inset. The arrow indicates the direction of increasing strains during main deviatoric experiments. 
(b) Evolution of the ratio of second anisotropy modulus with bulk modulus $A_2/B$ as scaled with the deviatoric fabric $\Fdev$ for five different volume fraction as shown in the inset. 
The solid line passing through the data represents Eq.\ (\ref{eq:A_2}) divided by $B$.}
\end{figure}

In Fig.\ \ref{dPdepsdev}, we plot the non-dimensional pressure increment $\delta \pkstar$ against the strain amplitude, 
when the material is subjected to small deviatoric perturbations $\delta \epsd$, 
to measure the first anisotropy modulus $A_1$ as defined in Eq.\ (\ref{eqnarray:moduli}), 
in given anisotropic configurations, as in Fig.\ \ref{dPdepsv}. 
Since the material is in an anisotropic state, an increment in deviatoric strain leads to 
a change in volumetric stress, along with shear stress. 
The slope of the curves, $A_1$, increases with the 
previous shear strain the system has experienced, going from small values in the initial isotropic configuration, 
to an asymptotic limit.

We are interested in the ratio $A_1/B$. In this ratio, the dependence of isotropic fabric $\Fv$ cancels out, all that remains is a pure dependence on $\Fdev$.
In Fig.\ \ref{A1},  we plot the variation of $A_1/B$, with $\Fdev$ for packings with different volume fractions $\nu_i$ as shown in the inset. 
Besides the fluctuations, 
the data collapse on a unique curve irrespective of volume fraction and pressure, 
that is, once a state has been achieved, a measurement of the overall anisotropy modulus is associated with a unique $\Fdev$. 
An increasing trend of $A_1/B$ with the fabric shows up.
As the deviatoric fabric decreases with volume fraction (see Fig.\ \ref{fdev}), 
this leads to lower values of the scaled anisotropy modulus for denser systems. 
In conclusion, we have a linear relation between for the first anisotropy modulus $A_1$:
\begin{equation}
\label{eq:A_1}
A_1= \frac{\delta{\pkstar}}{\delta{\epsd}}\bigg|_{\delta\epsiso=0} = \CAo B \Fdev,
\end{equation}
where $B$ is the bulk modulus, $\Fdev$ is the deviatoric part of fabric, and $\CAo\approx0.66$ is a fit parameter presented in Table \ref{moduli_param}. 

In Fig.\ \ref{dtaudepsv} we plot the stress response of the material $\delta \sigmadkstar$ to isotropic perturbation 
$3\delta \epsiso$, for the same anisotropic initial configurations as in Fig.\ \ref{dPdepsv}, 
to measure the second anisotropy modulus $A_2$ as defined in Eq.\ (\ref{eqnarray:moduli}). 
Similarly to $A_1$, the slope of the elastic curves, i.e.,\ $A_2$, 
increases with the previous shear strain the system has felt, starting form zero until an asymptotic limit is reached.
In Fig.\ \ref{A2}, we plot the variation of $A_2/B$, with $\Fdev$ for different volume fractions $\nu_i$ as shown in the inset. 
Data show a very similar trend to what observed in Fig.\ \ref{A1} and besides the fluctuations, a collapse of data is observed.\footnote[6]{A large data scatter is present in both figures Figs.\ \ref{A1} and \ref{A2}, 
which increases for increasing deviatoric fabric $\Fdev$. This is possibly due to other factors that may contribute to the evolution of the anisotropy moduli that are not considered in the present work.}
Again we can relate $A_2$ to deviatoric fabric as:
\begin{equation}
\label{eq:A_2}
A_2 =\frac{\delta{\sigmadkstar}} {3\delta\epsiso}\bigg|_{\delta\epsd=0} = \CAt B \Fdev.
\end{equation}
The equality between the two fitting constants $\CAo\approx\CAt\approx1$ (see Table \ref{moduli_param}), 
states the symmetry of the stiffness matrix in Eq.\ (\ref{eqnarray:hypo}).

Eq.\ \ref{eq:A_1} and \ref{eq:A_2} provides an interesting, novel way to back-calculate the deviatoric structure
in a granular sample
via $F_{\rm dev} = A/B$, where $A$ and $B$ can be inferred from wave propagation experiments, 
while the direct measurement of fabric is still an open issue 
\cite{hall2010discrete, sperl2006experiments, jorjadze2013microscopic, khidas2012probing}. 

%

\subsubsection{Octahedral shear modulus $\Goct$}
\begin{figure}[!ht]
\centering
\subfigure[]{\includegraphics[scale=0.5,angle=270]{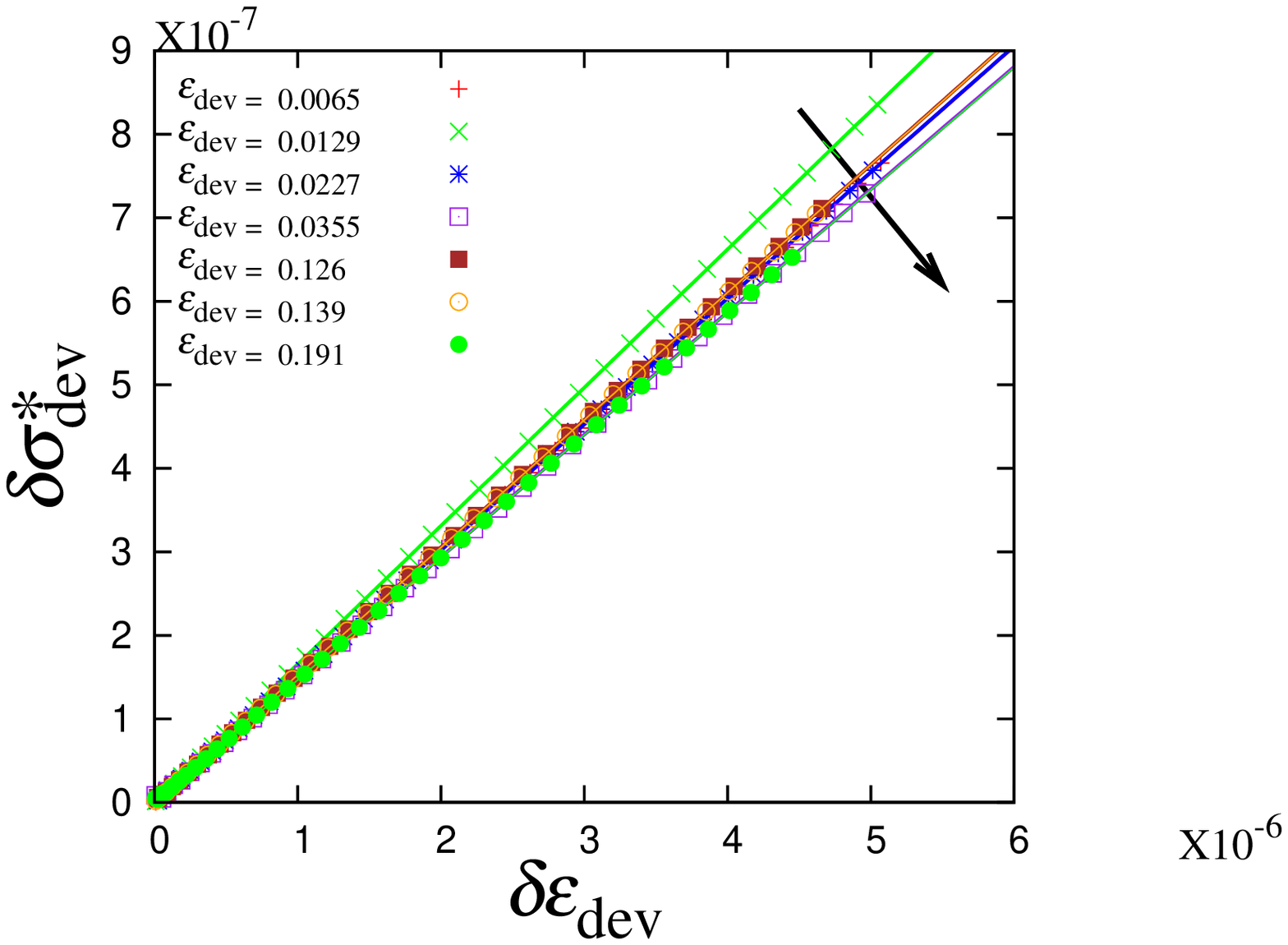}\label{dtaudepsdev}}
\subfigure[]{\includegraphics[scale=0.5,angle=270]{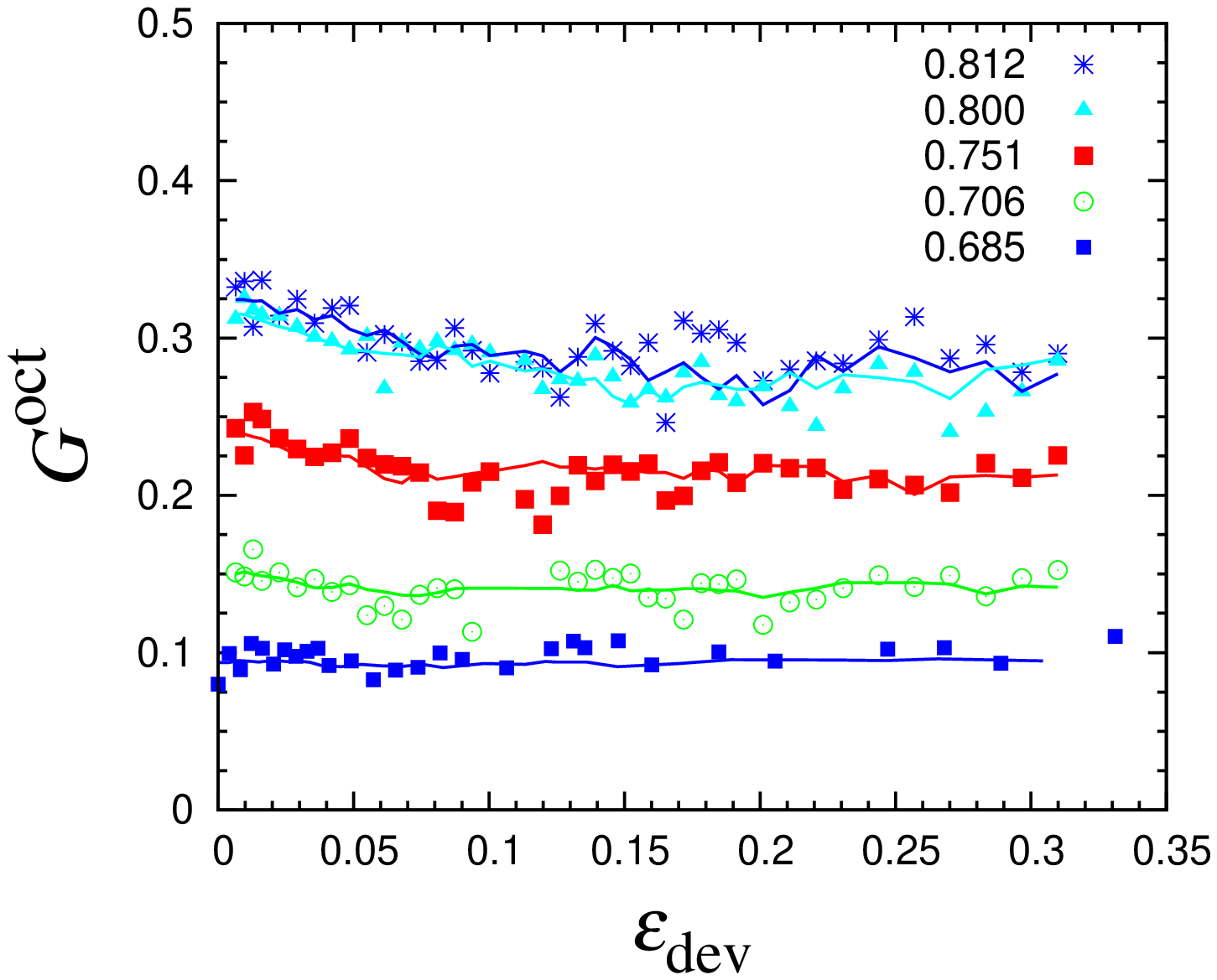}\label{shear}}\\
\subfigure[]{\includegraphics[scale=0.5,angle=270]{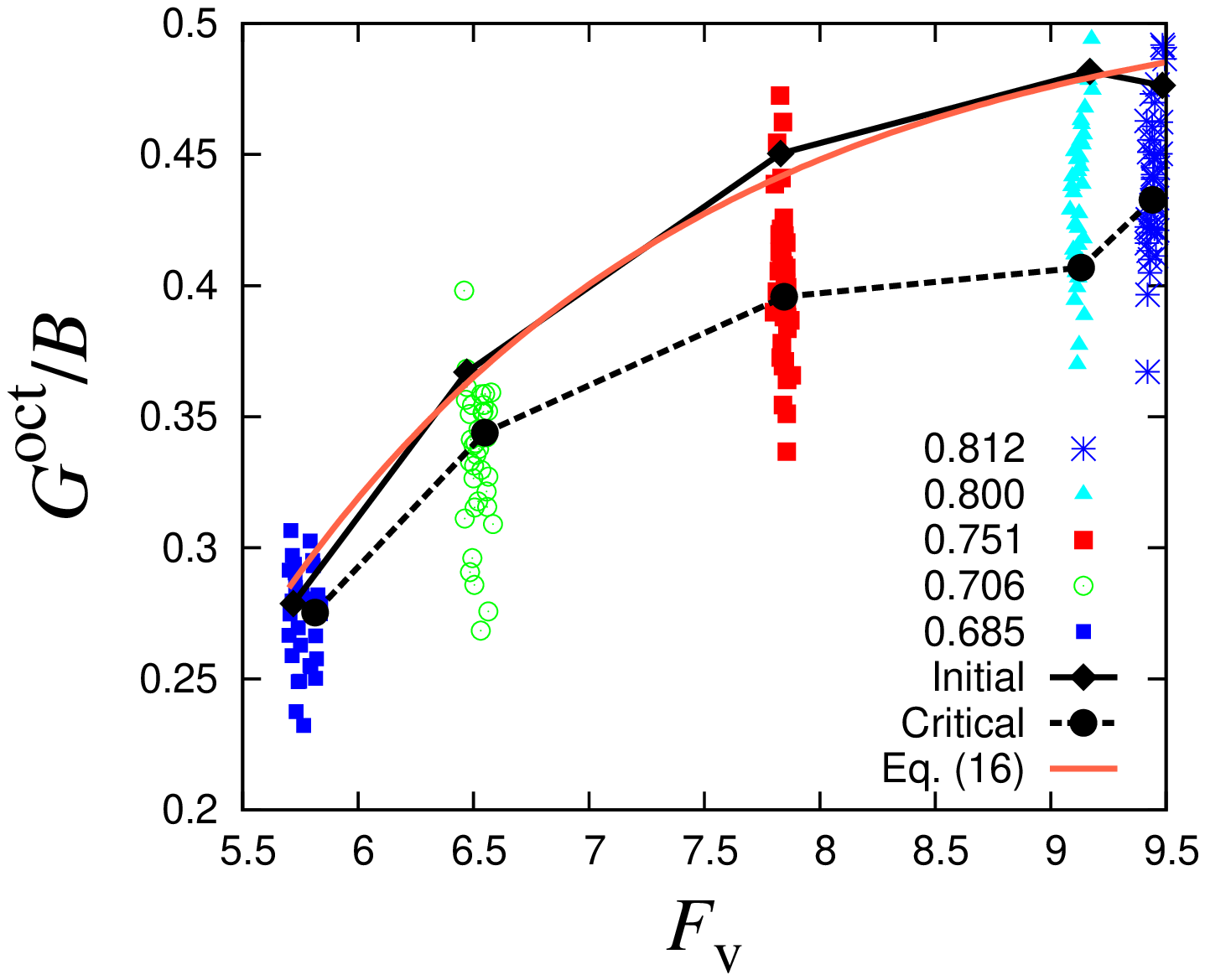}\label{shear_fricless_scaleratio_3}}
\subfigure[]{\includegraphics[scale=0.5,angle=270]{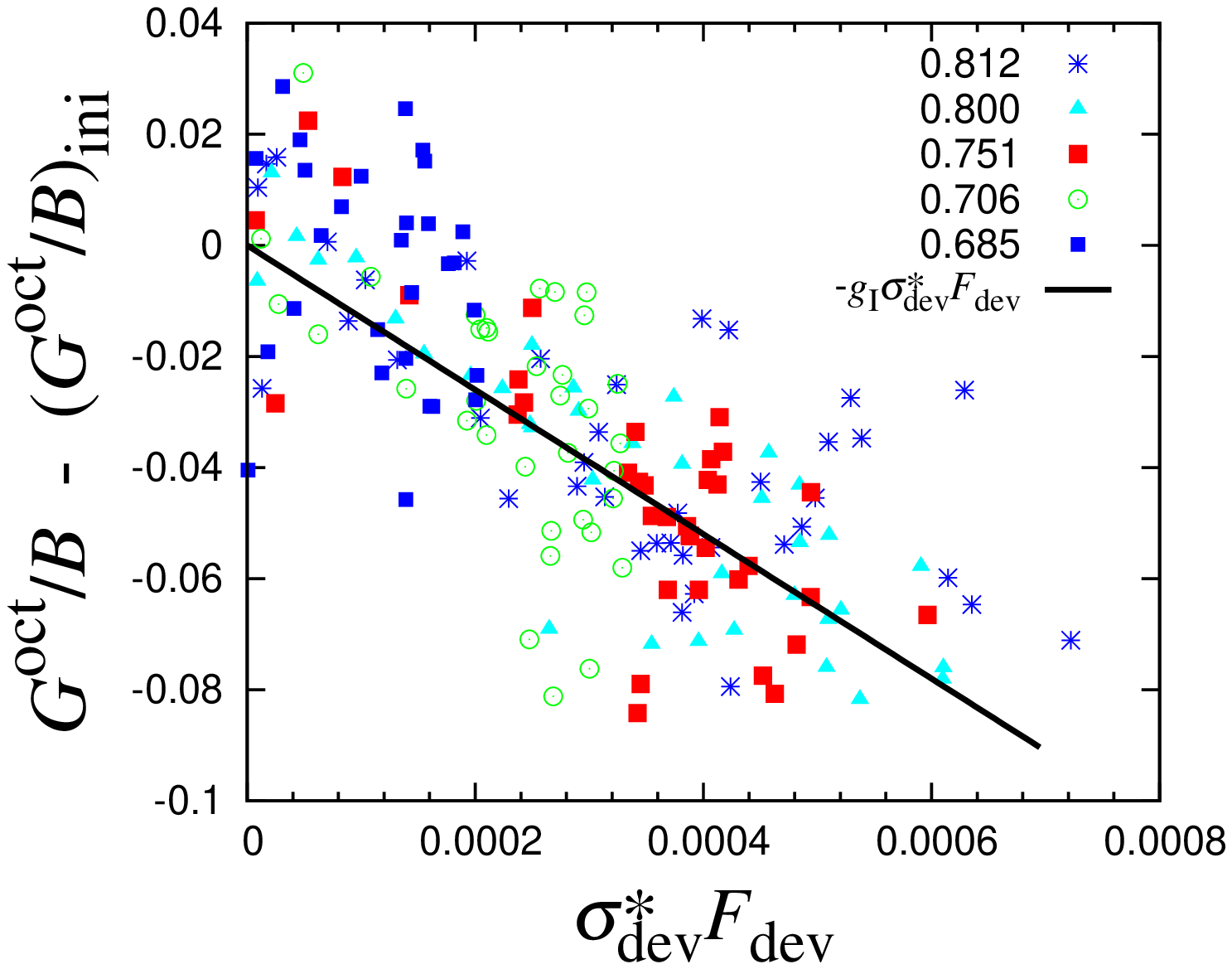}\label{shear_fricless_scaleratio_4}}
\caption{ (a) Change in shear stress $\delta \sigmadkstar$ versus strain amplitude during purely deviatoric perturbations $\delta \epsd$ 
for different states, with volume fraction $\nu = 0.706$, along the main path as shown in the inset. 
(b) Evolution of octahedral shear modulus $\Goct$ along the main deviatoric path $\epsd$ for five different volume fractions as shown in the inset. 
The corresponding lines passing through the data represents Eq.\ (\ref{eq:G}).
(c) Evolution of ratio of octahedral shear modulus and bulk modulus, i.e.,\ $\Goct/B$ with isotropic fabric $\Fv$, 
together with the averaged values at the initial (near isotropic state averaged over shear strain $\epsd \le 0.0065$) and the steady state (averaged dataset in the steady state), as given in the legend. 
Note that the difference between initial and steady state increases with denser systems. 
The solid orange line passing through the isotropic dataset represents Eq.\ (\ref{eq:iso_GB}).
(d) Evolution of the ratio of octahedral shear modulus and bulk modulus when its initial value, i.e.,\ $\Goct/B - \left(\Goct/B\right)_\mathrm{ini}$ is subtracted,
plotted using Eq.\ (\ref{eq:G}), for five different volume fractions as shown in the inset.}
\label{shear_all}
\end{figure}

In Fig.\ \ref{dtaudepsdev}, we plot the shear stress response $\delta \sigmadkstar$ of the material 
when the initial configurations in Fig.\ \ref{dPdepsv} are subjected to small deviatoric perturbations $\delta \epsd$. 
The octahedral shear modulus $\Goct$ is then measured, 
as defined in Eq.\ (\ref{eqnarray:moduli}). 
The slope of the curves for different initial configurations slightly decreases 
with the deviatoric state of the system, 
and saturates for high deformation $\epsd$, when the steady state is reached.
Fig.\ \ref{shear} shows the variation of $\Goct$ against shear strain $\epsd$. 
$\Goct$ starts from a finite value in the initial configuration, related to the 
isotropic contact network, and slightly decreases with increasing strain,
with different rates for different volume fractions. 
The behavior of $\Goct$ differs from that observed 
for the bulk modulus in the inset of Fig.\ \ref{bulkplot}: 
the shear resistance consistently decreases with shear strain
and no transition between initial decrease/increase is observed,
meaning that a factor other than $\Fv$ influences the change of $\Goct$ during the deviatoric path.
Similarly to what done for $A_{1}$ and $A_{2}$, we look at the ratio of the shear modulus with the bulk modulus $\Goct/B$ plotted against the isotropic fabric $\Fv$ in Fig.\ \ref{shear_fricless_scaleratio_3}. The ratio 
increases with increasing $\Fv$, with higher values in the initial state than in the steady state (data are averaged over shear strain $\epsd \le 0.0065$
to get the initial value and in the steady state to get the final one). 
The isotropic ratio $\left(\Goct/B\right)_\mathrm{ini}$ increases with $\Fv$, following the power law:
\begin{equation}
\label{eq:iso_GB}
\left(\Goct/B\right)_\mathrm{ini} = \left(\Goct/B\right)_\mathrm{max}\left[ 1- \exp{\left(\frac{\Fv-\Fv^\mathrm{min}}{\Fv^\mathrm{\alpha}}\right)} \right], 
\end{equation}
where $\left(\Goct/B\right)_\mathrm{max}\sim0.51$ represents the maximum value of ratio $\Goct/B$ for large $\Fv$ (or volume fraction),
$\Fv^\mathrm{min} \sim 4.2$ is the volumetric fabric at the jamming transition, presented in section\ \ref{sec:bulksection},
$\Fv^\mathrm{\alpha} \sim 1.9$ is the rate of growth of $\left(\Goct/B\right)_\mathrm{ini}$, when the numerical data is extrapolated to the jamming transition, where $\left(\Goct/B\right)_\mathrm{ini}=0$. 
This is in agreement with previous studies that find an upper limit equal $0.5$ for the ratio between the shear and bulk moduli \cite{kruyt2010micromechanical,somfai2007critical, ellenbroek2009jammed, magnanimo2008characterizing}.
In the limit of high $\Fv$, the granular assembly becomes highly coordinated and practically follows the affine approximation that predicts a constant value 
for the ratio $\Goct/B$ \cite{walton1987effective}. 
Here, a qualitatively similar behavior is observed for the values in the steady state, approaching a saturation ratio lower than the isotropic one.

Next, in Fig.\ \ref{shear_fricless_scaleratio_4}, we subtract the initial value $\left(\Goct/B\right)_\mathrm{ini}$
from $\Goct/B$ and assume that $\Fv$ does not change during the deviatoric deformation.
Interesting, we find that in this case the deviatoric microstructure alone is not able 
to capture the variation of the modulus along the shear path, 
but both stress $\pmb{\sigma}$ and fabric $\mathbf {F}$ seem to influence the incremental shear response,
in agreement with findings in \cite{zhao2013unique}. 
We relate the decrease of $\Goct$ to the deviatoric components of stress and fabric via:
\begin{equation}
\label{eq:G}
\Goct = \frac{\delta{\sigmadkstar}}{\delta{\epsd}}\bigg|_{\delta\epsiso=0} = B\left[ \left(\frac{\Goct}{B}\right)_\mathrm{ini} -\CGo \sigmadkstar\Fdev \right]. 
\end{equation}
where $\sigmadkstar$ is the non-dimensional shear stress, $\Fdev$ is the deviatoric fabric 
and $\CGo \approx86$ is a fit parameter reported in Table \ref{moduli_param}. 
Two contributions of the fabric to the shear stiffness can be recognized -- 
isotropic and deviatoric. The overall contribution is multiplicative proportional to $B$, due to the isotropic contact network, changing very little with deviatoric strain. 
In the bracket, the first term gives the resistance of the material in the initial isotropic configuration, whereas the second part only depends on 
the deviatoric (state) variables and characterizes the evolution of the shear modulus 
with deviatoric strain. That is, given the initial isotropic configuration, the corresponding $\Goct$ is known \cite{digby1981effective, walton1987effective, magnanimo2008characterizing}; 
on the other hand, the deviation from isotropic to anisotropic network of such configuration uniquely defines the reduction in the shear stiffness. 
The joint invariant of deviatoric stress and fabric $\sigmadkstar\Fdev$ as proposed in \cite{thornton2010quasistatic,zhao2013unique}, 
able to capture the evolution of the ratio of the elastic moduli along the whole undrained path, not only in the steady state, as seen in Fig.\ \ref{shear_fricless_scaleratio_4}.
\footnote[7]{Such a split between isotropic and deviatoric fabric influence applies to this specific deformation path, where the volume is conserved. 
Additional terms may enter when non volume-conserving deformation paths are considered. 
A very similar behavior is observed when the definition in Eq.\ (\ref{eq:scaledFabricdefn}) is employed for the deviatoric fabric. }
No more relation with volumetric quantities needs to be considered, 
as the evolution of $\sigmadkstar\Fdev$ depends on the volume fraction of the sample $\nu_i$. 

Note that when $\Goct$ is plotted against Eq.\ (\ref{eq:G}) in Fig.\ \ref{shear_fricless_scaleratio_3}, a deviation from the fitting law is observed for each 
volume fraction, showing that extra correction terms might be needed for a more accurate description. 
This is neglected in this preliminary work. 
It is interesting to point out that the isotropic fabric has different effects in case of the anisotropy moduli $A_1, A_2$ and $\Goct$, 
as in the former two cases $\Fv$, through $B$, is multiplied to $\Fdev$ and contributes to the growth of the moduli from zero to the asymptotic values,
while in the latter case $\Fv$ defines mostly the initial values of $\Goct$ via the bulk modulus, but does not affect the further decrease.

In the next section, we use the evolution equation for the fabric as predicted from Eq.\ (\ref{eq:Fveqn}), 
and the relations between the elastic moduli and the stress and fabric, 
to predict an independent deformation experiment, namely the cyclic shear deformation, i.e.,\ reverse shear after a large deviatoric strain.

\section{Prediction of undrained cyclic shear test}
\label{sec:predict}

In this section, the constitutive model is presented, involving the elastic moduli measured and calibrated in section\ \ref{sec:perturb}, and the plastic response of the material 
under large strain. The model is then used to predict the material response under cyclic shear, involving reversal.

\begin{figure}[htb]
\centering
\subfigure[]{\includegraphics[scale=0.48,angle=270]{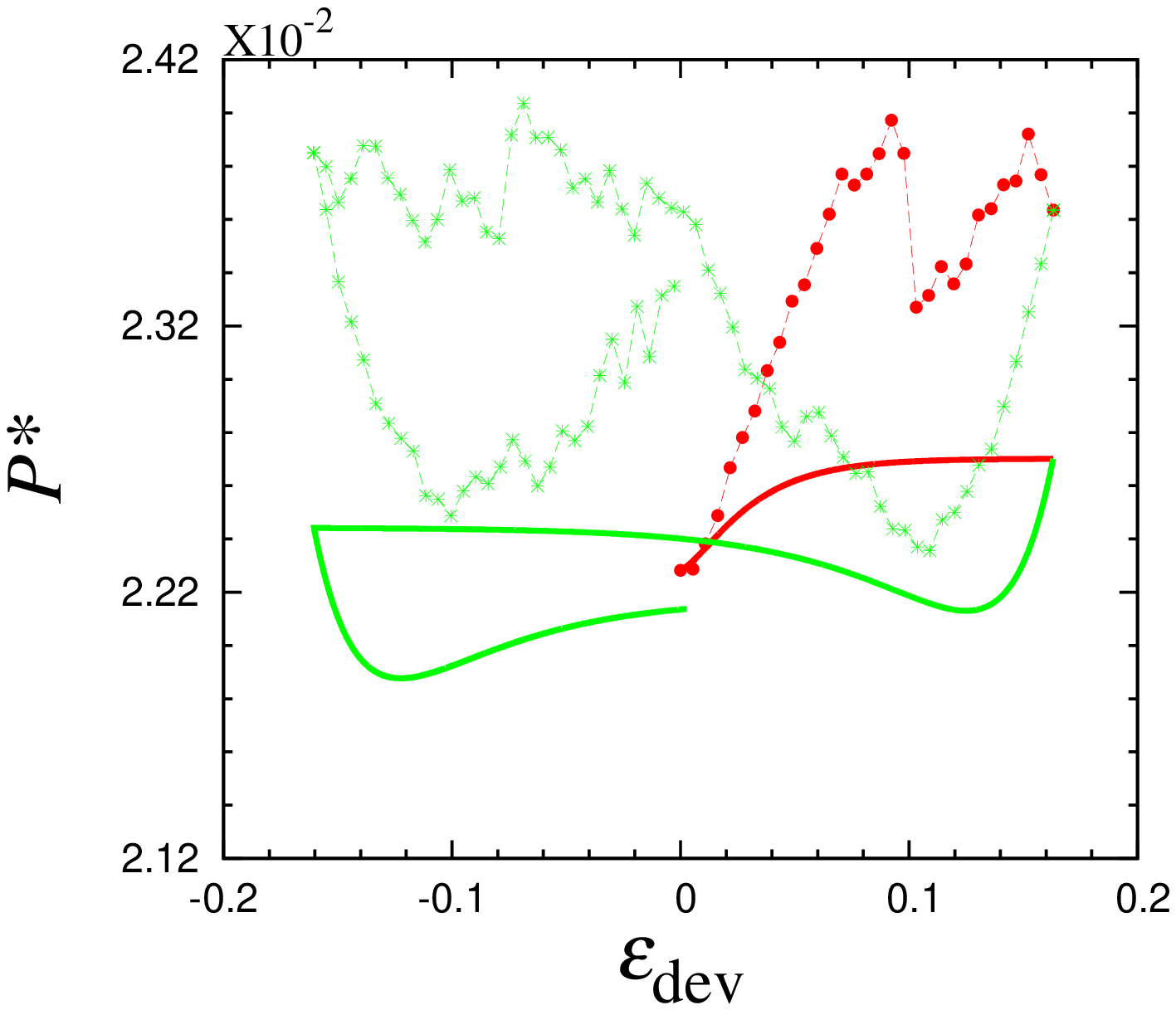}\label{Pcycle}}
\subfigure[]{\includegraphics[scale=0.48,angle=270]{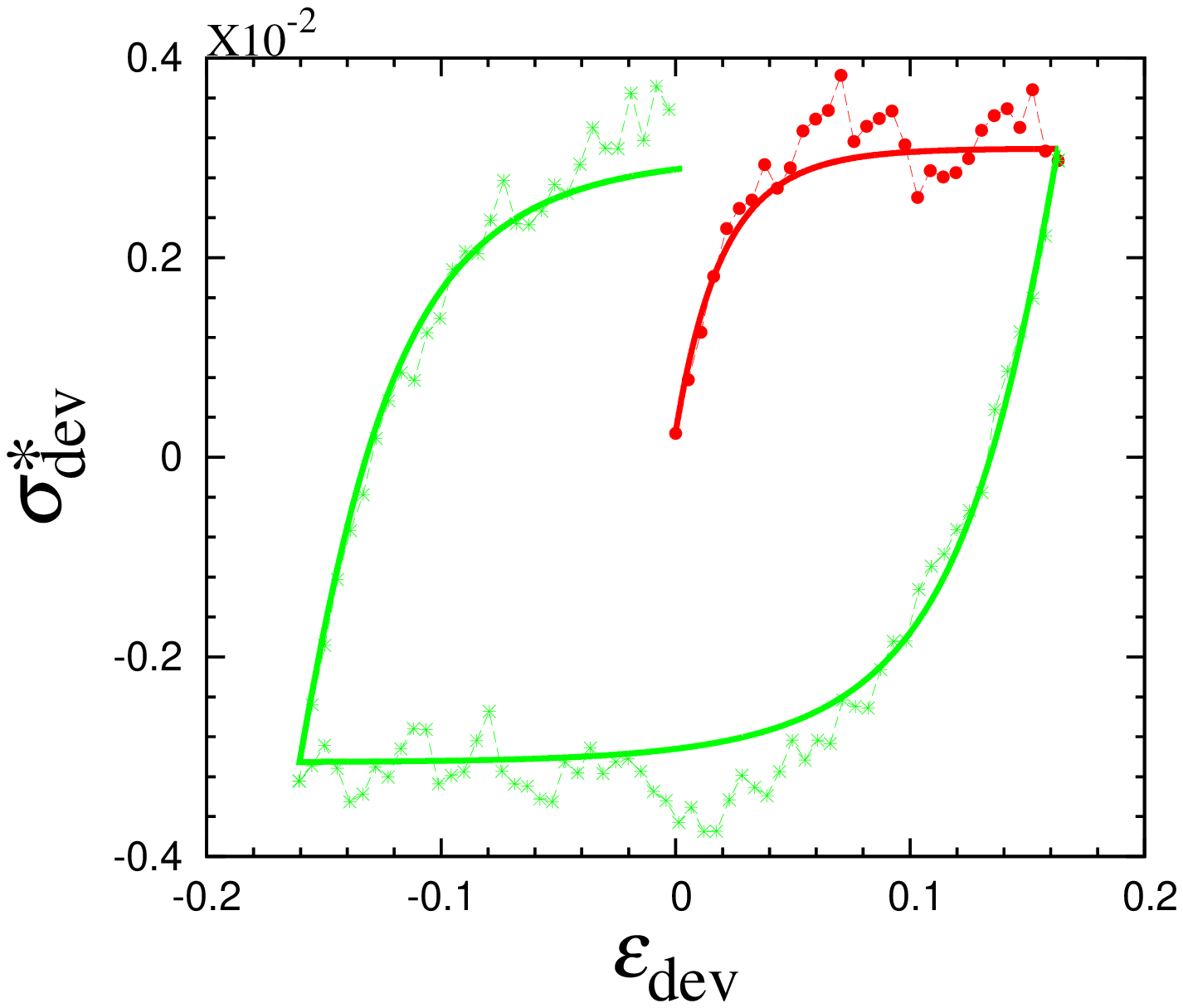}\label{taucycle}}\\
\subfigure[]{\includegraphics[scale=0.48,angle=270]{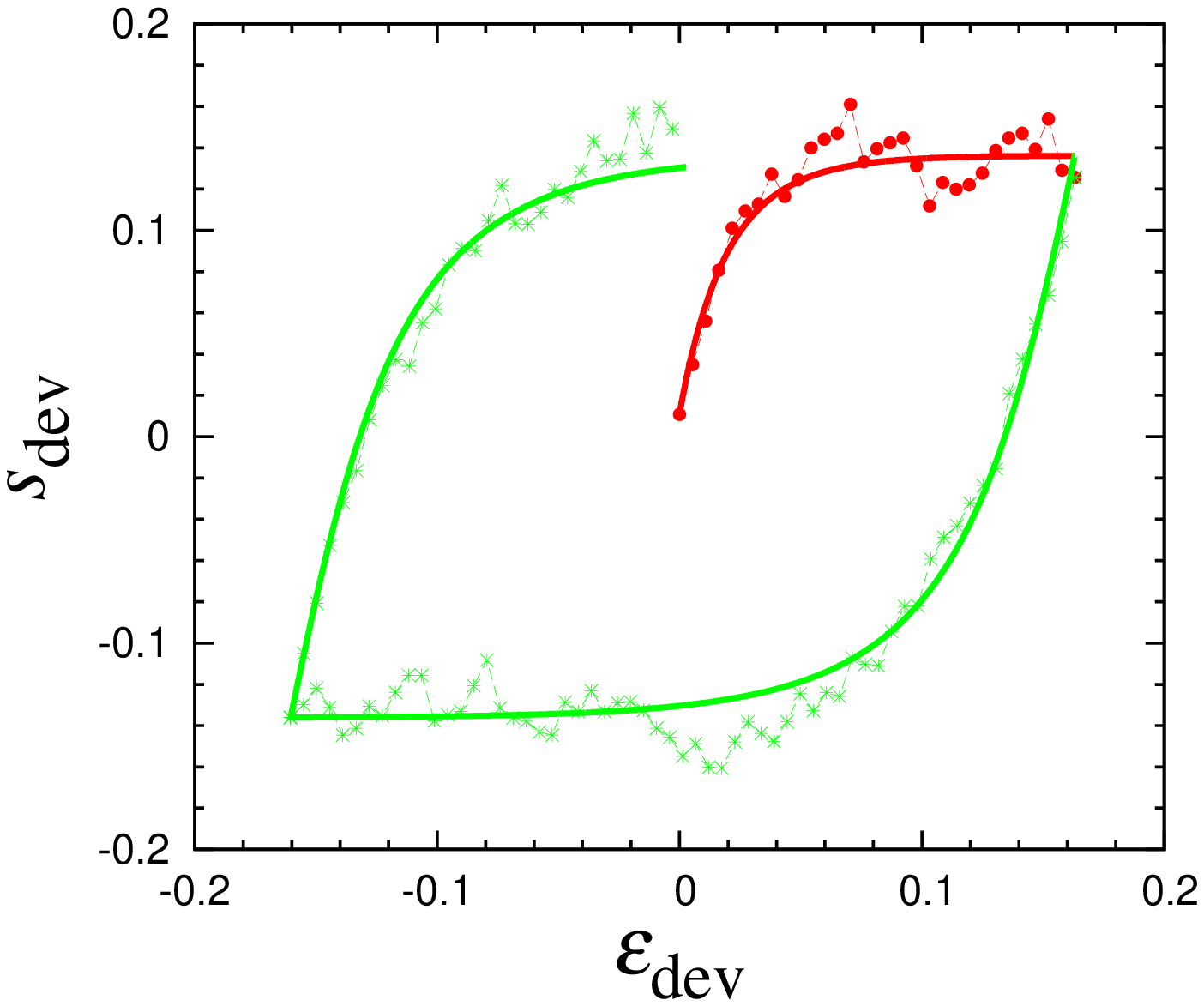}\label{sdevcycle}}
\subfigure[]{\includegraphics[scale=0.48,angle=270]{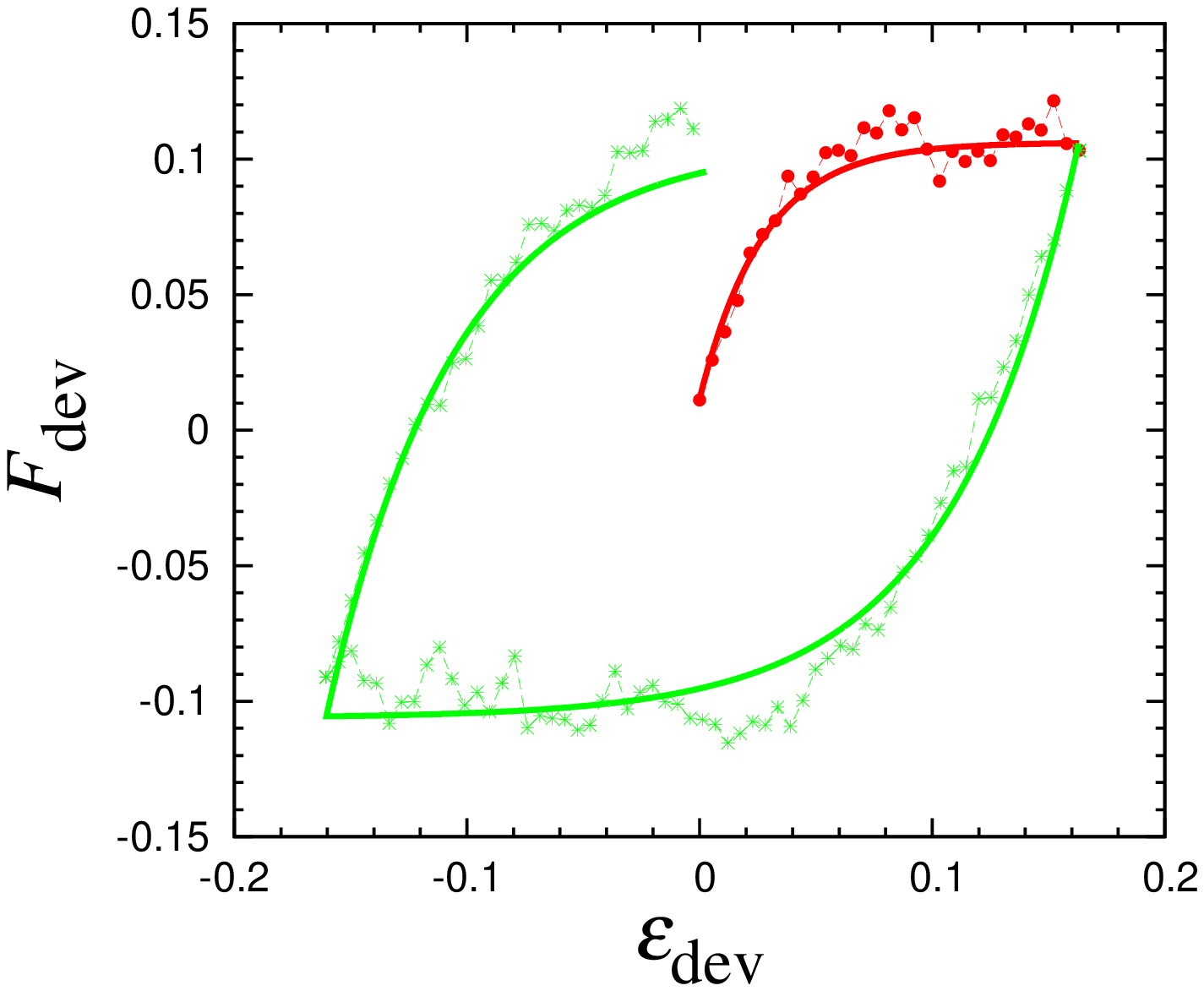}\label{fdevcycle}}
\caption{Evolution of (a) pressure $\pkstar$, (b) stress $\sigmadkstar$, (c) normalized stress $\sd$, and (d) deviatoric fabric $\Fdev$ with shear strain $\epsd$ 
during cyclic shear at constant volume $\nu=0.711$, starting from an initial isotropic configuration. 
The values of $\sdmax$, $\Fdmax$ and $\betaF$ for $\nu=0.711$ are $0.167$, $0.124$ and $40.04$ respectively, taken directly from the relations proposed in Refs.\ \cite{imole2013hydrostatic, kumar2013effect}. 
The red `$\bullet$' data points are the DEM simulation data over which the calibration of moduli was done, while the green `$*$' data points represents unloading (reversal) and re-loading.
The solid line is the prediction of the DEM observations using Eqs.\ (\ref{eqnarray:delevolutionprop}).}
\label{cyclic}
\end{figure}

\subsection{Calibration: Constitutive Model with Anisotropy}
We introduce here a constitutive model as proposed in Refs.\ \cite{luding2011local,magnanimo2011local, kumar2013effect, imole2013hydrostatic, krijgsman2013cyclic}, extended to three dimensions, 
that takes into account the evolution of fabric, independently of stress: 
\begin{eqnarray}
\delta{\pkstar} &=&   B 3\delta{\epsiso} + AS_{\sigma} \delta{\epsd}     \nonumber  , \\
 \delta{\sigmadkstar}  &=&  A 3\delta{\epsiso} +  G^{\rm oct}S_{\sigma}\delta{\epsd}, \nonumber \\
\delta{\Fdev}   &=& \betaF \mathrm{sign}\left( \delta \epsd \right) \Fdmax S_F  \delta{\epsd}.
\label{eqnarray:delevolutionprop}
\end{eqnarray}
In its simple, reduced form, the model involves only three moduli $B$, $A$ and $\Goct$, defined in the previous section in Eqs.\ (\ref{eq:B}) - (\ref{eq:G}). 
Due to $A$, the model provides a cross coupling between the two types of stress and strain 
in the model, namely the isotropic stress $\pkstar$ and shear stress $\sigmadkstar$ reacting to both
isotropic ($\epsiso$) and deviatoric ($\epsd$) strains. $\Fdev$ evolves differently from stress, 
as the rate of change with deviatoric strain can be (and in many cases is) different than the respective rate for the shear stress evolution.
Note that additional terms (cross coupling of fabric with strain) 
might be needed for the incremental evolution of $\delta{\Fdev}$ in Eqs.\ (\ref{eqnarray:delevolutionprop}), 
due to the observations from Fig.\ \ref{dFdevdepsvB_longgg}, where $\Fdev$ and $\Fv$ change also with $\epsiso$ and $\epsd$, respectively. 
However, 
both cross terms
appear to be more activated in the highly
anisotropic state, with values of the out-of-plane fabric considerably smaller
than out-of-plane stress - but this has to be confirmed by other deformation
paths also, i.e., we claim that some features are related to the
specific deformation path proposed here. 
If a dependence between stiffness and fabric 
similar to what proposed in Eq.(\ref{eq:A_1}) and (\ref{eq:A_2}) is assumed, previous arguments also lead to the conclusion
that the out-of-plane stiffness terms developing during plane strain and neglected in Eq.(\ref{eqnarray:hypo}) must be small 
compared to $B, A$ and $G$.
As a conclusion, for the sake of simplicity, 
both evolution of cross-coupling fabric terms and out-of-plane stiffness are neglected in the present work, 
and postponed to future investigations, 
for the description of arbitrary deformation paths.
The use of non-frictional particles is another possible reason
for the simplest model to work astonishingly well - so the general
model is expected to show all contributions for arbitrary deformation,
in the presence of friction.

$S_{\sigma} = S/S_{\sigma}^I$, with $S = (1- \sd/\sdmax)$ is a measure of the stress isotropy with normalized shear stress ratio $\sd=\sigmadkstar/\pkstar$, and 
$S_{\sigma}^I$ is the initial stress isotropy at the start of a new deformation direction and/or after relaxation. $1-S_{\sigma}$ is the measure for the probability of plastic events. 
Similarly, $S_F = (1- \Fdev/\Fdmax) /S_F^I $ is the fabric isotropy, and $S_F^I$ is the initial fabric isotropy at the start of a new deformation direction and/or after relaxation. 
$\sdmax$ and $\Fdmax$ represent the maximum (saturation) 
values of normalized shear stress ratio $\sd$ and deviatoric fabric $\Fdev$, respectively, 
and $\betaF$ is the rate of change in $\Fdev$ at smaller strains (as shown in Fig.\ \ref{fdev}).

%

It is worthwhile to point out that the definitions of $S_{\sigma}$ and $S_F$ are different to those used in Refs.\ \cite{luding2011local,magnanimo2011local}, as both $S_\sigma$ and $S_F$ are now 
scaled by the initial reference value and can take values between 0 and 1.
Due to $S_\sigma$ and $S_F$, the incremental response of the material is purely elastic, after relaxation or at strain reversal, 
with the elastic moduli evolving, 
as given by Eqs.\ (\ref{eq:B}) -- (\ref{eq:G}), as functions of the momentary stress and structure states. 
At reversal, the probability for plastic deformation drops to zero
and plastic events -- as related to the approach to steady state --
only occur after relatively large strain, that is the reversal stiffness is not affected.
Due to $S_\sigma$ and $S_F$, the incremental response of the material in the large-strain steady state ($S=0$) becomes elastic ($S=1$), 
just when the strain is reverted, or after relaxation (which is allowed before the probes).
Due to the dependence of the elastic moduli on the stress/fabric state, the model involves non-linear elasticity
in its present form (without contact non-linearities), 
while plasticity due to rearrangements is entirely associated to $S_\sigma$. On the other hand, 
the equation that describes the evolution of fabric is ``purely plastic'', as there is no change in fabric ($\delta \Fv=0$), in the elastic regime,
when no contact opening/closing and no multi-particle rearrangement 
happens.\footnote[8]{We want to point out here the difference between the non-linear elasticity built up along the main deviatoric path and the incremental elasticity, related to the small perturbations.
Lets select two states A-B along the deviatoric path as indicated by points in Fig.\ \ref{schematic}, the incremental measured elastic response (moduli) is different between states 
A and B as it depends on stress and fabric, that is the stiffness matrix in Eq.\ (\ref{eqnarray:hypo}), 
varies non-linearly with $\epsd$. 
On the other hand, when the incremental strain $\delta\epsd$ is applied to each state (e.g.,\ A or B), the incremental response is linearly elastic (by definition of incremental)
and becomes plastic for high $\delta\epsd$, as rearrangements happen and the moduli in that given state go from elastic to plastic.}
Thus the rate $\beta_F$ is associated to changes of structure 
with deviatoric (shear) strain amplitude (not rate); changes are 
becoming more and more probably in the steady state.
%
%

Now, we can predict an independent experiment, by using Eqs.\ (\ref{eqnarray:delevolutionprop}), 
and the relations for the four moduli $B$, $A$ and $\Goct$ with microscopic quantities 
given by Eqs.\ (\ref{eq:B}) -- (\ref{eq:G}) with the numerical scaling factors from Table \ref{moduli_param} (starting from $B$, we can calculate the other moduli using the ratio). 
Moreover, four other parameters $\sdmax$, $\Fdmax$ and $\betaF$ are needed to fully solve the coupled Eqs.\ (\ref{eqnarray:delevolutionprop}). 
The dependence of $\sdmax$, $\Fdmax$ and $\betaF$ on volume fraction $\nu$, 
is well described by the exponential decay relation proposed in Refs.\ \cite{imole2013hydrostatic, kumar2013effect}, 
where constant values, as given in Fig.\ \ref{cyclic} are used, as the volume is conserved during the cyclic shear test, as discussed next.
\subsection{Prediction: (Undrained) cyclic shear test}
We choose an initial isotropic configuration, with volume fraction  $\nu = 0.711$ and apply deviatoric (volume conserving) shear for one cycle:
loading, unloading and final re-loading, to recover the initial box configuration. 
Fig.\ \ref{cyclic} shows the evolution of pressure $\pkstar$, 
shear stress $\sigmadkstar$, shear stress ratio $\sd$ and deviatoric fabric $\Fdev$ with deviatoric shear strain $\epsd$
for one cycle, compared with the prediction using Eqs.\ (\ref{eqnarray:delevolutionprop}). 
Since the initial configuration is isotropic, the shear stress 
$\sigmadkstar$ and $\Fdev$ start from zero and approach saturation values (with fluctuations) at large strains. 
During reversal, both drop with a soft response from their respective saturation value and decrease with unloading strain, crossing their 
zero values at different strain levels, and finally reach their steady state with negative signs. 
This supports the need of independent descriptions for the evolution of stress and fabric. Finally, re-loading is applied to reach the initial box configuration. 

The qualitative behavior of pressure $\pkstar$ is similar in simulations and model, 
going from a finite initial value to saturation with much less pronounced variations, since the deformation path is volume conserving. 
It is also interesting that the final state after the complete cycle, which corresponds to the initial box configuration, is highly 
anisotropic (non-zero stress $\sigmadkstar$ and deviatoric fabric $\Fdev$). 

Both, the shear stress $\sigmadkstar$ and deviatoric fabric $\Fdev$, as well as their soft responses during strain reversal are well predicted by the model. 
$\pkstar$ increases during loading $\epsd$ by $\sim9\%$ and saturates at large strains. After reversal, 
$\pkstar$ drops because of opening and release of contacts and then increases again with unloading strain. 
Although $\pkstar$ is not quantitatively predicted by Eqs.\ (\ref{eqnarray:delevolutionprop}), the qualitative behavior is captured by the model, 
which requires a correction as proposed by Krijgsman and Luding \cite{krijgsman2013cyclic}. 
The concept of a history dependent jamming point, introduced by Kumar \textit{et al.\ }\cite{kumar2014memory}, is capable of capturing the behavior of $\pkstar$ quantitatively, however, this goes beyond the scope of this study.

Eqs.\ (\ref{eqnarray:delevolutionprop}) provide a set of equations able to 
describe the volumetric/deviatoric behavior of a granular assembly, in terms of  
stress and fabric. 
Once the initial state and the deformation path are defined, 
the evolution of isotropic fabric can be determined (using the coordination number and the fraction of rattlers) along the deformation path. 
The knowledge of isotropic and deviatoric fabric and the incremental relations in Eqs.\ (\ref{eqnarray:delevolutionprop}) 
allow for the definitions of the moduli at each incremental step. 
Given also the probabilities for the plastic events ($1-S_\sigma$ and $1-S_F$), the coupled system can be solved. 
That is, the characterization of the initial state is the information needed to fully 
describe the behavior of the material along a general deformation path, defined in terms of strain, 
since the incremental evolution equations for both stress and structure are given.

In the case of granular matter, the concept of a (homogeneous) material point in a continuum model is 
debated and many studies have been devoted to the introduction of a length scale in the constitutive model, 
starting from the Cosserat brothers, see \cite{lakes1995experimental, maugin2010mechanics, cosserat1909theorie} among others. 
Here we limit ourselves and state that a finite-size system is always needed, 
in order to calibrate any continuum model. That is, any model interpretation works only between the upper/lower bounds of infinite system 
and particle scale. When a finite-size system is considered an elastic range can always be detected, such that rearrangements 
happen (see section\ \ref{sec:howsmall}) with negligible(tiny) probability for very small strain, 
and an elasto-plastic framework could then make sense. Here, we introduce a local rate-type model in Eqs.\ (\ref{eqnarray:delevolutionprop}), 
and identify elasticity as the unique initial, static, configuration, 
from which the (incrementally irreversible) evolution of stress and structure follows. 
Our choice is to reduce elasticity to a ``punctual range'', as plastic deformations (which include irreversible opening/closing of contacts by large scale rearrangements) will dominate for large deformations.
Dynamics and kinetic fluctuations, leading to relaxation, are not considered here, but also needs to be taken into account, see e.g.,\ \cite{jiang2009granular}.

\section{Summary and Outlook}

In a triaxial box, the four elastic moduli that describe the incremental, elastic constitutive 
behavior of an anisotropic granular material in terms of volumetric/deviatoric components, namely 
the bulk modulus $B$, the two anisotropic moduli $A_1, A_2$ and the octahedral shear modulus 
$\Goct$, can be measured by applying small strain
perturbations to relaxed states that previously experienced a large strain, volume conserving (undrained) shear path.
A connection between the macroscopic elastic response and the micromechanics
is established, by considering both stress and fabric tensors, $\pmb{\sigma}$ and $\mathbf F$, respectively. 
While the bulk modulus $B$ depends on the isotropic contact network $\Fv$, 
the deviatoric component of the fabric tensor $\Fdev$ is the fundamental state variable needed to properly 
model the ratios between the (cross-coupling) anisotropic and bulk moduli. 
When the deviatoric stress and strain are appropriately scaled
(normalized), we find that the moduli reduce to three relevant ones,
i.e. $A=A_1=A_2=a F_{\rm dev} B$.
The anisotropy moduli are related 
to both deviatoric and isotropic fabric, as the whole contact network 
determines how the system will react to a perturbation. 
Surprisingly, when the shear resistance $\Goct$ is considered, 
both the contact network and the deviatoric stress 
determine the incremental behavior of the assembly. 
When the initial response is subtracted, the residual ratio $\Goct/B - \left(\Goct/B\right)_\mathrm{ini}$ 
scales with the deviatoric state of the system, through the product $\sigmadkstar\Fdev$.
For strain amplitude larger than $10^{-4}$, rearrangements in the sample take place and the behavior deviates from elastic (reversible). 
The effect of increasing amplitude of isotropic/deviatoric strain perturbations on isotropic/deviatoric stress and fabric is investigated, in the case of nearly isotropic states and steady states at various different densities.
For very small strain, the initial (linear) elastic regime, visible in the stress response, is associated to zero change in fabric.
For higher strain amplitude applied to nearly isotropic state, plasticity comes into the play, and the incremental stress-strain relation deviates from linear as soon as the contact network changes. 
In the case of steady state, deviatoric strain can only induce fluctuations around the saturation value for both stress and fabric. 
Large volumetric strain induce substantial modifications, 
as the sample previously subjected only to volume-conserving deformation, experiences now large volume changes.
In the limit of large strain, the tangential moduli of the stress-strain and fabric-strain curves (see Fig.\ \ref{schematic}) are recovered.
The relation between particle rearrangements and macro-scale plasticity is a present object of investigation, as well as the transition between local/global plastic regimes. 
As first important result, thanks to the independent study on elasticity, 
our study provides a new way to indirectly characterize the granular structure. 
Once the moduli in a given isotropic/anisotropic configuration, 
have been measured through wave propagation experiments, they can be uniquely associated with the internal fabric. 
However, we do not expect the proportionality
to remain constant for different materials.

As further step, a simple constitutive model is introduced that involves anisotropy, as proposed
in Refs.\ \cite{luding2011local, magnanimo2011local}. The non-linear elastic behavior is established and 
the irreversible/ plastic contribution is introduced via empirical probabilities for plastic events, 
that require more research and theoretical support. 
The dependence of the model parameters on volume fraction and polydispersity
has been analyzed in previous extensive work \cite{imole2013hydrostatic, kumar2013effect}. 
Here, by using the new relations for the elastic moduli, 
we are able to integrate the increments at each state along 
a generic deformation path. Hence 
we can predict the evolution of pressure, shear stress and fabric for large strain, and also at and after reversal. 
The method is first calibrated and then applied to a volume conserving (undrained) shear cycle. 
When the prediction 
is compared with numerical simulations, quantitative agreement is found for the deviatoric 
field variables. 
The most notable feature of soft but different reversal
responses of shear stress and fabric is well captured; the pressure response amplitude is underestimated by the present model.

This study concerns a seemingly unrealistic material of spheres without friction and interacting with linear contact forces to exclude effects that are due to contact non-linearity, friction and/or non-sphericity. 
This allows to unravel the peculiar interplay of stress with microstructure. 
However, the work should be extended to more realistic cases involving particle shape, friction, and non-linear contact behavior.
We expect that friction will not completely change qualitatively the observed relations between stiffness and fabric state, but possibly will add new effects to be explored in the future; 
the deviatoric fabric and the moduli are expected to change quantitatively when tangential forces are included.
On the other hand, non-linearity at contacts will introduce an extra pressure-dependence
for the moduli, as already shown by many authors (see e.g. \cite{digby1981effective, walton1987effective, chang1993micromechanical, magnanimo2008characterizing}
in the case of Hertzian interactions). Speculating about the effects of shape goes beyond the scope of this study.
A similar analysis is already in progress to check the influence of polydispersity on the 
relation between elastic stiffness and microstructure, as polydispersity affects 
the contact network, the structure, and the orientation of contacts \cite{goncu2010constitutive, goncu2013effect, kumar2013effect}.

Future work will focus on the extension of our small perturbation approach to elasto-plasticity, 
by using concepts like e.g.\ the Gudehus response envelope \cite{gudehus1979comparison, masin2012asymptotic}. 
Other theoretical approaches involve ideas proposed by Einav \cite{einav2012unification}, or by Jiang and Liu \cite{jiang2009granular}, 
for which our results can provide a microscopically based calibration of parameters, but details are not discussed here. 
The information obtained for the pure elastic range can then be used to decouple the plastic contribution, associated with rearrangements, 
and to study the flow rule. 
The validation of the present analysis with experimental data is another important goal. Nevertheless the issue of measuring
fabric from laboratory experiments is far from solved, even though big advances have been made in recent years using photoelasticity, and 
microtomography CT-scans \cite{hall2010discrete, sperl2006experiments, jorjadze2013microscopic, bi2011jamming}. 
A partial validation is anyway possible when measuring the residual dependence of the elastic response from 
variables other than stress and porosity \cite{ezaoui2009experimental}, by means of acoustic measurements \cite{khidas2012probing}. 
The behavior after more than one cycle deserves further investigation, 
from both simulational and theoretical points of view, to detect features like creep, liquefaction and 
ratcheting, analyzed in preliminary works \cite{magnanimo2011local} with constant elastic moduli and for many cycles \cite{kumar2014memory}.
Finally, a general tensor formulation that
allows for highly different orientations of strain rate, stress and fabric is an open issue
but can be inspired by the works of Thornton \cite{thornton2010quasistatic} and Zhao \& Guo \cite{zhao2013unique}.

\section*{Acknowledgement}
We thank Itai Einav (University of Sydney) and Mario Liu (University of T\"{u}bingen) for scientific discussions. 
Critical comments and reviews from O. I. Imole, M. B. Wojtkowski, F. G\"{o}nc\"{u}, J. Ooi and M. Ramaioli are gratefully acknowledged. 
This work is financially supported by the European Union funded Marie Curie Initial Training Network, FP7 (ITN-238577), see {http://www.pardem.eu/} for more information.

\bibliographystyle{abbrv}

\end{document}